\documentclass[11pt, prd]{revtex4}

\usepackage{amsmath}
\usepackage{amssymb}
\usepackage[dvips]{graphicx,color}

\begin{document}

\title{
Resurgence in sine-Gordon quantum mechanics: \\
Exact agreement between multi-instantons and uniform WKB
}

\author{Tatsuhiro Misumi}
\email{misumi(at)phys.akita-u.ac.jp}
\affiliation{Department of Mathematical Science, Akita University, 
1-1 Tegata Gakuen-machi, Akita 010-8502, Japan\\
Research and Education Center for Natural Sciences, 
Keio University, 4-1-1 Hiyoshi, Yokohama, Kanagawa 223-8521, Japan
}

\author{Muneto Nitta}
\email{nitta(at)phys-h.keio.ac.jp}
\affiliation{Department of Physics, and Research and Education 
Center for Natural Sciences, 
Keio University, 4-1-1 Hiyoshi, Yokohama, Kanagawa 223-8521, Japan}

\author{Norisuke Sakai}
\email{norisuke.sakai(at)gmail.com}
\affiliation{Department of Physics, and Research and Education 
Center for Natural Sciences, 
Keio University, 4-1-1 Hiyoshi, Yokohama, Kanagawa 223-8521, Japan}

\begin{abstract}
We compute multi-instanton amplitudes in the sine-Gordon quantum 
mechanics (periodic cosine potential) by integrating out 
quasi-moduli parameters corresponding to separations of 
instantons and anti-instantons.
We propose an extension of Bogomolnyi--Zinn-Justin prescription 
for multi-instanton configurations and an appropriate subtraction 
scheme. 
We obtain the multi-instanton contributions 
to the energy eigenvalue of the lowest band at the zeroth order 
of the coupling constant. 
For the configurations with only instantons (anti-instantons), 
we obtain unambiguous results. 
For those with both instantons and anti-instantons, we obtain 
results with imaginary parts, which depend on the path of 
analytic continuation. 
We show that the imaginary parts of the multi-instanton amplitudes 
precisely cancel the imaginary parts of the Borel resummation 
of the perturbation series,
and verify that our results completely agree with those based 
on the uniform-WKB calculations,
thus confirming the resurgence : 
divergent perturbation series combined with the nonperturbative 
multi-instanton contributions conspire to give unambiguous results. 
We also study the neutral bion contributions in the 
${\mathbb C}P^{N-1}$ model on ${\mathbb R}^1\times S^{1}$ 
with a small circumference, taking account of the relative 
phase moduli between the fractional instanton and anti-instanton.
We find that the sign of the interaction potential depends on 
the relative phase moduli, and that both the real and imaginary parts resulting 
from quasi-moduli integral of the neutral bion get quantitative 
corrections compared to the sine-Gordon quantum mechanics. 
\end{abstract}

\maketitle

\newpage


\section{Introduction}
\label{sec:Intro}

In the recent study on quantum field theories and quantum mechanics, 
topologically neutral soliton molecules, 
which are locally composed of (fractional) instantons and anti-instantons,
have been attracting a great deal of attention in relation to 
the IR-renormalon\cite{Unsal:2007vu, Unsal:2007jx, Shifman:2008ja, 
Poppitz:2009uq, Anber:2011de, Poppitz:2012sw,Argyres:2012vv, 
Argyres:2012ka, Dunne:2012ae, Dunne:2012zk, Dabrowski:2013kba, 
Dunne:2013ada, Cherman:2013yfa, Basar:2013eka, Dunne:2014bca, 
Cherman:2014ofa, Behtash:2015kna, Bolognesi:2013tya, Misumi:2014jua, Misumi:2014raa, 
Misumi:2014bsa, Nitta:2015tua,Nitta:2014vpa,Shermer:2014wxa, 
Dunne:2015ywa}. 
Imaginary ambiguities arising in amplitudes of such topologically neutral configurations 
can cancel out those arising in non-Borel-summable perturbative 
series (IR-renormalon) in quantum theories under certain conditions 
on the spacetime manifold
\cite{Argyres:2012vv, Argyres:2012ka, Dunne:2012ae, Dunne:2012zk, 
Dabrowski:2013kba, Dunne:2013ada, Cherman:2013yfa, Basar:2013eka, 
Dunne:2014bca, Cherman:2014ofa,Bolognesi:2013tya,Misumi:2014jua,  
Misumi:2014raa, Misumi:2014bsa, 'tHooft:1977am, Fateev:1994ai, Fateev:1994dp}.
In field theories on compactified spacetime with a small compact dimension, 
these objects are termed as ``bions"
\cite{Argyres:2012vv, Argyres:2012ka, Dunne:2012ae, Dunne:2012zk}.
It is expected that full semi-classical expansion including perturbative and non-perturbative sectors as bions, which is called ``resurgent" expansion
\cite{Ec1, Marino:2007te, Marino:2008ya, Marino:2008vx, Pasquetti:2009jg, Drukker:2010nc,
Aniceto:2011nu, Marino:2012zq, Hatsuda:2013gj, Schiappa:2013opa, Hatsuda:2013oxa,
Aniceto:2013fka, Santamaria:2013rua, Kallen:2013qla, Honda:2014ica,
Grassi:2014cla, Sauzin, Kallen:2014lsa,
Couso-Santamaria:2014iia, Honda:2014bza, Aniceto:2014hoa, 
Couso-Santamaria:2015wga, Honda:2015ewa, Hatsuda:2015owa,
Aniceto:2015rua, Dorigoni:2015dha}, leads to unambiguous and self-consistent definition 
of field theories in the same manner as the conjecture in quantum mechanics \cite{Bogomolny:1980ur, 
ZinnJustin:1981dx, ZinnJustin:1982td, ZinnJustin:1983nr,ZinnJustin:2004ib, ZinnJustin:2004cg, 
Jentschura:2010zza}.

The resurgence in theoretical physics was at first investigated
in the matrix model and topological string theory
\cite{Marino:2007te, Marino:2008ya, Marino:2008vx, Pasquetti:2009jg, Marino:2012zq, 
Hatsuda:2013oxa}.
Then, the study on the topic has been extended to ABJM theory
\cite{Drukker:2010nc, Hatsuda:2013gj, Honda:2014ica, Kallen:2014lsa},
string and supersymmetric gauge theories
\cite{Aniceto:2011nu, Grassi:2014cla, Couso-Santamaria:2014iia, Aniceto:2014hoa},
and general quantum systems
\cite{Schiappa:2013opa, Aniceto:2013fka, Santamaria:2013rua, Kallen:2013qla,
Sauzin, Honda:2014bza, Couso-Santamaria:2015wga, Honda:2015ewa,
Hatsuda:2015owa, Aniceto:2015rua, Dorigoni:2015dha}.
Bions and resurgence in non-SUSY field theories, especially in the low-dimensional models, 
have been extensively investigated for the ${\mathbb C}P^{N-1}$ model
\cite{Dunne:2012ae, Dunne:2012zk, Dabrowski:2013kba,
Bolognesi:2013tya, Misumi:2014jua,Shermer:2014wxa}, 
the Grassmann sigma model \cite{Misumi:2014bsa, Dunne:2015ywa}, 
the principal chiral model 
\cite{Cherman:2013yfa,Cherman:2014ofa,Nitta:2015tua}, 
and the $O(N)$ model \cite{Nitta:2014vpa, Dunne:2015ywa}.
According to these studies, the leading-order renormalon ambiguity 
$\mp i\pi e^{-2S_{I}/N}$ arising in non-Borel-summable 
perturbative series, which corresponds to the 
singularity closest 
to the origin on the Borel plane, 
is compensated by the amplitude of neutral bions. 
On the other hand, it is expected but not verified that the 
ambiguities corresponding to singularities further from the 
origin ($\mp i\pi e^{-4S_{I}/N}$, $\mp i\pi e^{-6S_{I}/N}$,...) 
are cancelled by amplitudes of bion molecules with more than 
four instanton constituents.

In the case of quantum mechanics, not only the sector of zero 
instanton charge but also those of nonzero instanton charge 
contribute to physical observables such as the energy levels. 
The authors in Refs.~\cite{ZinnJustin:1981dx, ZinnJustin:1982td, 
ZinnJustin:1983nr,ZinnJustin:2004ib, ZinnJustin:2004cg, 
Jentschura:2010zza} investigated quantum mechanics with several 
types of potential including the sine-Gordon type.
They showed that the leading instanton contributions are consistent 
with the perturbative calculation,
and conjectured the explicit equation connecting perturbative 
and instanton contributions, 
which they call the generalized quantization condition.
Recently the authors in Refs.~\cite{Dunne:2013ada, Dunne:2014bca} 
adopted the uniform-WKB method based on the boundary condition, 
which is equivalent to the quantization condition in 
Refs.~\cite{ZinnJustin:1981dx, ZinnJustin:1982td, 
ZinnJustin:1983nr, ZinnJustin:2004ib, ZinnJustin:2004cg, 
Jentschura:2010zza}, and pointed out the general relation 
between perturbative and non-perturbative contributions.
Explicit calculations of multi-instanton amplitudes at each 
configuration level are expected to clarify the structure of 
resurgence and to verify the conjectured relation between 
perturbative and non-perturbative contributions 
\cite{ZinnJustin:1981dx, ZinnJustin:1982td, ZinnJustin:1983nr, 
ZinnJustin:2004ib, ZinnJustin:2004cg, Jentschura:2010zza,
Escobar-Ruiz:2015nsa, Dunne:2013ada, Dunne:2014bca}.

In this paper, we focus on a quantum mechanical system with the 
sine-Gordon potential, and we calculate the multi-instanton 
amplitude by explicitly integrating quasi moduli parameters 
corresponding to separations of instanton-constituents in a 
semi-classical limit, in comparison with the uniform WKB calculations 
\cite{Dunne:2013ada, Dunne:2014bca, ZinnJustin:2004ib, ZinnJustin:2004cg}.
We adopt an extension of Bogomolnyi--Zinn-Justin prescription 
\cite{Bogomolny:1980ur, ZinnJustin:1981dx} for multi-instanton 
configurations with an appropriate subtraction scheme for divergent parts. 
We calculate contributions to the energy eigenvalue of the 
lowest band from each multi-instanton configuration in a semi-classical 
limit ($|g^{2}|\ll 1$).
For the configurations with only instantons (anti-instantons) 
such as 
$[\mathcal{I}\mathcal{I}]$, 
$[\mathcal{I}\mathcal{I}\mathcal{I}]$ and 
$[\mathcal{I}\mathcal{I}\mathcal{I}\mathcal{I}]$,
we have unambiguous results without imaginary parts. 
Here, we have denoted an instanton (anti-instanton) as 
$\mathcal{I}$ ($\bar{\mathcal{I}}$).
For configurations containing both instantons and anti-instantons 
such as  $[\mathcal{I}\bar{\mathcal{I}}]$, 
$[\mathcal{I}\mathcal{I}\bar{\mathcal{I}}]$ 
and $[\mathcal{I}\mathcal{I}\bar{\mathcal{I}}\bar{\mathcal{I}}]$, 
the results contain ambiguous imaginary parts, which depend 
on the path of analytic continuation. 
These imaginary parts correspond to the large-order behavior of 
perturbation series around the saddle point without the 
$[\mathcal{I}\bar{\mathcal{I}}]$ pair. 
For instance, we show explicitly that the imaginary part of 
the multi-instanton amplitude $[\mathcal{I}\mathcal{I}\bar{\mathcal{I}}]$ 
cancels the imaginary part of the Borel resummation 
of the large-order perturbation series around the nontrivial background 
with a single instanton $[\mathcal{I}]$. 
By investigating the uniform-WKB calculations in detail, we 
verify that all of our results agree completely 
with those based on the uniform-WKB calculations up to a four-instanton order.

While the sine-Gordon 
quantum mechanics is worth to study on its own, 
another strong motivation lies in its close relationship to 
small circumference limit of the two-dimensional 
${\mathbb C}P^{N-1}$ model on ${\mathbb R}^1\times S^{1}$ 
(circumference $L$) with 
the ${\mathbb Z}_N$-symmetric twisted boundary condition 
\cite{Dunne:2012ae, Dunne:2012zk}. 
However, we observe that some of the field configurations of 
the ${\mathbb C}P^{N-1}$ model are not faithfully represented 
by means of the sine-Gordon quantum mechanics. 
One can derive the sine-Gordon quantum mechanics from the 
two-dimensional ${\mathbb C}P^{N-1}$ model by applying 
the Scherk-Schwarz dimensional reduction, which requires 
a particular dependence of the phases of fields on the coordinate 
$x_2$ of compactified dimension ($0\le x_2 < L$).  
It is important to realize that only parts of field configuration 
of ${\mathbb C}P^{N-1}$ model can be consistent with this $x_2$ 
dependence. 
For instance, the BPS solution of two fractional instantons 
is not consistent with the Scherk-Schwarz reduction, and hence 
its small circumference limit cannot be described by the 
sine-Gordon quantum mechanics. 
On the other hand, two adjacent instantons in the sine-Gordon 
quantum mechanics are mutually non-BPS, although each individual 
instanton may be understood as a limit of BPS fractional 
instanton (with a different $x_2$ dependence). 
Even in the instanton and anti-instanton configurations, 
 ${\mathbb C}P^{N-1}$ model has a significant difference 
compared to the sine-Gordon quantum mechanics: 
The phase moduli of the fractional instantons in the 
${\mathbb C}P^{N-1}$ model are neglected in the sine-Gordon 
quantum mechanics. 
For the configuration of a neutral bion composed of a fractional instanton and an 
anti-fractional instanton,
we find that the interaction between them strongly depends on the 
relative phase of constituents.
We calculate the neutral bion contribution 
in the ${\mathbb C}P^{N-1}$ model, based on the interaction 
potential with the quasi moduli parameter corresponding 
to the relative phase between the fractional instanton and 
anti-instanton. 
We find that this calculation gives a correction factor compared 
to the neutral bion amplitude obtained in the sine-Gordon 
quantum mechanics \cite{Dunne:2012ae, Dunne:2012zk}.

This paper is organized as follows. 
In Sec.~\ref{sec:SG}, we review instantons and their interactions 
in the quantum mechanics with sine-Gordon potential and the 
Borel summation.
In Sec.~\ref{sec:IC} we calculate amplitudes of multi-instanton 
configurations in sine-Gordon quantum mechanics by integrating 
out the moduli parameters. 
In Sec.~\ref{sec:WKB} we discuss the results from the uniform 
WKB calculations, and show that they completely agree with the 
instanton moduli calculations. 
In Sec.~\ref{sec:CPN} we discuss the neutral bion contributions 
in the compactified ${\mathbb C}P^{N-1}$ model based on the 
interaction potential including the relative phase parameter.
Section \ref{sec:SD} is devoted to a summary and discussion.
In Appendix~\ref{app1} we give some details of four-instanton 
calculations. 



\section{Quantum mechanics with the sine-Gordon potential}
\label{sec:SG}

In this article, we focus on the sine-Gordon quantum mechanics
described by the Schr\"odinger equation 
\begin{equation}
H\psi(x)=-{1\over{2}}{d^{2}\over{dx^{2}}} \psi(x)\,+\, 
{1\over{8g^{2}}}\sin^{2} (2gx)\,\psi(x)\,=\, E\,\psi(x)\,,
\label{Seq1}
\end{equation}
where we follow the notation in Refs.~\cite{ZinnJustin:2004ib, Dunne:2013ada} 
except $g$ is replaced by $g^{2}$ here 
\footnote{In Ref.~\cite{Dunne:2014bca}, 
a different convention 
for the Schr\"odinger equation is adopted, but appears to be 
mixed up with those in Refs.~\cite{ZinnJustin:2004ib, Dunne:2013ada} 
and ours. 
Thus we follow the notation in 
Refs.~\cite{ZinnJustin:2004ib, Dunne:2013ada} in this article
for consistency.}.
The
Euclidian Lagrangian for the sine-Gordon 
quantum mechanics is given by \footnote{
We can compare our convention with that in Ref.~\cite{Manton:2004tk}: 
their coordinate variable is $\phi=4gx$, and their Euclidean 
Lagrangian is $L_{M}=16g^2L$. 
}
\begin{equation}
L\,=\, {1\over{2}}\left({dx\over{dt}}\right)^{2} \,+\, V(x), 
\qquad 
V(x)={1\over{8g^{2}}} \sin^{2} (2gx) \,. 
\label{eq:lagrangian}
\end{equation}
In the $g^{2}\to 0$ limit, it reduces to the Schr\"odinger 
equation of the harmonic oscillator.

The energy eigenvalues of periodic potentials split 
into bands of states. 
Within each band, they are labeled by the Bloch angle 
$\theta\in[0, \pi]$ defined by 
\begin{equation}
\psi\left(x+\frac{\pi}{2g}\right)\,=\, e^{i\theta}\psi(x)\,.
\label{eq:bloch-angle}
\end{equation}
In this article, we are interested in the lowest band, 
although excited bands can be treated similarly. 
The energy eigenvalue $E$ of the lowest band can be expressed 
in terms of the path-integral 
\begin{equation}
E 
= \lim_{\beta\to\infty}\frac{-1}{\beta}{\rm Tr} e^{-\beta H}
= \lim_{\beta\to\infty}\frac{-1}{\beta}
\int_{x(t=-\beta/2)=x(t=\beta/2)} Dx(t) \,
 e^{-S+iQ\theta}.
\label{eq:path-integral}
\end{equation}
For weak coupling, the path-integral has contributions 
$E_{\rm pert}(g^2)$ around the perturbative vacuum, 
as well as contributions $\triangle E$ from nonperturbative 
saddle points 
\begin{equation}
E 
= E_{\rm pert}(g^2)+ \triangle E .
\label{eq:pert-nonpert}
\end{equation}

Perturbation series in powers of coupling constant $g^2$ in 
quantum field theories or in quantum mechanics are extremely 
useful, but are usually factorially divergent 
\begin{equation}
E_{\rm pert}(g^2) 
= \sum_{K=0}^{\infty} a_K (g^2)^K, 
\qquad 
a_K \sim K! \; .
\label{eq:perturbation-series}
\end{equation}
It is useful to define the Borel transform $B_{\rm pert}(t)$  
\begin{equation}
B_{\rm pert}(t) 
= \sum_{K=0}^{\infty} \frac{a_K}{K!} t^K. 
\label{eq:borel-transform}
\end{equation}
The Borel resummation ${\mathbb E}_{\rm pert}(g^2)$ of the 
divergent series $E_{\rm pert}(g^2)$ is defined as an integral 
of the Borel transform along 
the positive real axis in the complex Borel plane $t$ 
\begin{equation}
{\mathbb E}_{\rm pert}(g^2) 
= \int_0^\infty {dt} 
e^{-t} B_{\rm pert}(g^2t). 
\label{eq:borel-resum}
\end{equation}

If the factorially divergent series is alternating, the Borel 
transform has no singularities along the positive 
real $t$ axis, and the Borel resummation becomes well-defined 
(
Borel-summable). 
For the potential with degenerate minima such as the 
sine-Gordon quantum mechanics, however, the perturbation 
series is non-alternating factorially divergent. 
In that case, the Borel transform is convergent with the 
finite radius of convergence, but the Borel resummation is 
ill-defined because of singularities in the complex Borel plane. 
Since the series become alternating and the Borel resummation 
 ${\mathbb E}_{\rm pert}(g^2)$ is unambiguous for $-g^2>0$, 
we can analytically continue it from $-g^2>0$ to the physical 
region $g^2>0$ to obtain a real analytic function 
${\mathbb E}_{\rm pert}(g^2)$. 
If there is no complex singularities, we obtain a branch cut 
along the positive real axis of complex $g^2$ plane. 
The imaginary part ${\rm Im} {\mathbb E}_{\rm pert}(g^2)$ at 
$g^2>0$ is related to the large-order behavior ($K\gg 1$) of 
perturbation series $E_{\rm pert}(g^2)$ in 
Eq.(\ref{eq:perturbation-series}) through the dispersion relation 
\cite{ZinnJustin:1989mi} 
\begin{equation} 
a_{K} \,\approx\, {-1\over{\pi}}\int_{0}^{\infty} 
\,d g^2
 \,\frac{[{\rm Im} {\mathbb E}_{\rm pert}(g^2)]}{(g^2)
^{K+1}}.
\label{eq:dispersion_rel}
\end{equation}
This large-order behavior corresponds to the singularities 
of the Borel transform $B_{\rm pert}(g^2t)$ in the complex 
Borel plane $t$. 
Of course this ambiguous (path-dependent) imaginary part 
is unacceptable, and should disappear, since the energy 
eigenvalue $E$ should be real, and ambiguity due to the choice 
of path is unphysical. 
In fact, it has been found that the leading term of the 
imaginary ambiguities is cancelled by the contributions 
from non-perturbative saddle points associated with neutral 
objects composed of instantons 
\cite{Dunne:2013ada, Dunne:2014bca}. 
This phenomenon 
is called the resurgence 
of the perturbation series.

Let us now consider non-perturbative saddle points. 
By rescaling the variable 
\begin{equation}
2gx = y, 
\label{eq:rescaled_variable}
\end{equation}
the Euclidean Lagrangian in Eq.(\ref{eq:lagrangian}) can be rewritten as 
\begin{equation}
L\,=\, {1\over{8g^2}}\left({dy\over{dt}}\right)^{2} \,+\, V, \qquad 
V={1\over{8g^{2}}} \sin^{2} (y) \,. 
\label{eq:y_lagrangian}
\end{equation}
and 
the instanton number as a topological charge may be defined by 
\begin{equation}
Q=\frac{1}{\pi}\int_{-\infty}^{\infty} dt \, \frac{dy}{dt} \, .
\label{eq:instanton-number}
\end{equation}
Single instanton solution ($Q=1$) is given by\footnote{
We take the branch $-\pi/2\le \arctan y \le \pi/2$. }
\begin{equation}
y_{\mathcal I}(t)=2\arctan e^{t-t_0}+n\pi, 
\quad n\in {\mathbb Z}, 
\label{eq:instanton}
\end{equation}
whereas single anti-instanton solution ($Q=-1$) is given by 
\begin{equation}
y_{\bar {\mathcal I}}(t)=2\arctan e^{-(t-t_0)}+(n-1)\pi, 
\quad n\in {\mathbb Z}, 
\label{eq:anti-instanton}
\end{equation}
with the Euclidean action 
\begin{equation}
S_{\mathcal I}\, = \, {1\over{2 g^{2}}}\,. 
\label{eq:instanton_action}
\end{equation}
The moduli parameter $t_0$ is a zero mode (moduli) associated to the 
breakdown of translation, representing the location of the 
(anti-)instanton. 
For even $n$, the solutions (\ref{eq:instanton}) and 
(\ref{eq:anti-instanton}) satisfy the following BPS equation\footnote{
One should note that the periodicity $\pi/g$ of the (anti-)BPS 
equation (\ref{eq:bps-bound}) and (\ref{eq:anti-bps-bound}) is 
twice as large as the periodicity 
$\pi/(2g)$ of the Lagrangian in Eq.(\ref{eq:lagrangian}). 
} 
saturating the BPS bound for $S$ 
\begin{equation}
\frac{dy}{dt}=\sin y, \quad 
S=\int_{-\infty}^{\infty} dt \frac{dy}{dt} \sin y. 
\label{eq:bps-bound}
\end{equation}
For odd $n$, they satisfy the anti-BPS equation 
saturating the anti-BPS bound for $S$ 
\begin{equation}
\frac{dy}{dt}=-\sin y, \quad 
S=-\int_{-\infty}^{\infty} dt \frac{dy}{dt} \sin y. 
\label{eq:anti-bps-bound}
\end{equation}

By integrating over the translational zero mode $t_0$, 
one finds the contribution $\triangle E
^{(1,0)}$ of single 
instanton $[\mathcal{I}]$ to the energy as 
\begin{equation}
\triangle E
^{(1,0)} = -[\mathcal{I}]= 
- \left({e^{-S_{I}}\over{\sqrt{\pi g^{2}}}}\right)\,
e^{i\theta}\,.  
\end{equation}

Suppose, for instance, we have a BPS instanton in 
Eq.(\ref{eq:instanton}) with $n=0$ and wish to 
place another instanton or anti-instanton to its right, we are 
forced to take either instanton with $n=1$ in Eq.(\ref{eq:instanton}) 
or anti-instanton with $n=1$ in Eq.(\ref{eq:anti-instanton}), 
both of which are anti-BPS configurations. 
Therefore two successive (anti-)instantons are inevitably non-BPS. 
The energy of the non-BPS configuration of two successive 
instantons should be more than the sum of individual instanton 
energies. They are found to repel each other with the 
potential~\cite{Manton:2004tk} for large separations $R\gg 1$ 
\begin{equation}
V_{\mathcal{I} \mathcal{I}}(R)\,=\, {2\over{g^{2}}} \exp[-R]\,. 
\label{eq:interaction-I-I}
\end{equation}
The non-BPS configuration of successive instanton and 
anti-instantons are found to attract each other with the 
potential for large separations $R\gg 1$  
\begin{equation}
V_{\mathcal{I} \bar{\mathcal{I}}}(R)\,=\, -{2\over{g^{2}}} \exp[-R]\,.
\label{eq:interaction-I-antiI}
\end{equation}

For later convenience, we introduce the uniform-WKB ansatz 
by following Ref.~\cite{Dunne:2013ada}.
With the coordinate variable $y$ in 
Eq.(\ref{eq:rescaled_variable}), Eq.(\ref{Seq1}) 
can be rewritten as 
\begin{equation}
-g^{4}{d^{2}\over{dy^{2}}} \psi(y)\,+\, 
{1\over{16}}\sin^{2} (y)\,\psi(y)\,=\, {g^{2}\over{2}}E\,\psi(y)\,.
\label{Seq2}
\end{equation}
We define the potential as $U(y)\equiv{1\over{16}}\sin^{2} (y)$.
By using the parabolic cylinder function $D_{\nu}(z)$ 
satisfying the differential equation 
\begin{equation}
{d^{2}\over{dz^2}}D_{\nu}(z)\,+\,\left( \nu+{1\over{2}}
-{z^{2}\over{4}} \right)D_{\nu}(z) \,=\,0, 
\end{equation}
we introduce an ansatz for the wave 
function\cite{alvarez, langer, cherry, millergood, galindo} 
\begin{equation}
\psi(y)\,=\, {D_{\nu}(u(y)/g)\over{\sqrt{u'(y)}}}\,,
\label{Dansatz}
\end{equation}
where the parameter $\nu=E-1/2$ is the shift of energy eigenvalue 
$E$ from the ground state energy of the harmonic oscillator 
($g^2=0$ limit). 
Then the Schr\"odinger equation (\ref{Seq2}) becomes 
\begin{equation}
U(y) - {1\over{4}}u^{2}(u')^{2}-{g^{2}E\over{2}}
+g^{2}\left( \nu+{1\over{2}} \right)(u')^{2}
+{g^{4}\over{2}}\sqrt{u'}\left( {u''\over{(u')^{3/2}}} \right)' = 0, 
\label{eq:nonlinear_eq_u}
\end{equation}
with $u'\equiv du/dy$. 
In the $g^{2} \to 0$ limit, Eq.(\ref{eq:nonlinear_eq_u}) just 
reduces to $4U(y) = u^{2}(u')^{2}$, whose solution $u_{0}(y)$ is 
\begin{equation}
u_{0}(y)^{2}\,=\, 4\int ^{y}_{0}\sqrt{U}dy \,=\, 
2\sin^{2}{y\over{2}} \,\,\,\,\,\,\,\,\,\to\,\,\,\,\,\,\,\,\,
u_{0}(y)\,=\,\sqrt{2}\sin{y\over{2}}\,,
\end{equation}
which gives the zeroth-order argument of the parabolic 
cylinder function in Eq.(\ref{Dansatz}) and solves the Schr\"odinger 
equation of the harmonic oscillator.



\section{Multi-instanton amplitudes in Sine-Gordon quantum mechanics}
\label{sec:IC}

\subsection{General setting}
\label{sec:general}

In this section we calculate multi-instanton amplitudes in 
sine-Gordon quantum mechanics. 
We need to integrate over the distances $R$ between instantons 
and (anti-)instantons as quasi-moduli. 
Since the interaction (\ref{eq:interaction-I-I}) and 
 (\ref{eq:interaction-I-antiI}) between instantons and 
(anti-)instantons vanish at large distances, we need to 
regulate the integral by introducing a factor $\epsilon$ 
into the effective potential
\begin{equation}
V[R] \,=\, \pm {2\over{g^{2}}}\exp(-R) \,+\, \epsilon R\,,
\label{pot2}
\end{equation}
where $+$ is for the instanton-instanton repulsive interaction 
and $-$ for the instanton--anti-instanton attractive interaction. 
The regularization parameter $\epsilon$ can be identified as the 
number $N_{\rm f}$ of fictitious fermions\cite{Bogomolny:1980ur,
ZinnJustin:1981dx,ZinnJustin:1982td,
Jentschura:2010zza,Jentschura:2011zza,Dunne:2012ae}. 
After subtracting divergences, we need to take the limit 
$\epsilon\to 0$.

Even after eliminating the divergence arising from large 
separations ($R\to\infty$), we have another 
source of divergence for the case of the attractive 
instanton--anti-instanton interaction: the integrand 
$\exp[ {2\over{g^{2}}}\exp(-R) - \epsilon R]$ becomes 
divergent as $g^2\to +0$, contrary to the repulsive case. 
Therefore the moduli-integral gets divergent contributions 
from small $R$ regions, and is ill-defined in the 
semi-classical limit ($|g^{2}|\ll 1$). 
This is why we need to introduce the Bogomolnyi--Zinn-Justin 
(BZJ) prescription \cite{Bogomolny:1980ur,ZinnJustin:1981dx}: 
We first regard $-g^{2}$ as real positive ($-g^{2}>0$) to 
make the integral well-defined in the semi-classical limit, 
and then we analytically continue $-g^{2}>0$ back to 
$g^{2}>0$ in the complex $g^{2}$ plane at the end of the 
calculation.

The energy eigenvalue of the lowest band has contributions 
 $\triangle E^{(n,m)}$ from the amplitude 
$[\mathcal{I}\cdot\cdot\cdot\bar{\mathcal{I}}]_{\rm all}$ 
of $n$-instanton and $m$-anti-instanton configuration as 
\begin{equation}
\triangle E^{(n,m)} 
\,=\, - [\mathcal{I}\mathcal{I}\cdot\cdot\cdot 
\bar{\mathcal{I}}\bar{\mathcal{I}}]_{\rm all}\,,
\end{equation}
\begin{equation}
 [\mathcal{I}\mathcal{I}\cdot\cdot\cdot \bar{\mathcal{I}}
\bar{\mathcal{I}}]_{\rm all}
\,=\, 
 \left({e^{-S_{\mathcal I}}\over{\sqrt{\pi g^{2}}}}\right)^{n+m}\,
e^{i(n-m)\theta}\,  \int dR_{1}dR_{2}...dR_{n+m-1} \, 
e^{-V[R_{1}]-V[R_{2}]-\cdot\cdot\cdot V[R_{n+m-1}]} \,,
\end{equation}
where $ [\mathcal{I}\mathcal{I}\cdot\cdot\cdot \bar{\mathcal{I}}
\bar{\mathcal{I}}]_{\rm all}$ stands for the sum of 
configurations which can be composed of $n$ instantons and $m$ 
anti-instantons in all possible orderings. 
As shown in Ref.~\cite{ZinnJustin:2004ib}, the contribution 
contains $e^{iQ\theta}$ with $Q=n-m$ being the instanton 
charge since the Bloch angle $\theta$ shows up in a 
topological term $iQ\theta$ in the Euclidian action in 
Eq.(\ref{eq:path-integral}).
We perform the quasi-moduli integral taking only interactions 
between neighboring instantons among the ($n+m-1$) 
instantons. 
We should perform this multi-integral in the 
semi-classical region $|g^{2}|\ll 1$, and subtract 
the divergent parts appropriately at each level of the 
multi-integral. 
We will evaluate them explicitly from the next subsection.

\subsection{2 instantons}

\begin{figure}[htbp]
\begin{center}
 \includegraphics[width=0.4\textwidth]{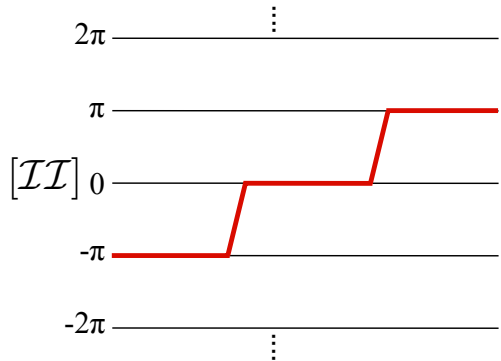}
\end{center}
\caption{A schematic figure of an example of two instanton 
configurations $[\mathcal{I}\mathcal{I}]$.
Each horizontal line stands for the vacuum in the 
sine-Gordon potential.
}
\label{fig:II}
\end{figure}
The amplitude of two instantons shown in Fig.~\ref{fig:II} 
is obtained as
\begin{align}
[\mathcal{I}\mathcal{I}]e^{-2i\theta}\xi^{-2}\,=\,
\int_{0}^{\infty}dR
\exp\left(-{2\over{g^2}}e^{- R}-\epsilon R\right)
&\,=\,
\left({g^{2}\over{2}}\right)^{\epsilon} \int_{0}^{2/g^{2}}ds\,
e^{-s} s^{\epsilon-1} 
\nonumber\\[5pt] 
&\,\overset{|g^{2}|\ll 1}{\longrightarrow}\,
\left({g^{2}\over{2}}\right)^{\epsilon}\Gamma(\epsilon) 
\nonumber\\[5pt] 
&\,=\,
-\left(\gamma\,+\,\log{2\over{g^{2}}}\right)\,+\, 
O\left(\frac{1}{\epsilon}\right)\,+\, O(\epsilon)\,,
\label{IIa}
\end{align}
where $\gamma$ is the Euler constant and $\xi$ is an instanton 
factor defined by 
\begin{equation}
\xi\equiv e^{-S_{I}}/\sqrt{\pi g^{2}} 
= e^{-1/(2g^{2})}/\sqrt{\pi g^{2}}.
\end{equation}
Here we have neglected terms of order $O(g^2)$ or higher. 
To simplify the formula, we divide the amplitude by $\xi^{2}$ 
and $e^{2i\theta}$.
Precisely speaking, the interaction energy between instantons 
at small separation $R \ll 1$ may not be precisely represented 
by the potential in Eq.(\ref{pot2}). 
However, our result is unchanged as long as $|g^{2}|\ll 1$ is 
satisfied. 
We need to subtract the divergent term $O(1/\epsilon)$ 
while the $O(\epsilon)$ term disappears in the $\epsilon \to 0$ limit.
The contribution from this amplitude to the energy eigenvalue 
of the lowest band is then given by
\begin{equation}
\triangle E^{(2,0)} \,=\, e^{2i\theta} \xi^{2}\, 
\left(\gamma\,+\,\log{2\over{g^{2}}}\right)\,,
\label{II}
\end{equation}
where the superscript $(2,0)$ stands for two-instanton and 
zero--anti-instanton amplitude.
We note that the contribution from the two anti-instanton 
amplitude is obtained by replacing $e^{2i\theta}$ by $e^{-2i\theta}$.

\subsection{1 instanton $+$ 1 anti-instanton}
\begin{figure}[htbp]
\begin{center}
 \includegraphics[width=0.4\textwidth]{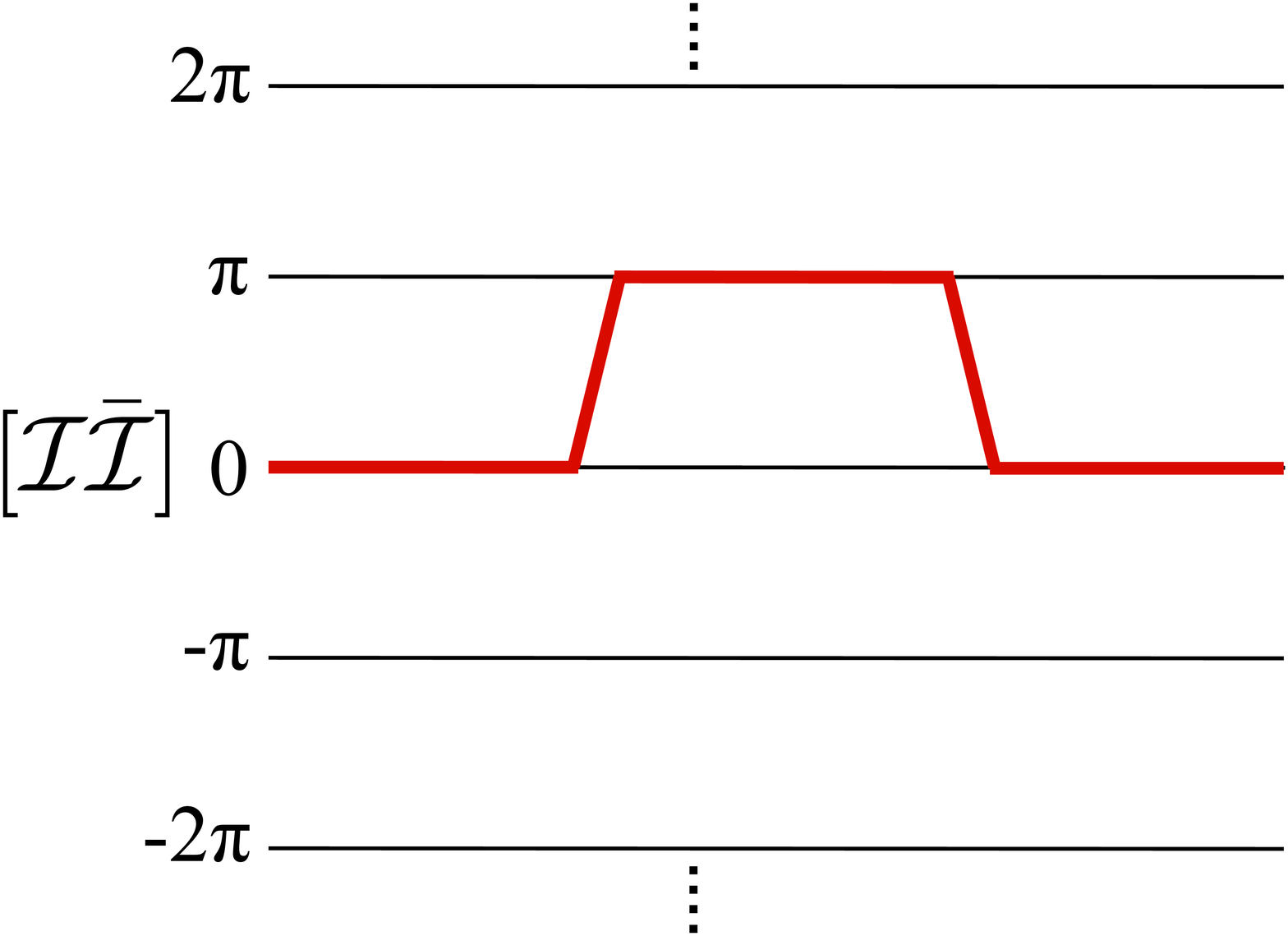}
  \includegraphics[width=0.4\textwidth]{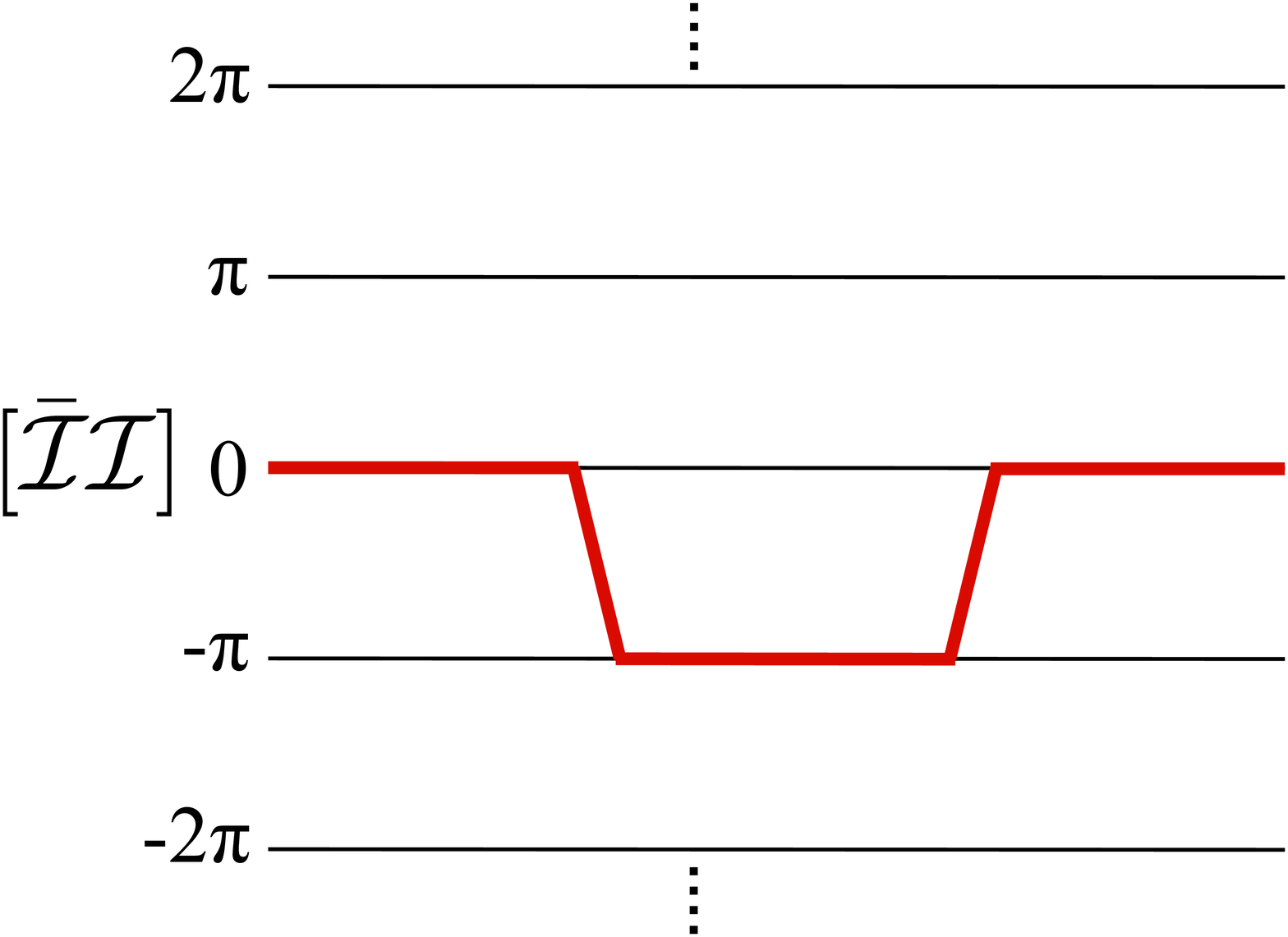}
\end{center}
\caption{A schematic figure of an example of one-instanton 
and one anti-instanton amplitude 
($[\mathcal{I}\bar{\mathcal{I}}], [\bar{\mathcal{I}}\mathcal{I}]$).
Each horizontal line stands for the vacuum in the sine-Gordon 
potential.
}
\label{fig:IIbar}
\end{figure}

The amplitude of one instanton and one anti-instanton 
amplitude is composed of two configurations 
$[\mathcal{I}\bar{\mathcal{I}}]$ and 
$[\bar{\mathcal{I}}\mathcal{I}]$, as shown in 
Fig.~\ref{fig:IIbar}.
In these cases, the interaction between the two constituents 
is attractive, and the quasi moduli integral is ill-defined. 
Therefore we introduce the Bogomolnyi--Zinn-Justin (BZJ) 
prescription \cite{Bogomolny:1980ur,ZinnJustin:1981dx}:
we first 
evaluate the integral by taking $-g^{2}>0$, and then we 
analytically continue the result from $-g^{2}>0$ back to 
$g^{2}>0$ in the complex $g^{2}$ plane. 
This procedure provides the imaginary ambiguity depending 
on the path
of the analytic continuation as 
$-g^{2}=e^{\mp i\pi}g^{2}$.

The amplitude of one-instanton and one anti-instanton 
configuration $[\mathcal{I}\bar{\mathcal{I}}]$ corresponding 
to the left of Fig.~\ref{fig:IIbar} is obtained as
\begin{align}
[\mathcal{I}\bar{\mathcal{I}}]\xi^{-2}&\,=\,
\int_{0}^{\infty}dR\,
\exp\left(-{2\over{-g^2}}e^{- R}-\epsilon R\right)
\overset{|g^{2}|\ll 1}{\longrightarrow}\,
\left({-g^{2}\over{2}}\right)^{\epsilon}\Gamma(\epsilon) 
\nonumber\\[5pt] 
&\,\overset{-g^{2}= e^{\mp i\pi}g^{2}}{\longrightarrow}\,
-\left(\gamma\,+\,\log{2\over{e^{\mp i\pi}g^{2}}}\right)  
\,+\, O\left(\frac{1}{\epsilon}\right)\,+\, O(\epsilon)
\nonumber\\[5pt] 
&\,=\,
-\left(\gamma\,+\,\log{2\over{g^{2}}}\right) \mp i\pi   
\,+\, O\left(\frac{1}{\epsilon}\right)\,+\, O(\epsilon)\,,
\label{eq:I-barI-1}
\end{align}
where we perform the integral in the first line by considering 
$-g^{2}>0$,  and in the second line analytically continue 
$-g^{2}>0$ back to $g^{2}>0$ in the complex $g^{2}$ plane 
\cite{Bogomolny:1980ur,ZinnJustin:1981dx}. 
The third line shows a two-fold ambiguous expression 
of $-g^{2}$ depending on the path of analytic continuation 
as $-g^{2}=e^{\mp i\pi}g^{2}$.
As with the two-instanton case, we have subtracted the 
divergent part $O(1/\epsilon)$ while the $O(\epsilon)$ term 
disappears in the $\epsilon \to 0$ limit.

Another amplitude of one-instanton and one anti-instanton 
configurations $[\bar{\mathcal{I}}\mathcal{I}]$ corresponding 
to the right of Fig.~\ref{fig:IIbar} turns out to give identical 
contribution as that in Eq.(\ref{eq:I-barI-1}). 
The total contribution is given by the sum of them with 
dropping $O(1/\epsilon)$ and $O(\epsilon)$ terms, 
\begin{equation}
([\mathcal{I}\bar{\mathcal{I}}]\,+\,[\bar{\mathcal{I}}\mathcal{I}])
\xi^{-2}\,=\,
-2\left(\gamma\,+\,\log{2\over{g^{2}}}\right)\,\mp\,  2i\pi\,.
\end{equation}
Its contribution to the energy eigenvalue of the lowest band 
is then given by
\begin{equation}
\triangle E^{(1,1)} \,=\,  \xi^{2}\, 
\left[ 2\left(\gamma\,+\,\log{2\over{g^{2}}}\right)\,\pm\,  
2i\pi\right]\,.
\label{IIbar}
\end{equation}

If the resurgence idea is valid, this imaginary ambiguity 
should cancel the imaginary ambiguity of the non-Borel 
summable divergent series of the perturbative contribution. 
\begin{equation}
{\rm Im}[\triangle E^{(1,1)}] +  {\rm Im}[E_{\rm pert}]=0.
\label{Imag_cancel}
\end{equation}
If we insert Eqs.(\ref{IIbar}) and (\ref{Imag_cancel}) into 
the dispersion relation in Eq.(\ref{eq:dispersion_rel}), we 
should be able to reproduce the large-order behavior of the 
perturbation series 
\begin{align} 
a_{k} &=
{-1\over{\pi}}\int_{0}^{\infty} \,d g^2 \,
{{\rm Im} [E_{\rm pert}(g^2)]\over{(g^2)^{k+1}}}
\,=\, {1\over{\pi}}\int_{0}^{\infty} \,d g^2 \,
{{\rm Im} [\triangle E^{(1,1)}]\over{(g^2)^{k+1}}}
\nonumber\\
&\,=\, -{1\over{\pi}}\int_{0}^{\infty} \,d(g^2) \,
{ 2 e^{-1/g^2}\over{(g^2)^{k+2}}}
\,=\, -{2\over{\pi}} \,k!\, \quad\quad\quad\quad\quad\,( k\geq 2)\,,
\end{align}
in accordance with the leading large-order behavior of the 
perturbation series \cite{ZinnJustin:1981dx}. 
Thus we find that the imaginary ambiguity of the instanton--anti-instanton 
amplitude correctly cancels the imaginary ambiguity of the Borel 
resummation of the (non-Borel summable) perturbation series.

\subsection{3 instantons}
\label{sec:3instanton}

\begin{figure}[htbp]
\begin{center}
 \includegraphics[width=0.4\textwidth]{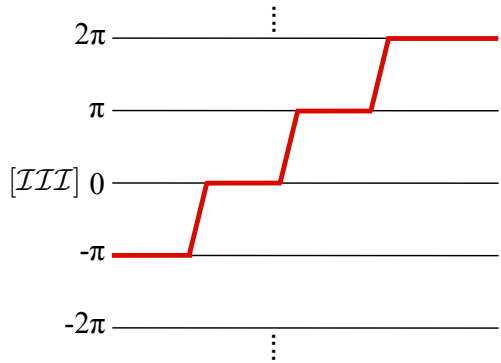}
\end{center}
\caption{A schematic figure of an example of three-instanton 
amplitudes 
($[\mathcal{I}\mathcal{I}\mathcal{I}]$).
Each horizontal line stands for the vacuum in the sine-Gordon potential.}
\label{fig:III}
\end{figure}
For the three-instanton amplitude shown in Fig.~\ref{fig:III}, 
we have two quasi modulus $R_1, R_2$ corresponding to the 
separations between adjacent instantons. 
For multiple moduli integral of each given configuration, we 
need to specify subtraction scheme explicitly, and propose the 
following: 
\begin{enumerate}
\item 
Enumerate possible ordering of quasi moduli integrations, such as 
$\int dR_1\int dR_2$ and $\int dR_2 \int dR_1$ for the three instanton 
case. 
\item
 Subtract possible poles like $1/\epsilon$ for the first integration, 
and then perform the next integration successively, and retain 
the finite piece. 
\item
 Average the results of all possible orderings. 
\end{enumerate}

Incorporating the repulsive potentials between adjacent instantons, 
we obtain the three instanton amplitude as\footnote{ 
One should note that the double integral appears formally a 
product of the single moduli integral in Eq.(\ref{IIa}), but 
our subtraction scheme specify the successive subtraction 
which gives a different result from the product of the result 
of single moduli integral in Eq.(\ref{IIa}). }
\begin{align}
[\mathcal{I} \mathcal{I}\mathcal{I} ]e^{-3i\theta}\xi^{-3}&\,=\,
\int_{0}^{\infty}dR_{1}\int_{0}^{\infty}dR_{2}
\exp\left[-{2\over{g^2}}(e^{- R_{1}}+e^{- R_{2}})-\epsilon(R_{1}+R_{2})\right]
\nonumber\\[5pt] 
&\,\overset{|g^{2}|\ll 1}{\longrightarrow}\,
\left({g^{2}\over{2}}\right)^{\epsilon}\Gamma(\epsilon)
\left[\left({g^{2}\over{2}}\right)^{\epsilon}\Gamma(\epsilon)-{1\over{\epsilon}}\right] 
\nonumber\\[5pt] 
&\,=\,
{3\over{2}}\left(\gamma+\log{2\over{g^{2}}}\right)^2 \,+ \,
{\pi^2\over{12}} \,+\, O\left(\frac{1}{\epsilon}\right)\,+\, O(\epsilon)\,.
\end{align}
In the three instanton case, we have two possible orderings of the multi integral, 
each of which gives the identical contribution. 
Then the three instanton contribution to the energy eigenvalue 
of the lowest band is given by 
\begin{equation}
\triangle E^{(3,0)} \,=\,  -e^{3i\theta}\,\xi^{3}\, 
\left[ {3\over{2}}\left(\gamma+\log{2\over{g^{2}}}\right)^2 \,
+ \,{\pi^2\over{12}}\right]\,.
\label{III}
\end{equation}
We note that the contribution from the three anti-instanton 
amplitude is obtained by replacing $e^{3i\theta}$ by $e^{-3i\theta}$.

\subsection{2 instantons $+$ 1 anti-instanton}

\begin{figure}[htbp]
\begin{center}
 \includegraphics[width=0.4\textwidth]{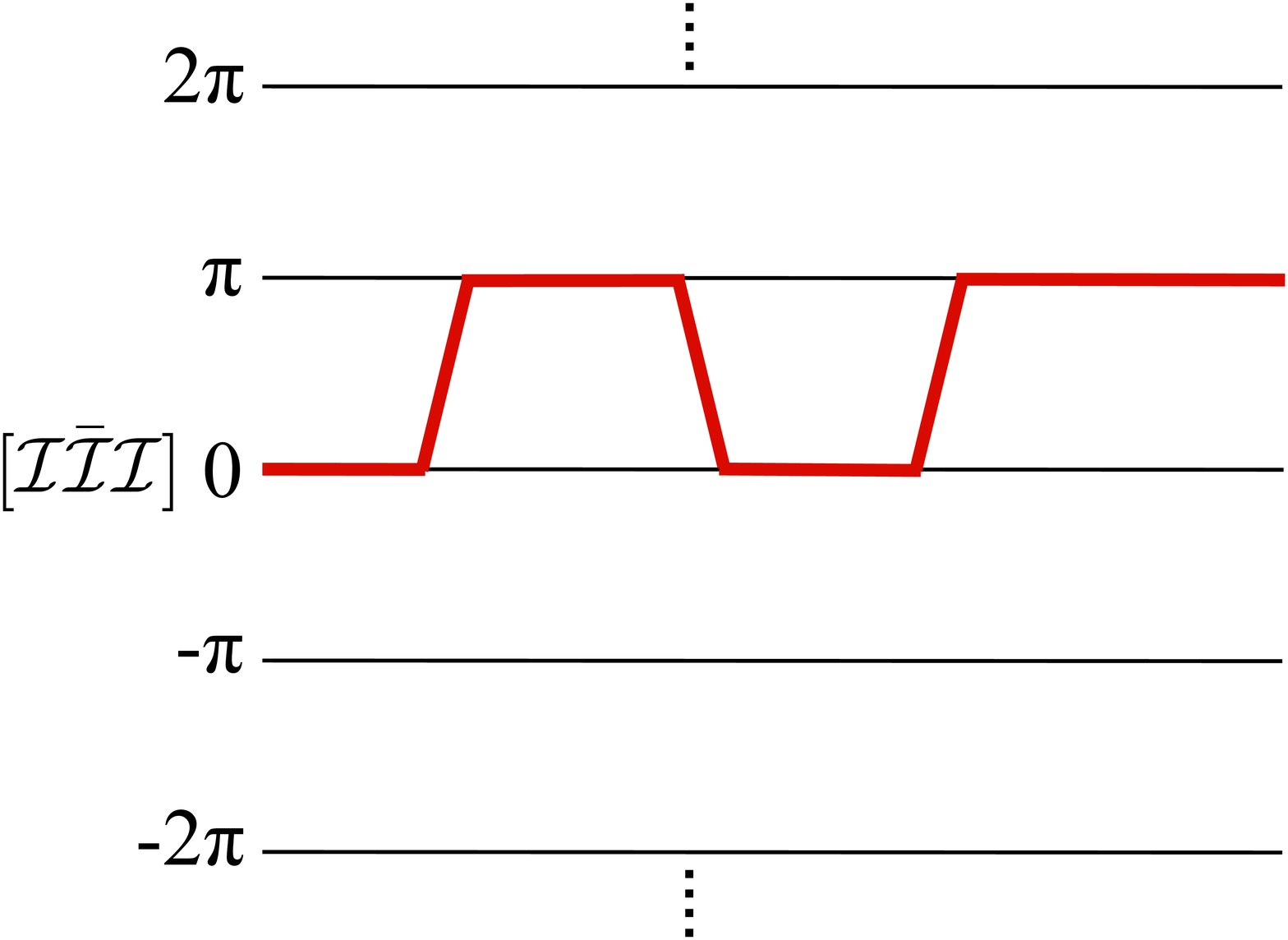}\\
  \includegraphics[width=0.4\textwidth]{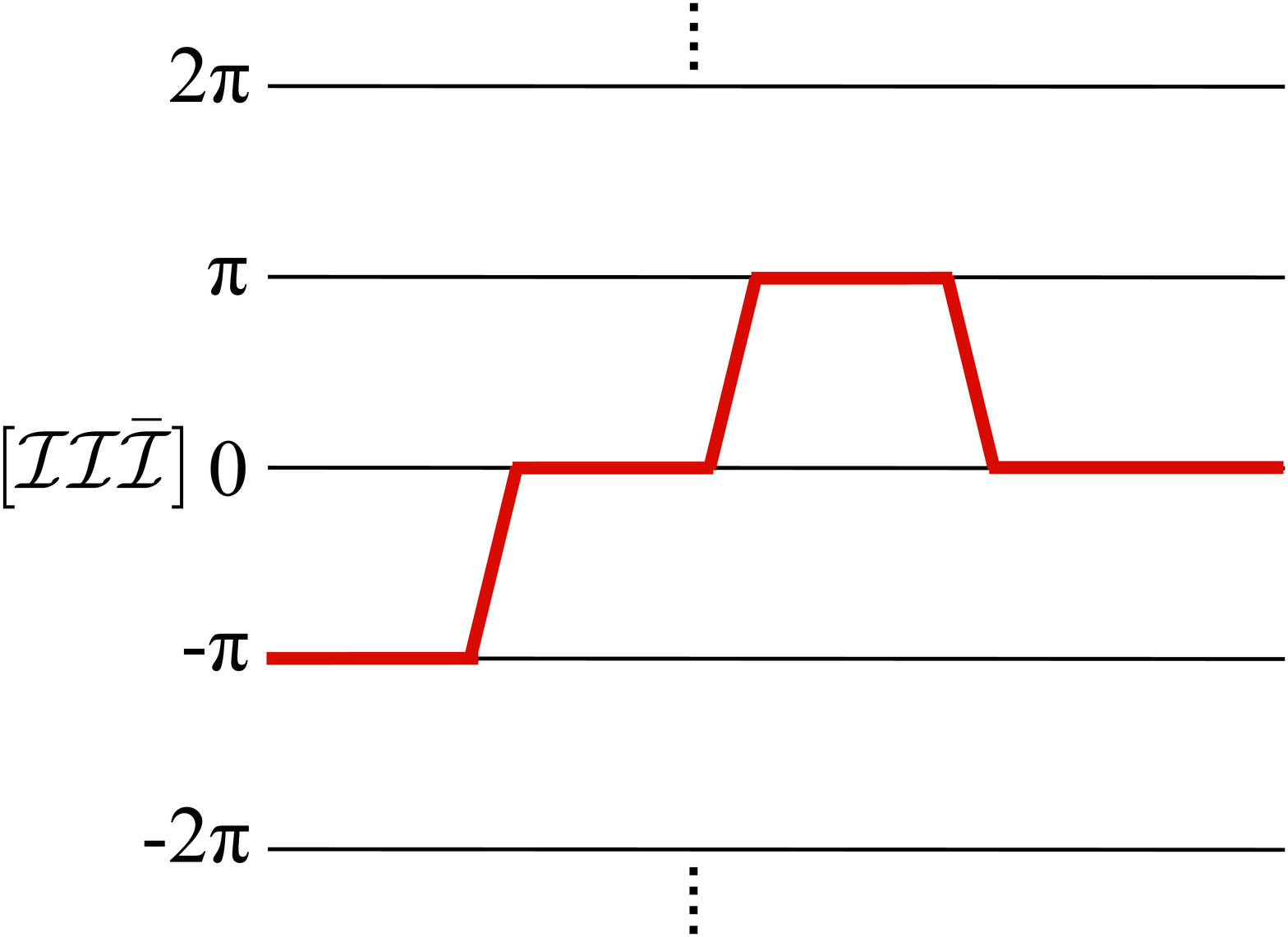}
   \includegraphics[width=0.4\textwidth]{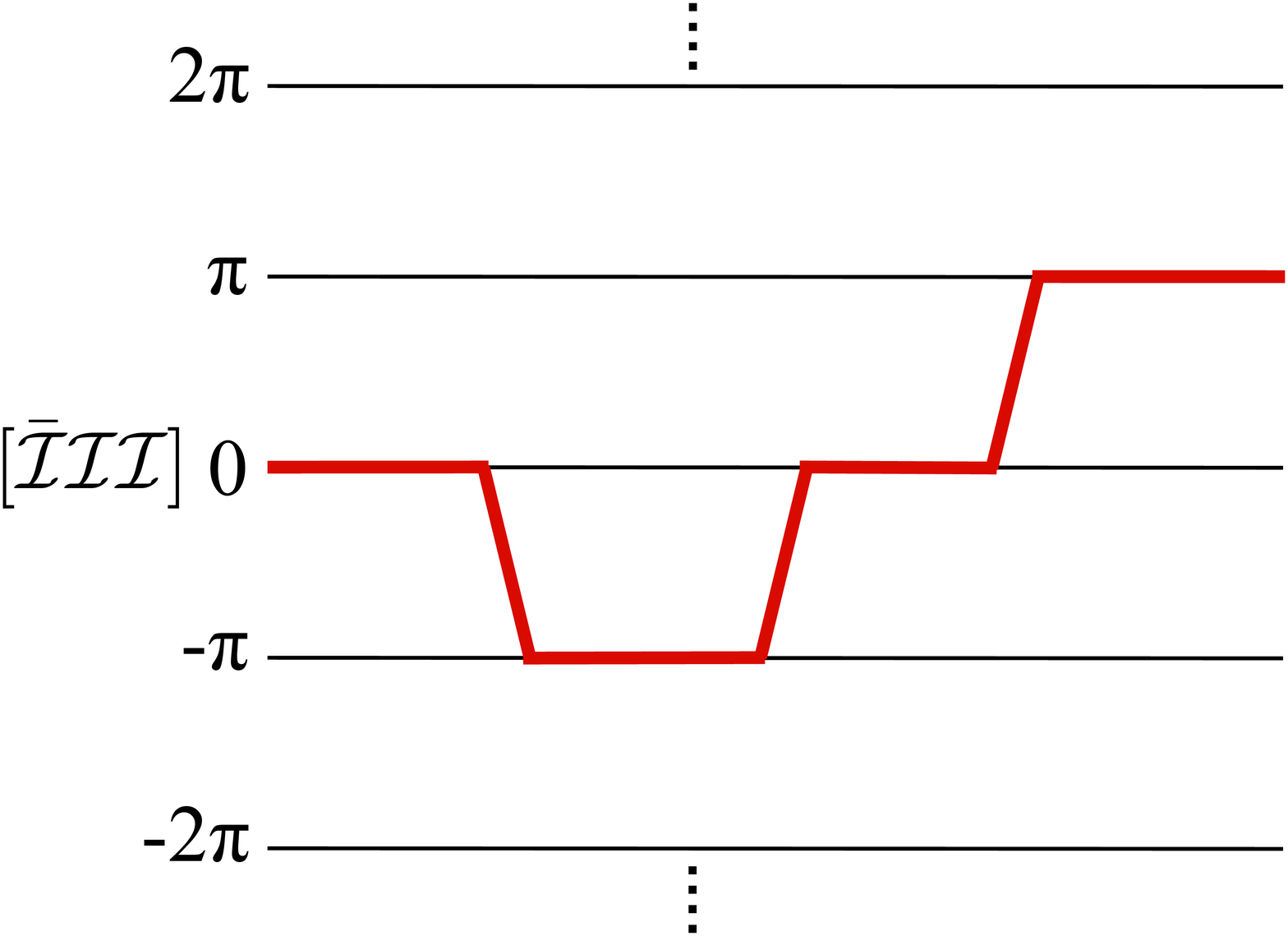}
\end{center}
\caption{A schematic figure of an example of two-instanton 
and one--anti-instanton amplitudes 
($[\mathcal{I} \bar{\mathcal{I}}\mathcal{I}], 
[\mathcal{I} \mathcal{I}\bar{\mathcal{I}}], 
[\bar{\mathcal{I}}\mathcal{I} \mathcal{I}]$).
Each horizontal line stands for the vacuum in the 
sine-Gordon potential.}
\label{fig:IIIbar}
\end{figure}

The two-instanton and one--anti-instanton amplitudes consist 
of three types of configurations, as shown in Fig.~\ref{fig:IIIbar}.
The first one is $[\mathcal{I} \bar{\mathcal{I}}\mathcal{I}]$,
where the anti-instanton is sandwiched between two instantons. 
For this type of configuration, adjacent constituent 
(anti-)instantons attract each other. 
Therefore we first take $-g^2>0$ in order to apply the BZJ 
prescription to the integral, and follow our subtraction 
prescription 
\begin{align}
[\mathcal{I} \bar{\mathcal{I}}\mathcal{I} ]e^{-i\theta}\xi^{-3}&\,=\,
\int_{0}^{\infty}dR_{1}dR_{2}
\exp\left[-{2\over{-g^2}}(e^{- R_{1}}+e^{- R_{2}})
-\epsilon(R_{1}+R_{2})\right]
\nonumber\\[5pt] 
&\,\overset{|g^{2}|\ll 1}{\longrightarrow}\,
\left({-g^{2}\over{2}}\right)^{\epsilon}\Gamma(\epsilon)
\left[\left({-g^{2}\over{2}}\right)^{\epsilon}
\Gamma(\epsilon)-{1\over{\epsilon}}\right] 
\nonumber\\[5pt] 
&\,\overset{-g^{2}=g^{2} e^{\mp i\pi} }{\longrightarrow}\,
{3\over{2}}\left(\gamma+\log{2\over{g^{2}}}\right)^2 \,
- \,{17\pi^2\over{12}} 
\pm 3i\pi\left(\gamma+\log{2\over{g^{2}}}\right)
 \,+\,O\left(\frac{1}{\epsilon}\right)\,+\, O(\epsilon)\,,
\end{align}
where we again subtract $1/\epsilon$ in the first integral,
before integrating the second quasi-moduli.

The second type of configuration is 
$[\mathcal{I} \mathcal{I}\bar{\mathcal{I}}]$,
where one pair of constituents is repulsive and the other 
attractive. 
In order to apply the BZJ prescription only to the attractive 
part of the interaction, we temporarily distinguish two 
coupling constants $g^{2}$ for the repulsive interaction
and $\tilde{g}^{2}$ for the attractive interaction, 
and apply the BZJ prescription to $\tilde{g}^{2}$, but not to $g^{2}$. 
After analytic continuation, we identify $\tilde{g}^{2}$ as the 
original $g^{2}$ at the end of the calculation. 
The moduli-integral of $[\mathcal{I} \mathcal{I}\bar{\mathcal{I}}]$ 
is given by 
\begin{equation}
[\mathcal{I}\mathcal{I} \bar{\mathcal{I}} ]e^{-i\theta}\xi^{-3}\,
=\, \int_{0}^{\infty}dR_{1}dR_{2}\,
\exp\left[-{2\over{g^2}}e^{- R_{1}}-{2\over{-\tilde{g}^2}}
e^{- R_{2}}-\epsilon(R_{1}+R_{2})\right]\,.
\end{equation}
For this configuration 
$[\mathcal{I} \mathcal{I}\bar{\mathcal{I}}]$, 
we have two possible orderings of moduli integral. 
The first ordering is to integrate over $R_2$ and 
to subtract the $1/\epsilon$ pole, before integrating 
over $R_1$. We call this ordering as $F_1(g^2)$ 
\begin{align}
F_{1}(g^{2})
&\,\overset{|g^{2}|,|\tilde{g}^{2}| \ll 1}{\longrightarrow}\,
\left({g^{2}\over{2}}\right)^{\epsilon}\Gamma(\epsilon)
\left[\left({-\tilde{g}^{2}\over{2}}\right)^{\epsilon}
\Gamma(\epsilon)-{1\over{\epsilon}}\right] 
\nonumber\\[5pt] 
&\,\overset{-\tilde{g}^{2}= \tilde{g}^{2} e^{\mp i\pi}}
{\longrightarrow}\,
{3\over{2}}\left(\gamma+\log{2\over{g^{2}}}\right)^2 \,
- \,{5\pi^2\over{12}} 
\pm 2i\pi\left(\gamma+\log{2\over{g^{2}}}\right) \,+\,
O\left(\frac{1}{\epsilon}\right) + O(\epsilon) \,.
\end{align}
We call the result of another ordering as $F_2(g^2)$, which 
is given by integrating over $R_1$ first and then over 
$R_2$ 
\begin{align}
F_{2}(g^{2})
&\,\overset{|g^{2}|,|\tilde{g}^{2}| \ll 1}{\longrightarrow}\,
\left({-\tilde{g}^{2}\over{2}}\right)^{\epsilon}\Gamma(\epsilon)
\left[\left({g^{2}\over{2}}\right)^{\epsilon}\Gamma(\epsilon)-{1\over{\epsilon}}\right] 
\nonumber\\[5pt] 
&\,\overset{-\tilde{g}^{2}= \tilde{g}^{2} e^{\mp i\pi}}{\longrightarrow}\,
{3\over{2}}\left(\gamma+\log{2\over{g^{2}}}\right)^2 \,+ \,{\pi^2\over{12}} 
\pm i\pi\left(\gamma+\log{2\over{g^{2}}}\right) \,+\,
O\left(\frac{1}{\epsilon}\right) \,+\, O(\epsilon) .
\end{align}
In the final expressions for $F_1, F_2$, we implicitly put 
back $\tilde{g}^{2}$ to $g^2$ after analytic continuation. 
We obtain the amplitude of 
$[\mathcal{I}\mathcal{I} \bar{\mathcal{I}}]$ as 
an average of $F_{1}(g^{2})$ and $F_{2}(g^{2})$ with dropping 
$O(1/\epsilon)$ and $O(\epsilon)$ terms 
\begin{align}
[\mathcal{I}\mathcal{I} \bar{\mathcal{I}} ]e^{-i\theta}\xi^{-3}
&\,=\,
(F_{1}\,+\,F_{2})/2
\nonumber\\
&\,=\, 
{3\over{2}}\left(\gamma+\log{2\over{g^{2}}}\right)^2 \,
- \,{\pi^2\over{6}} 
\pm {3\over{2}}i\pi\left(\gamma+\log{2\over{g^{2}}}\right)\,.
\label{eq:IIbarI}
\end{align}
The third type of configuration is 
$[\bar{\mathcal{I}}\mathcal{I} \mathcal{I}]$, 
which gives the identical result as the second type 
in Eq.(\ref{eq:IIbarI}) : 
$[ \bar{\mathcal{I}}\mathcal{I}\mathcal{I} ]=
[\mathcal{I}\mathcal{I} \bar{\mathcal{I}}]$. 

By taking the sum of all three types of configurations, we 
end up with 
\begin{align}
([\mathcal{I} \bar{\mathcal{I}}\mathcal{I} ]\,
+\,[\mathcal{I} \mathcal{I} \bar{\mathcal{I}} ]
\,+\,[\bar{\mathcal{I}}\mathcal{I} \mathcal{I}])e^{-i\theta}\xi^{-3}
\,=\,
{9\over{2}}\left[\left(\gamma+\log{2\over{g^{2}}}\right)^2 \,
- \,{7\pi^2\over{18}} 
\pm {4\over{3}}i\pi\left(\gamma+\log{2\over{g^{2}}}\right)\right]\,.
\end{align} 
Its contribution to the energy eigenvalue of the lowest band 
is then given by 
\begin{equation}
\triangle E^{(2,1)} \,=\,  -{9\over{2}}e^{i\theta}\,\xi^{3} 
\left[\left(\gamma+\log{2\over{g^{2}}}\right)^2 \,- \,{7\pi^2\over{18}} 
\pm {4\over{3}}i\pi\left(\gamma+\log{2\over{g^{2}}}\right)\right]\,.
\label{IIIbar}
\end{equation}
We note that the two--anti-instanton and one-instanton 
contribution is obtained by replacing $e^{i\theta}$ by $e^{-i\theta}$.

Resurgence implies that the two--anti-instanton and one-instanton 
amplitude should correspond to the large-order behavior of the 
perturbation series around the one-instanton saddle point. 
Using the cancellation between imaginary ambiguities of Borel 
resummed perturbation series around one-instanton saddle point 
and of the two--anti-instanton and one-instanton amplitude, 
the large-order behavior of the perturbation series around 
the one instanton saddle point can be estimated from 
the imaginary part of (\ref{IIIbar}) by means of the dispersion 
relation as
\begin{align} 
a_{k} &\,\approx\, {1\over{\pi}}\int_{0}^{\infty} \,d g^2 
\,{{\rm Im}[\triangle E^{(2,1)}e^{-i\theta}/\xi]
\over{(g^2)^{k+1}}}
\,=\, {6\over{\pi}}\int_{0}^{\infty} \,dg^2 
\,{ e^{-1/g^2}(\gamma+\log(2/g^2))\over{(g^2)^{k+2}}}
\nonumber\\
&\,=\, {6\over{\pi}} k!\,\left(\log 2\,+\,{s(k+1,2)\over{k!}} \right) 
\quad\quad\quad\quad\quad\,( k\geq 2)\,,
\label{ak3}
\end{align}
where $s(k+1,2)$ is the Stirling number of the first kind, 
which is the solution of the recurrence relation 
$s(k+1,2)=(k+1)s(k,2)+k!$. The first few numbers are given as 
$s(k+1,2)=3, 11, 50, 274, 1764, 13068, 109584$ for $k=2,3,4,5,6,7,8$.
In Ref.~\cite{ZinnJustin:1981dx}, the large-order perturbative 
series around one instanton is numerically calculated as 
\begin{equation}
a_{k}^{\rm pert.} \,=\,  {6\over{\pi}} k!\,
\left(\gamma\,+\,\log2k\right)\,+\,O\left({\log k\over{k}}\right)\,,
\end{equation}
which is consistent with Eq.~(\ref{ak3}) for large $k$.

\subsection{4 instantons}

\begin{figure}[htbp]
\begin{center}
 \includegraphics[width=0.4\textwidth]{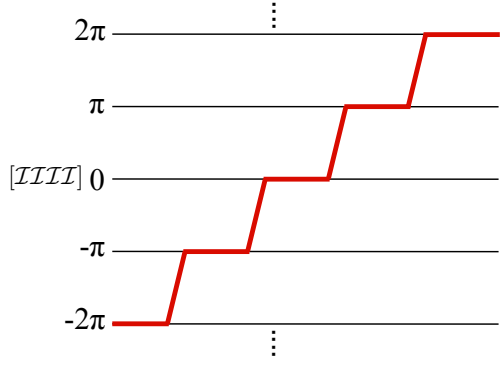}
\end{center}
\caption{A schematic figure of an example of the four-instanton 
amplitude $[\mathcal{I} \mathcal{I}\mathcal{I}\mathcal{I}]$.
Each horizontal line stands for the vacuum in the sine-Gordon potential.}
\label{fig:IIII}
\end{figure}

For the four-instanton amplitude, 
the interaction between each pair of instantons is repulsive, 
as shown in Fig.~\ref{fig:IIII}. 
Since all the possible orderings of multi-moduli integral has 
the identical contribution, 
we easily obtain the amplitude using the polygamma function defined 
as $\psi^{(2)}(z) \equiv d^{3} \log\Gamma(z)/dz^{3}$ 
\begin{align}
[\mathcal{I} \mathcal{I}\mathcal{I}\mathcal{I} ]e^{-4i\theta}
\xi^{-4}&\,=\,\int_{0}^{\infty}dR_{1}dR_{2}dR_{3}\,
\exp\left[-{2\over{g^2}}(e^{- R_{1}}+e^{- R_{2}}+e^{-R_{3}})
-\epsilon(R_{1}+R_{2}+R_{3})\right]
\nonumber\\[5pt] 
&\,\overset{|g^{2}| \ll 1}{\longrightarrow}\,
\left({g^{2}\over{2}}\right)^{\epsilon}\Gamma(\epsilon)
\left[\left({g^{2}\over{2}}\right)^{\epsilon}\Gamma(\epsilon)
\left\{\left({g^{2}\over{2}}\right)^{\epsilon}\Gamma(\epsilon)
-{1\over{\epsilon}}\right\}
+ {\gamma+\log(2/g^2)\over{\epsilon}}\right]
\nonumber\\[5pt] 
&\,=\,
-{8\over{3}}
\left[
\left(\gamma+\log{2\over{g^{2}}}\right)^3 \,
+ \,{\pi^2\over{8}}\left( \gamma+\log{2\over{g^{2}}}\right) 
-{1\over{16}}\psi^{(2)}(1)
\right]
\,+\,O\left(\frac{1}{\epsilon}\right)\,+\, O(\epsilon)\,,
\end{align}
where we have first subtracted $1/\epsilon$ in the first integral 
$dR_3$, and subtracted $-(\gamma+\log(2/g^{2}))/\epsilon$ in 
the next integral $dR_2$, and finally dropped the 
$O(1/\epsilon)$ term in the integral $dR_1$. 
Its contribution to the eigenvalue of the lowest band is then 
given by 
\begin{equation}
\triangle E^{(4,0)} \,=\,  {8\over{3}}e^{4i\theta}\,\xi^{4}
\left[\left(\gamma+\log{2\over{g^{2}}}\right)^3 \,
+ \,{\pi^2\over{8}}\left( \gamma+\log{2\over{g^{2}}}\right) 
-{1\over{16}}\psi^{(2)}(1)
\right]\,.
\label{IIII}
\end{equation}
The four anti-instanton contribution is obtained by replacing 
$e^{4i\theta}$ by $e^{-4i\theta}$.

\subsection{3 instantons $+$ 1 anti-instanton}

\begin{figure}[htbp]
\begin{center}
 \includegraphics[width=0.4\textwidth]{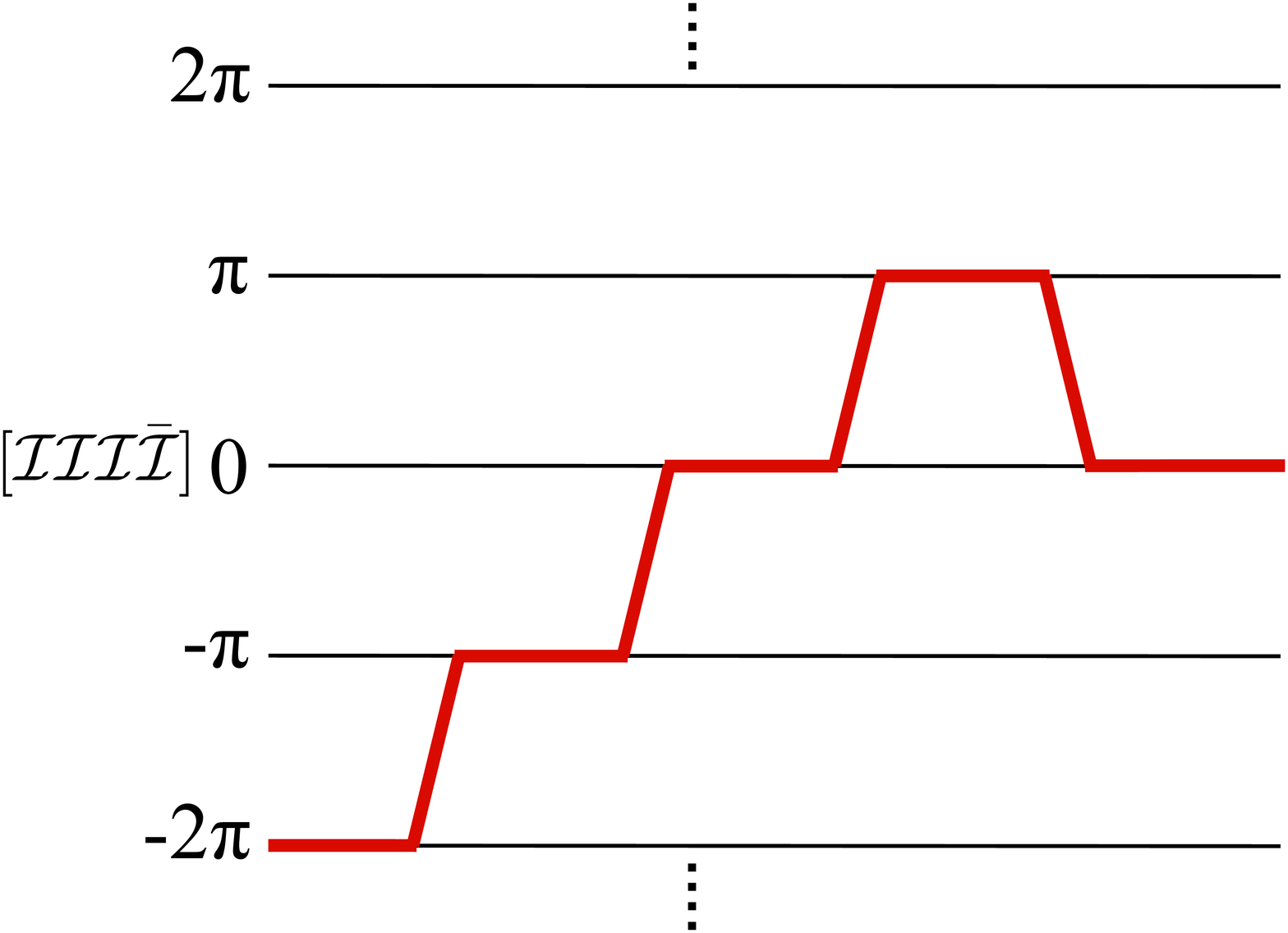}
  \includegraphics[width=0.4\textwidth]{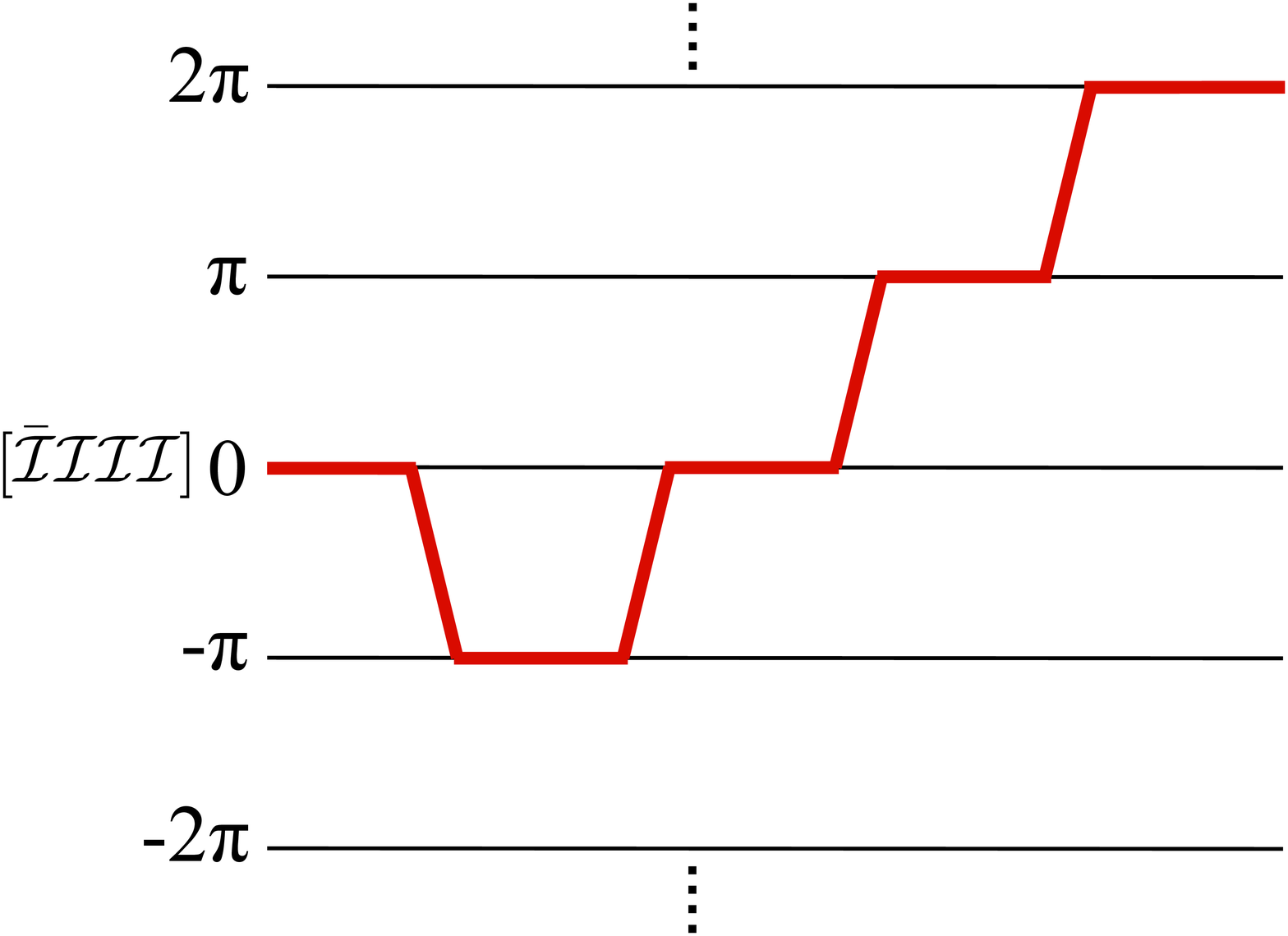}
   \includegraphics[width=0.4\textwidth]{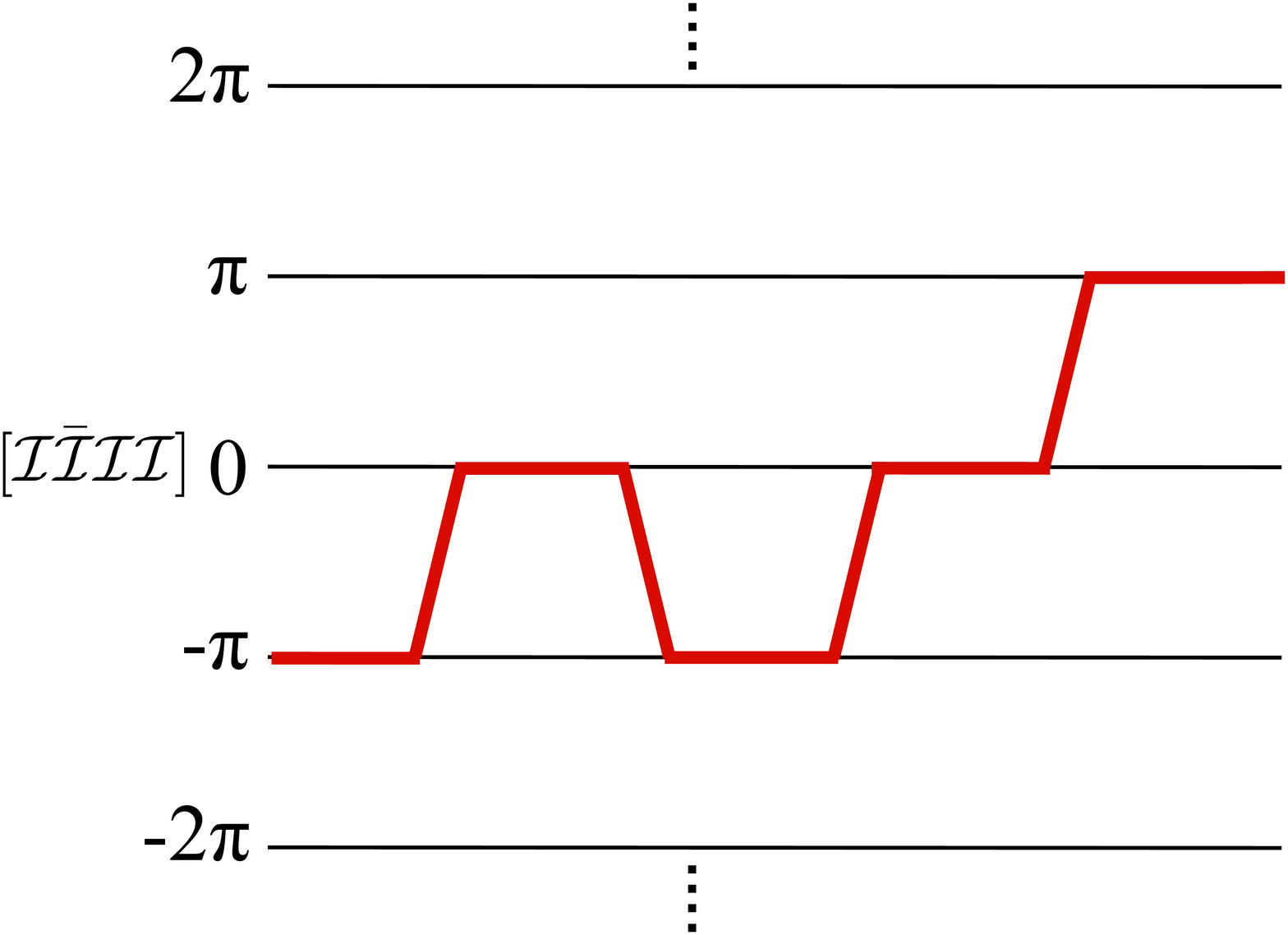}
     \includegraphics[width=0.4\textwidth]{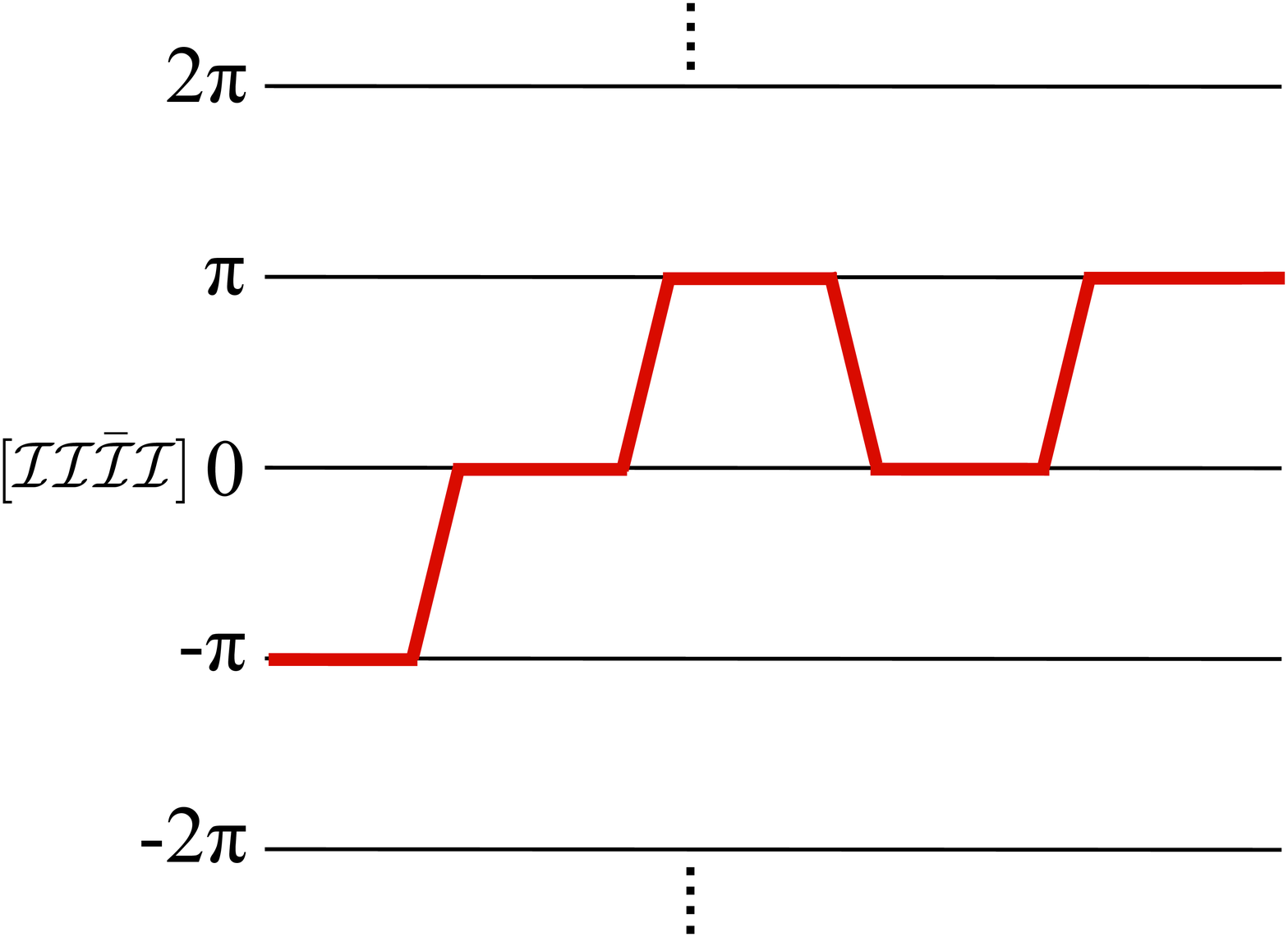}
\end{center}
\caption{A schematic figure of examples of  four-instanton 
amplitudes 
($[\mathcal{I}\mathcal{I}\mathcal{I}\bar{\mathcal{I}}]$,
$[\bar{\mathcal{I}}\mathcal{I}\mathcal{I}\mathcal{I}]$,
$[\mathcal{I}\bar{\mathcal{I}}\mathcal{I}\mathcal{I}]$ and
$[\mathcal{I}\mathcal{I}\bar{\mathcal{I}}\mathcal{I}]$).
Each horizontal line stands for the vacuum in the sine-Gordon potential.}
\label{fig:IIIIbar}
\end{figure}

The three-instanton and one--anti-instanton amplitudes 
consist of four types of configurations 
$[\mathcal{I}\mathcal{I}\mathcal{I}\bar{\mathcal{I}}]$, 
$[\bar{\mathcal{I}}\mathcal{I}\mathcal{I}\mathcal{I}]$, 
$[\mathcal{I}\bar{\mathcal{I}}\mathcal{I}\mathcal{I}]$ and 
$[\mathcal{I}\mathcal{I}\bar{\mathcal{I}}\mathcal{I}]$, 
as shown in Fig.~\ref{fig:IIIIbar}. 

In the first configuration 
$[\mathcal{I}\mathcal{I}\mathcal{I}\bar{\mathcal{I}}]$, 
two pairs of adjacent instantons have repulsive 
interactions, and the other is attractive. 
We again apply the BZJ prescription only to the integral for 
the pair with the attractive interaction, where we denote the 
coupling as $\tilde{g}^{2}$, evaluate the integral at 
$-\tilde{g}^{2}>0$, and analytically continue to 
$\tilde{g}^{2}>0$, while keeping $g^{2}>0$ for the repulsive 
interactions. The multi-moduli integral is given by 
\begin{equation}
{[\mathcal{I}\mathcal{I}\mathcal{I} \bar{\mathcal{I}}]
\over{e^{2i\theta}\xi^{4}}}\,=\,
\int_{0}^{\infty}dR_{1}dR_{2}dR_{3}
\,\exp\left[-{2\over{g^2}}e^{- R_{1}}-{2\over{g^2}}
e^{- R_{2}}-{2\over{-\tilde{g}^2}}e^{-R_{3}}
-\epsilon(R_{1}+R_{2}+R_{3})\right]\,.
\label{G123}
\end{equation}
Here we have three orderings of the multi-integral, 
distinguished by the ordering of $dR_3$ (attractive interaction) 
relative to $dR_1, dR_2$. 
We denote the results $G_1(g^2)$, $G_{2}(g^2)$, and $G_{3}(g^2)$ 
as the first, the second, and the third integral, 
respectively. 
Referring the calculations given in Appendix~\ref{app1}, we 
just show the result 
\begin{align}
G_{1}(g^2)&\,=\,
-{8\over{3}}
\left(\gamma+\log{2\over{g^{2}}}\right)^3 \,+ \,{7\over{6}}
\pi^{2}\left( \gamma+\log{2\over{g^{2}}}\right) 
+{1\over{6}}\psi^{(2)}(1)\,\mp\,4i\pi\left(\gamma
+\log{2\over{g^{2}}}\right)^2\,,
\nonumber\\[5pt] 
G_{2}(g^2)&\,=\,-{8\over{3}}
\left(\gamma+\log{2\over{g^{2}}}\right)^3 \,+ \,{1\over{6}}
\pi^{2}\left( \gamma+\log{2\over{g^{2}}}\right) 
+{1\over{6}}\psi^{(2)}(1)\,\mp\,i\pi\Big[{5\over{2}}
\left(\gamma+\log{2\over{g^{2}}}\right)^2 +{\pi^2\over{12}}\Big]\,
,
\nonumber\\[5pt] 
G_{3}(g^2)&\,=\,
-{8\over{3}}
\left(\gamma+\log{2\over{g^{2}}}\right)^3 \,- \,{1\over{3}}
\pi^{2}\left( \gamma+\log{2\over{g^{2}}}\right) 
+{1\over{6}}\psi^{(2)}(1)\,\mp\,i\pi\Big[{3\over{2}}
\left(\gamma+\log{2\over{g^{2}}}\right)^2 +{\pi^2\over{12}}\Big]\,.
\end{align}

We obtain the amplitude of 
$[\mathcal{I}\mathcal{I}\mathcal{I} \bar{\mathcal{I}}]$ as 
the average of $G_{1}(g^2)$, $G_{2}(g^2)$ and $G_{3}(g^2)$, 
\begin{align}
{[\mathcal{I}\mathcal{I}\mathcal{I} 
\bar{\mathcal{I}}]\over{e^{2i\theta}\xi^{4}}}
&\,=\,
(G_{1}(g^2)\,+\,G_{2}(g^2)\,+\, G_{3}(g^2))/3
\nonumber\\
&\,=\, 
-{8\over{3}}
\left(\gamma+\log{2\over{g^{2}}}\right)^3 \,
+ \,{1\over{3}}\pi^{2}\left( \gamma+\log{2\over{g^{2}}}\right) 
+{1\over{6}}\psi^{(2)}(1)\,\mp\,i\pi\Big[{8\over{3}}
\left(\gamma+\log{2\over{g^{2}}}\right)^2 +{\pi^2\over{18}}\Big]\,.
\label{eq:IIIbarI}
\end{align}

The amplitude of the second configuration 
$[\bar{\mathcal{I}}\mathcal{I}\mathcal{I}\mathcal{I}]$ 
is found to give identical result as the above configuration : 
$[ \bar{\mathcal{I}}\mathcal{I}\mathcal{I}\mathcal{I}]
=[\mathcal{I}\mathcal{I}\mathcal{I} 
\bar{\mathcal{I}}]$. 


In the third configuration 
$[\mathcal{I}\bar{\mathcal{I}}\mathcal{I}\mathcal{I}]$, 
the two pairs of adjacent instantons have attractive interactions, 
while the other is repulsive. 
We again apply the BZJ prescription only to the attractive pairs, 
by denoting the couplings as $\tilde{g}^{2}$. 
The multi-integral is given as 
\begin{equation}
{[\mathcal{I}\bar{\mathcal{I}}\mathcal{I}\mathcal{I}]\over
{e^{2i\theta}\xi^{4}}}
\,=\,\int_{0}^{\infty}dR_{1}dR_{2}dR_{3}\,
\exp\left[-{2\over{g^2}}e^{- R_{1}}-{2\over{-\tilde{g}^2}}
e^{- R_{2}}-{2\over{-\tilde{g}^2}}e^{-R_{3}}
-\epsilon(R_{1}+R_{2}+R_{3})\right]\,.
\label{G456}
\end{equation}

We have three orderings of the multi-integral, 
distinguished by the ordering of $dR_1$ (repulsive interaction) 
relative to $dR_2, dR_3$. 
We denote the results $G_4(g^2)$, $G_{5}(g^2)$, and $G_{6}(g^2)$ 
as the first, the second, and the third integral, 
respectively. 
Referring the calculations given in Appendix~\ref{app1}, we 
just show the result 
\begin{align}
G_{4}&\,=\,
-{8\over{3}}
\left(\gamma+\log{2\over{g^{2}}}\right)^3 \,+ 
\,{7\over{6}}\pi^{2}\left( \gamma+\log{2\over{g^{2}}}\right) 
+{1\over{6}}\psi^{(2)}(1)\,\mp\,i\pi\Big[4\left(\gamma
+\log{2\over{g^{2}}}\right)^2 +{\pi^2\over{6}}\Big]\,,
\nonumber\\[5pt] 
G_{5}&\,=\,
-{8\over{3}}
\left(\gamma+\log{2\over{g^{2}}}\right)^3 \,
+ \,{19\over{6}}\pi^{2}\left( \gamma+\log{2\over{g^{2}}}\right) 
+{1\over{6}}\psi^{(2)}(1)\,\mp\,i\pi\Big[{11\over{2}}
\left(\gamma+\log{2\over{g^{2}}}\right)^2 -{5\pi^2\over{12}}\Big]\,,
\nonumber\\[5pt] 
G_{6}&\,=\,
-{8\over{3}}
\left(\gamma+\log{2\over{g^{2}}}\right)^3 \,+ \,{14\over{3}}
\pi^{2}\left( \gamma+\log{2\over{g^{2}}}\right) 
+{1\over{6}}\psi^{(2)}(1)
\,\mp\,i\pi\Big[{13\over{2}}\left(\gamma+\log{2\over{g^{2}}}\right)^2 
-{11\pi^2\over{12}}\Big]\,.
\end{align}

We obtain the amplitude of 
$[\mathcal{I}\bar{\mathcal{I}}\mathcal{I}\mathcal{I} ]$ as 
the average of $G_{4}(g^{2})$, $G_{5}(g^{2})$ and $G_{6}(g^{2})$,
\begin{align}
{[\mathcal{I}\bar{\mathcal{I}}\mathcal{I}\mathcal{I}]\over{e^{2i\theta}\xi^{4}}}
&\,=\,
(G_{4}(g^{2})\,+\,G_{5}(g^{2})\,+\, G_{6}(g^{2}))/3
\nonumber\\
&\,=\, 
-{8\over{3}}
\left(\gamma+\log{2\over{g^{2}}}\right)^3 \,+ \,3\pi^{2}
\left( \gamma+\log{2\over{g^{2}}}\right) 
+{1\over{6}}\psi^{(2)}(1)\,\mp\,i\pi\Big[{16\over{3}}
\left(\gamma+\log{2\over{g^{2}}}\right)^2 
-{7\pi^2\over{18}}\Big]\,.
\label{eq:IbarIII}
\end{align}
The amplitude of the fourth configuration 
$[\mathcal{I}\mathcal{I}\bar{\mathcal{I}}\mathcal{I}]$ 
turns out to be identical to the above third configuration : 
$[\mathcal{I}\mathcal{I}\bar{\mathcal{I}}\mathcal{I}]
=[\mathcal{I}\bar{\mathcal{I}}\mathcal{I}\mathcal{I} ]$.

The sum of all the four configurations gives 
\begin{align}
&([\mathcal{I}\mathcal{I}\mathcal{I}\bar{\mathcal{I}}]+
[\bar{\mathcal{I}}\mathcal{I}\mathcal{I}\mathcal{I}]+
[\mathcal{I}\bar{\mathcal{I}}\mathcal{I}\mathcal{I}]+
[\mathcal{I}\mathcal{I}\bar{\mathcal{I}}\mathcal{I}])e^{-2i\theta}\xi^{-4}
\nonumber\\
&\,=\,
-{32\over{3}}
\left(\gamma+\log{2\over{g^{2}}}\right)^3 \,+ \,{20\over{3}}
\pi^{2}\left( \gamma+\log{2\over{g^{2}}}\right) 
+{2\over{3}}\psi^{(2)}(1)\,\mp\,i\pi\Big[16\left(\gamma
+\log{2\over{g^{2}}}\right)^2 
-{2\pi^2\over{3}}\Big]\,.
\end{align}
Therefore the contribution of three instantons and one 
anti-instanton to the energy eigenvalue of the lowest band 
is given by 
\begin{align}
&\triangle E
^{(3,1)} 
\nonumber\\
&\,=\,  e^{2i\theta}\,\xi^{4}
\left[{32\over{3}}\left(\gamma+\log{2\over{g^{2}}}\right)^3 \,
- \,{20\over{3}}\pi^{2}\left( \gamma+\log{2\over{g^{2}}}\right) 
-{2\over{3}}\psi^{(2)}(1)\,\pm\,i\pi\Big[16\left(\gamma
+\log{2\over{g^{2}}}\right)^2 
-{2\pi^2\over{3}}\Big]
\right]\,.
\label{IIIIbar}
\end{align}
We note that the three anti-instanton and one-instanton 
contribution is obtained by replacing $e^{2i\theta}$ by $e^{-2i\theta}$.

Resurgence implies that the three-instanton and one--anti-instanton 
amplitude should correspond to the large-order behavior of the 
perturbation series around the two-instanton saddle point. 
Using the cancellation between imaginary ambiguities of Borel 
resummed perturbation series around two-instanton saddle point 
and of the three-instanton and one--anti-instanton amplitude, 
the large-order behavior of the perturbation series around 
the two instanton saddle point can be estimated from 
the imaginary part of (\ref{IIIIbar}) by means of the dispersion 
relation as 
\begin{align} 
a_{k} &\,\approx\, {1\over{\pi}}\int_{0}^{\infty} \,dg^2 
\,{{\rm Im}[\triangle E^{(3,1)}e^{-2i\theta}/\xi^{2}]
\over{(g^2)^{k+1}}}
\nonumber\\
&\,=\, -{16\over{\pi}}\int_{0}^{\infty} \,dg^2 \,
{ e^{-1/g^2}\over{(g^2)^{k+2}}}
\left(    (\gamma+\log 2)^{2}-{\pi^{2}\over{24}} \,
- \, 2(\gamma+\log 2)\log(g^2)\,+\, \log^{2}(g^2) 
\right)\,.
\end{align}
This integral can be performed numerically, and 
the first few results are 
$a_{k}\,\sim\,-48.826, \,-191.16, \,  -919.05,\,-5273.0$ for $k=2,3,4,5$.
One should be able to check this large-order behavior, by 
performing the perturbation around the two-instanton saddle 
point to high orders.

\subsection{2 instantons $+$ 2 anti-instantons}
\begin{figure}[htbp]
\begin{center}
 \includegraphics[width=0.32\textwidth]{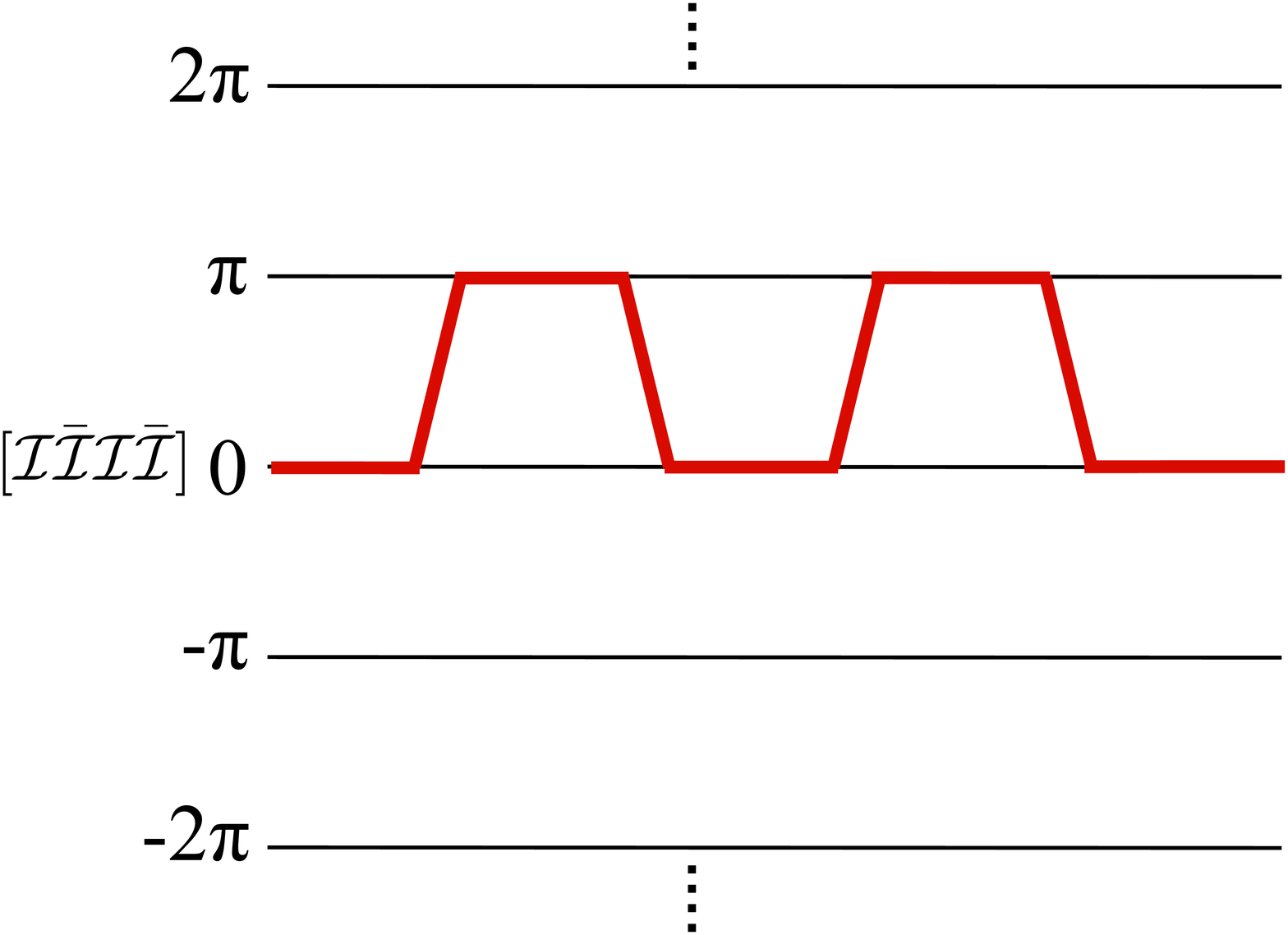}
  \includegraphics[width=0.32\textwidth]{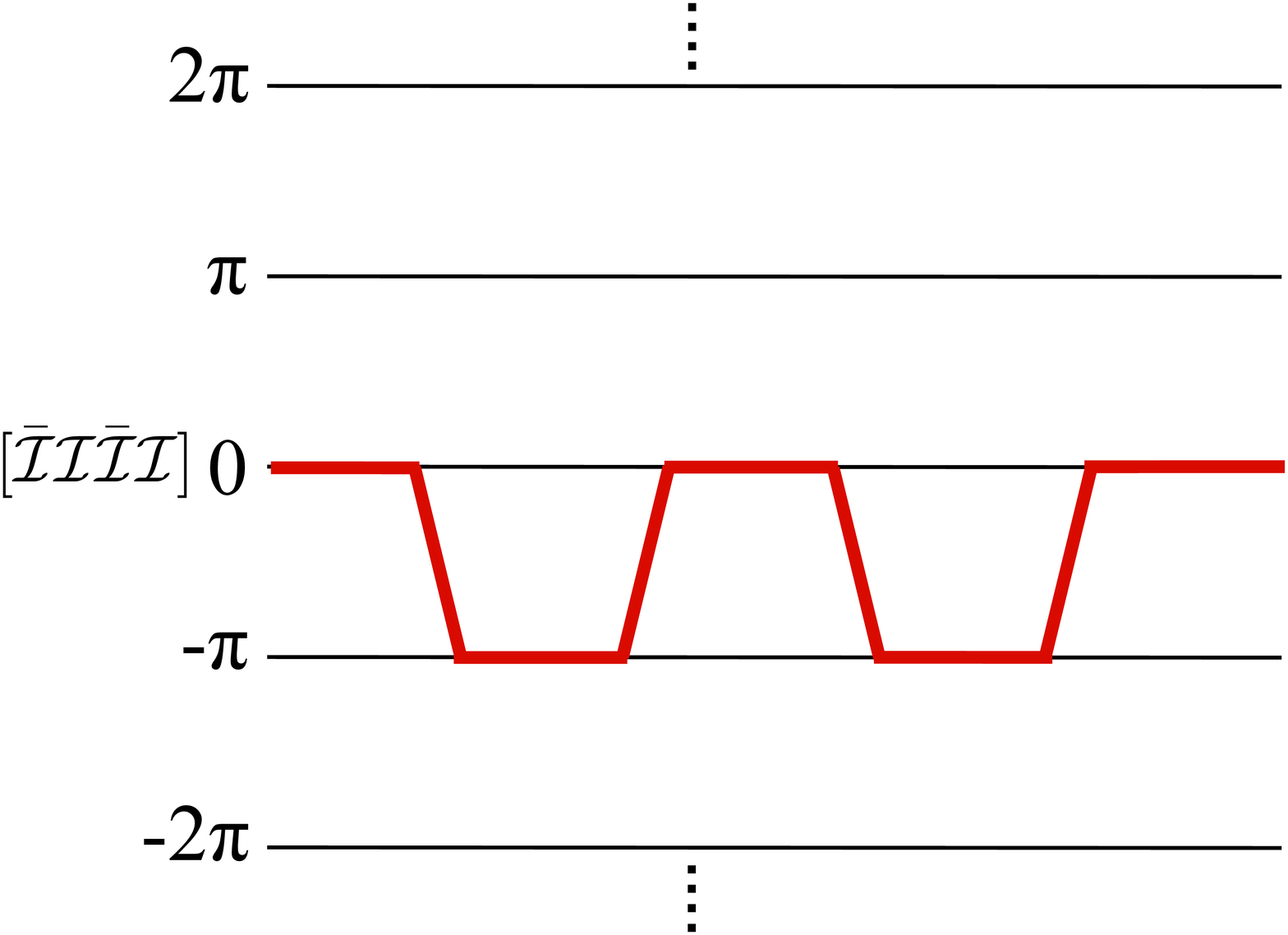}\\
   \includegraphics[width=0.32\textwidth]{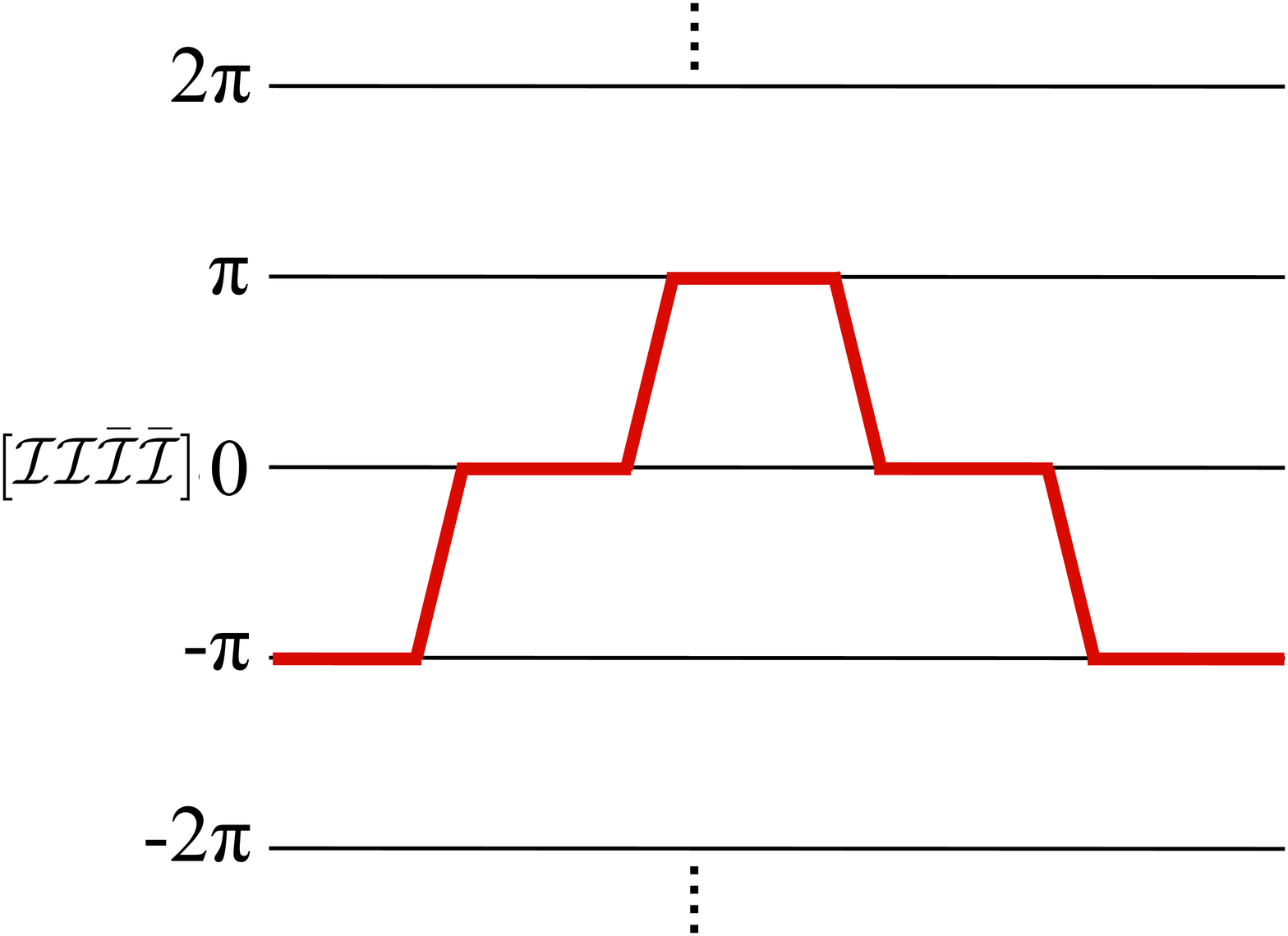}
     \includegraphics[width=0.32\textwidth]{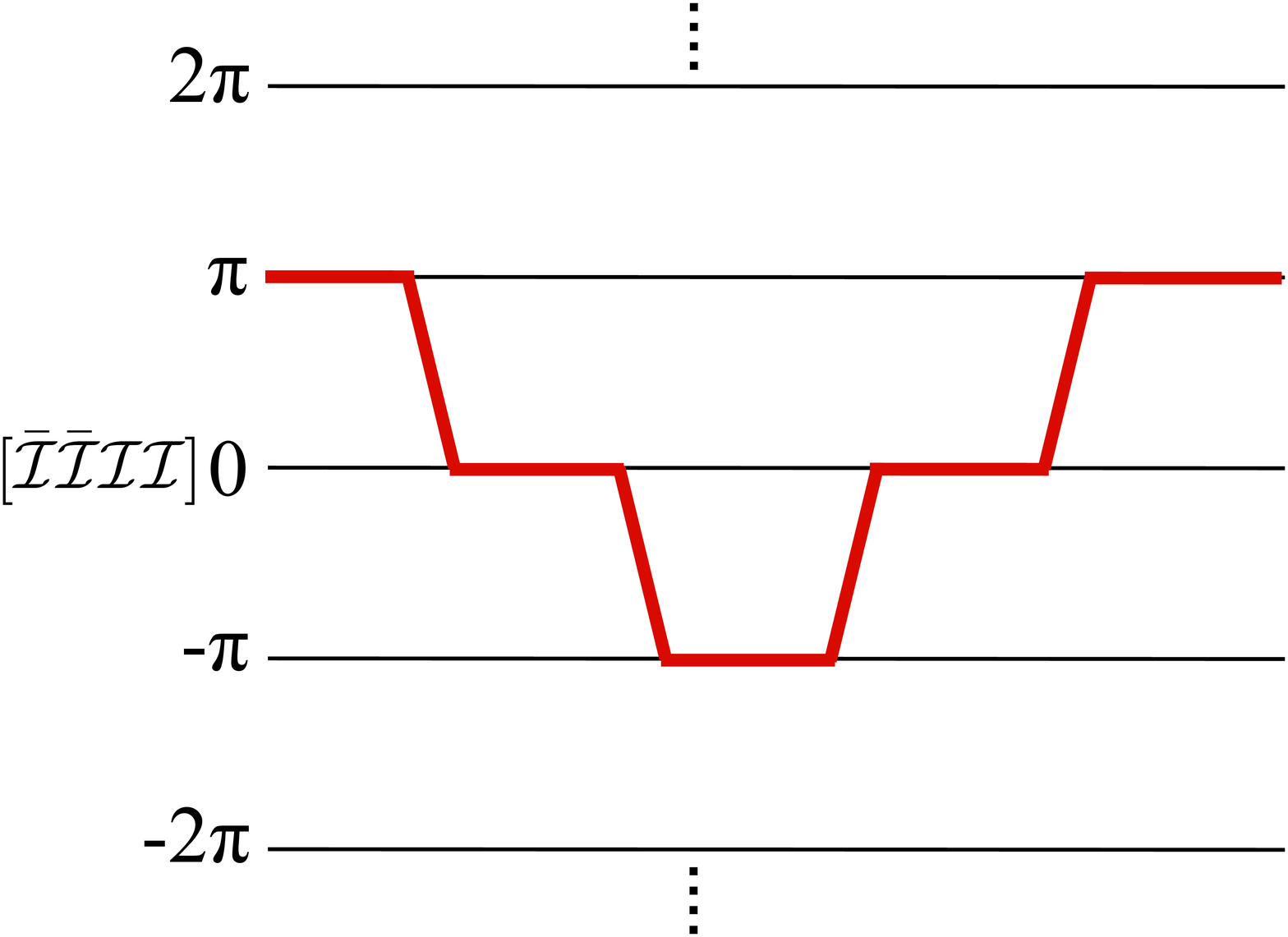}\\
      \includegraphics[width=0.32\textwidth]{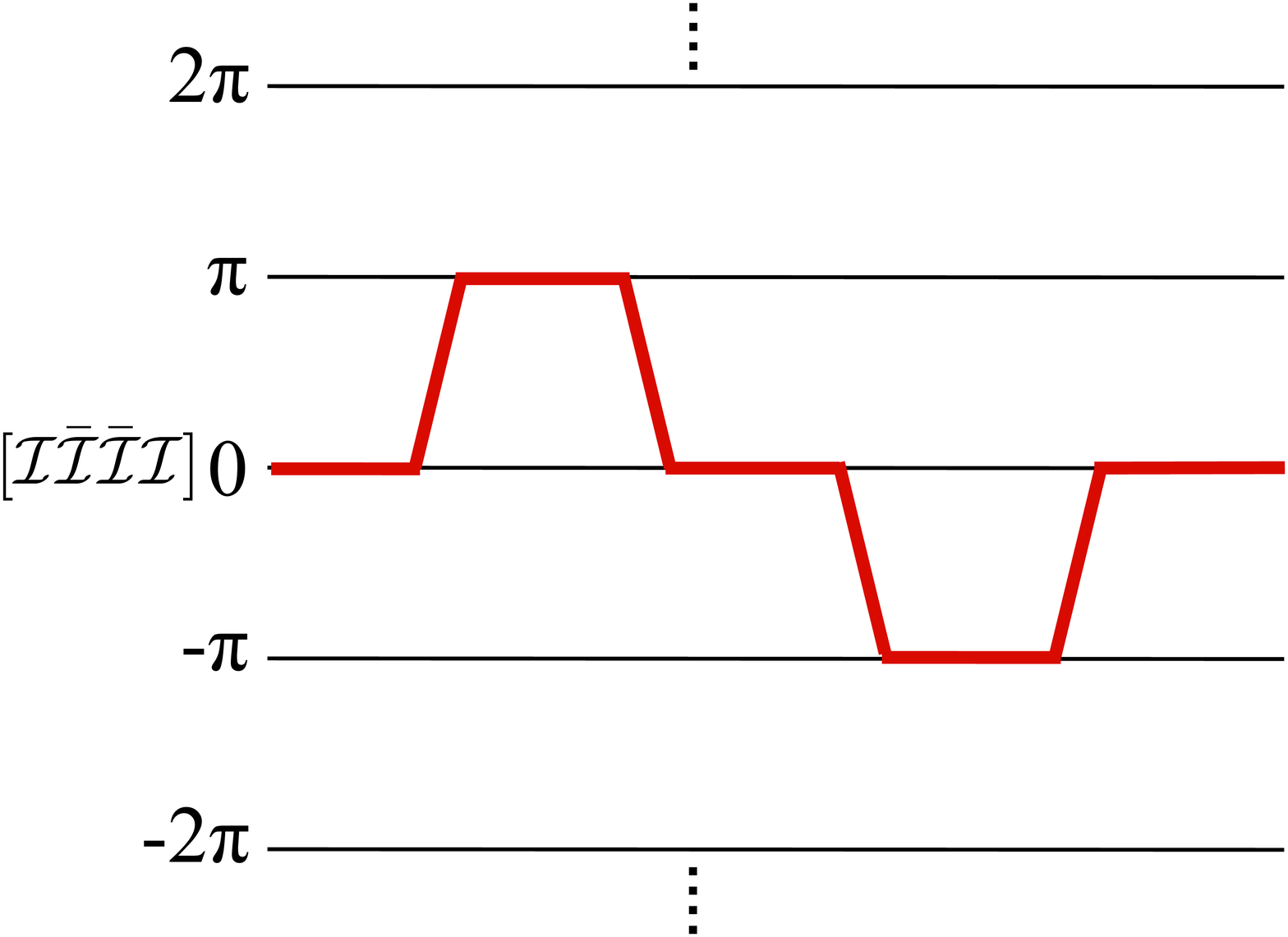}
     \includegraphics[width=0.32\textwidth]{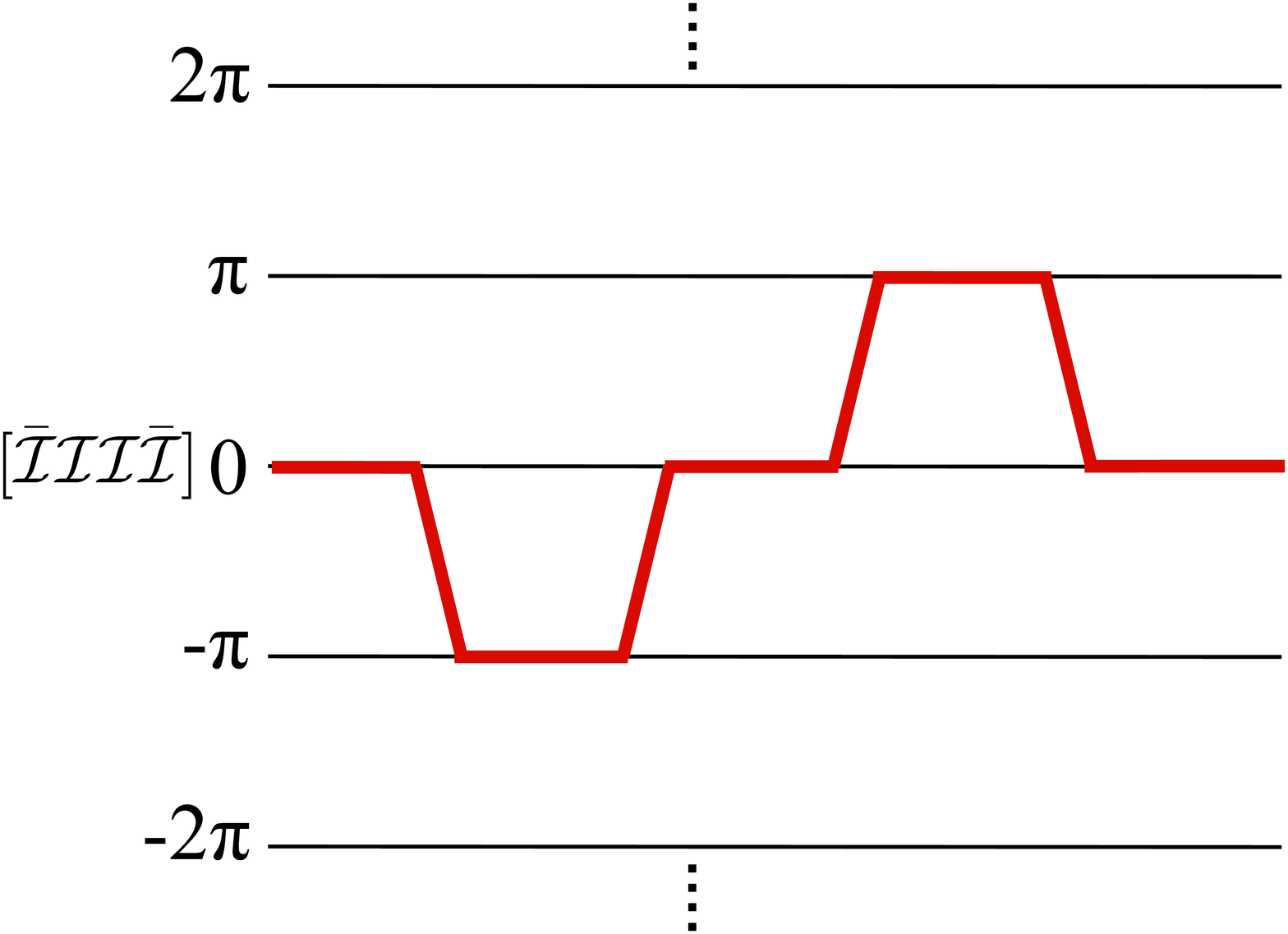}
\end{center}
\caption{A schematic figure of examples of  four-instanton 
amplitudes 
($[\mathcal{I}\bar{\mathcal{I}}\mathcal{I}\bar{\mathcal{I}}]$,
$[\bar{\mathcal{I}}\mathcal{I}\bar{\mathcal{I}}\mathcal{I}]$,
$[\mathcal{I}\mathcal{I}\bar{\mathcal{I}}\bar{\mathcal{I}}]$,
$[\bar{\mathcal{I}}\bar{\mathcal{I}}\mathcal{I}\mathcal{I}]$,
$[\mathcal{I}\bar{\mathcal{I}}\bar{\mathcal{I}}\mathcal{I}]$ and
$[\bar{\mathcal{I}}\mathcal{I}\mathcal{I}\bar{\mathcal{I}}]$).
Each horizontal line stands for the vacuum in the sine-Gordon potential.}
\label{fig:IIIbarIbar}
\end{figure}

The two-instanton and two-anti-instanton amplitudes consist of 
six types of configurations 
$[\mathcal{I}\bar{\mathcal{I}}\mathcal{I}\bar{\mathcal{I}}]$,
$[\bar{\mathcal{I}}\mathcal{I}\bar{\mathcal{I}}\mathcal{I}]$,
$[\mathcal{I}\mathcal{I}\bar{\mathcal{I}}\bar{\mathcal{I}}]$,
$[\bar{\mathcal{I}}\bar{\mathcal{I}}\mathcal{I}\mathcal{I}]$,
$[\mathcal{I}\bar{\mathcal{I}}\bar{\mathcal{I}}\mathcal{I}]$ and
$[\bar{\mathcal{I}}\mathcal{I}\mathcal{I}\bar{\mathcal{I}}]$, 
as shown in Fig.~\ref{fig:IIIbarIbar}.

The first one is $[\mathcal{I}\bar{\mathcal{I}}\mathcal{I}\bar{\mathcal{I}}]$,
where all three adjacent pairs of constituents have attractive 
 interactions. 
Thus we apply the simple BZJ prescription to all the integration. 
Since all the orderings of the multi integral have the same contributions, 
the amplitude can be easily calculated as 
\begin{align}
[\mathcal{I}\bar{\mathcal{I}}\mathcal{I}\bar{\mathcal{I}}]\xi^{-4}
&\,=\,
\int_{0}^{\infty}dR_{1}dR_{2}dR_{3}
\exp\left[-{2\over{-g^2}}(e^{- R_{1}}+e^{- R_{2}}+e^{-R_{3}})
-\epsilon(R_{1}+R_{2}+R_{3})\right]
\nonumber\\[5pt] 
&\,\overset{|g^{2}|\ll 1}{\longrightarrow}\,
\left({-g^{2}\over{2}}\right)^{\epsilon}\Gamma(\epsilon)
\left[\left({-g^{2}\over{2}}\right)^{\epsilon}\Gamma(\epsilon)
\left\{\left({-g^{2}\over{2}}\right)^{\epsilon}\Gamma(\epsilon)
-{1\over{\epsilon}}\right\}  
+ {\gamma+\log(2/g^2)\over{\epsilon}}\right]
\nonumber\\[5pt] 
&\,\overset{-g^{2}\to g^{2}e^{\mp i\pi}}{\longrightarrow}\,
-{8\over{3}}
\left(\gamma+\log{2\over{g^{2}}}\right)^3 \,+ \,{23\over{3}}
\pi^{2}\left( \gamma+\log{2\over{g^{2}}}\right) 
+{1\over{6}}\psi^{(2)}(1)
\nonumber\\
&\qquad\qquad\qquad\mp\,i\pi\Big[8\left(\gamma
+\log{2\over{g^{2}}}\right)^2 -{7\pi^2\over{3}}\Big]
\,+\, O\left(\frac{1}{\epsilon}\right)\,+\,O(\epsilon)\,.
\end{align}

The second configuration 
$[\bar{\mathcal{I}}\mathcal{I}\bar{\mathcal{I}}\mathcal{I}]$ 
gives identical result as the first one : 
$[\bar{\mathcal{I}}\mathcal{I}\bar{\mathcal{I}}\mathcal{I}]=
[\mathcal{I}\bar{\mathcal{I}}\mathcal{I}\bar{\mathcal{I}}]$. 

For the third 
$[\mathcal{I}\mathcal{I}\bar{\mathcal{I}}\bar{\mathcal{I}}]$ and
fourth  configurations 
$[\bar{\mathcal{I}}\bar{\mathcal{I}}\mathcal{I}\mathcal{I}]$, 
the two pairs of interactions are repulsive, but the other pair 
is attractive. 
Since these moduli-integrals are the same as that of 
$[\mathcal{I}\mathcal{I}\mathcal{I}\bar{\mathcal{I}}]$ in 
Eq.(\ref{eq:IIIbarI}), we obtain 
\begin{align}
&[\mathcal{I}\mathcal{I}\bar{\mathcal{I}}\bar{\mathcal{I}}] \xi^{-4}
= [\bar{\mathcal{I}}\bar{\mathcal{I}}\mathcal{I}\mathcal{I}]\xi^{-4}
\nonumber\\
&\,=\,
-{8\over{3}}
\left(\gamma+\log{2\over{g^{2}}}\right)^3 \,+ \,{1\over{3}}
\pi^{2}\left( \gamma+\log{2\over{g^{2}}}\right) 
+{1\over{6}}\psi^{(2)}(1)\,\mp\,i\pi\Big[{8\over{3}}
\left(\gamma+\log{2\over{g^{2}}}\right)^2 +{\pi^2\over{18}}\Big]\,.
\end{align}

For the fifth 
$[\mathcal{I}\bar{\mathcal{I}}\bar{\mathcal{I}}\mathcal{I}]$ and 
the sixth configurations 
$[\bar{\mathcal{I}}\mathcal{I}\mathcal{I}\bar{\mathcal{I}}]$, 
the two pairs of interactions are attractive, but the other pair 
is repulsive. 
Since these moduli-integrals are the same as 
$[\mathcal{I}\bar{\mathcal{I}}\mathcal{I}\mathcal{I}]$ in 
Eq.(\ref{eq:IbarIII}), we obtain 
\begin{align}
&[\mathcal{I}\bar{\mathcal{I}}\bar{\mathcal{I}}\mathcal{I}]\xi^{-4}
=[\bar{\mathcal{I}}\mathcal{I}\mathcal{I}\bar{\mathcal{I}}]\xi^{-4}
\nonumber\\
&\,=\,
-{8\over{3}}
\left(\gamma+\log{2\over{g^{2}}}\right)^3 \,+ \,3\pi^{2}\left( \gamma+\log{2\over{g^{2}}}\right) 
+{1\over{6}}\psi^{(2)}(1)\,\mp\,i\pi\Big[{16\over{3}}\left(\gamma+\log{2\over{g^{2}}}\right)^2 
-{7\pi^2\over{18}}\Big]\,.
\end{align}

The sum of all the six configurations gives 
\begin{align}
&([\mathcal{I}\bar{\mathcal{I}}\mathcal{I}\bar{\mathcal{I}}]+
[\bar{\mathcal{I}}\mathcal{I}\bar{\mathcal{I}}\mathcal{I}]+
[\mathcal{I}\mathcal{I}\bar{\mathcal{I}}\bar{\mathcal{I}}]+
[\bar{\mathcal{I}}\bar{\mathcal{I}}\mathcal{I}\mathcal{I}]+
[\mathcal{I}\bar{\mathcal{I}}\bar{\mathcal{I}}\mathcal{I}]+
[\bar{\mathcal{I}}\mathcal{I}\mathcal{I}\bar{\mathcal{I}}])\xi^{-4}
\nonumber\\
\,=\,
&-16
\left(\gamma+\log{2\over{g^{2}}}\right)^3 \,+ \,22\pi^{2}
\left( \gamma+\log{2\over{g^{2}}}\right) 
+\psi^{(2)}(1)\,\mp\,i\pi\Big[32\left(\gamma+\log{2\over{g^{2}}}\right)^2 
-{16\pi^2\over{3}}\Big]\,.
\end{align}

Therefore the contribution of two instantons and two 
anti-instantons to the energy eigenvalue of the lowest band 
is given by 

\begin{align}
&\triangle E
^{(2,2)} 
\nonumber\\
&\,=\,  \xi^{4}
\left[16
\left(\gamma+\log{2\over{g^{2}}}\right)^3 \,- \,22\pi^{2}\left( \gamma+\log{2\over{g^{2}}}\right) 
-\psi^{(2)}(1)\,\pm\,i\pi\Big[32\left(\gamma+\log{2\over{g^{2}}}\right)^2 
-{16\pi^2\over{3}}\Big]
\right]\,.
\label{IIIbarIbar}
\end{align}

According to the resurgence, the imaginary part of the 
two-instanton and two anti-instanton amplitudes should cancel 
the imaginary part of the Borel resummation of the divergent 
perturbation series. 
Using the dispersion relation, we can obtain the large-order 
behavior of the perturbation series corresponding to the two-instanton 
and two anti-instanton amplitude 
in Eq.~(\ref{IIIbarIbar}) 
\begin{align} 
a_{k} &\,\approx\, {1\over{\pi}}\int_{0}^{\infty} \,dg^2 
\,{{\rm Im}[\triangle E^{(2,2)}]
\over{(g^2)^{k+1}}}
\nonumber\\
&\,=\, -{32\over{\pi^2}}\int_{0}^{\infty} \,dg^2 
\,{ e^{-2/g^2}\over{(g^2)^{k+3}}}
\left(    (\gamma+\log 2)^{2}-{\pi^{2}\over{6}} \,
- \, 2(\gamma+\log 2)\log(g^2)\,+\, \log^{2}(g^2) 
\right)\,.
\label{4Iak}
\end{align}
This integral can be performed numerically, and
the first few results are 
$a_{k}\sim-2.4317, \,-7.0925, \,  -22.797,\,-82.273$ for $k=2,3,4,5$. 
Because of the factor $e^{-2/g^2}$ associated with the 
four-instanton action $2/g^2 = 4S_{I}$, the 
large-order ($k \gg 1$) behavior of $a_k$ in Eq.~(\ref{4Iak}) 
can be estimated with a constant $C$ as 
\begin{equation}
a_{k}  \,=\, \left({1\over{2}}\right)^{k+2} (k+1)!\, 
\left[ C\,+\,O(\log k /k)  \right] \,,
\end{equation}
exhibiting the $(1/2)^{k+2}$ factor besides the factorial 
growth $k!$. 
This large-order behavior corresponds to the singularity of 
the Borel transform $B_{\rm pert}(g^2t)$ at 
$t=2/g^2 = 4 S_{I}$ in the Borel plane 
\begin{equation}
B_{\rm pert}(g^2t)\,\approx\,C \sum_{k=0}^{\infty} 
\left({1\over{2}}\right)^{k+2}\,(k+1)!\, 
\left({(g^2t)^{k}\over{k!}}\right)
\,\to\, {C\over{(2-g^2t)^{2}}}\,.
\end{equation}
This is consistent with the known and expected results in 
quantum mechanics. 
We comment that we can also calculate the large-order perturbative 
behavior around the one-instanton and one--anti-instanton vacuum as 
$a_{k}\approx {1\over{\pi}}\int_{0}^{\infty} dg^{2} 
{\rm Im}[\triangle E^{(2,2)}\xi^{-2}]/(g^{2})^{k+1}$.


\subsection{General cases: $n$ instantons $+$ $m$ anti-instantons}

For general cases with $n$ instantons and $m$ anti-instantons,
we have ${}_{(n+m)} C_{n}$ configurations depending on the arrangement of constituents.
We can classify these ${}_{(n+m)} C_{n}$ configurations into classes based on
how many attractive and repulsive interactions are among the $n+m-1$ nearest-neighbor interactions.
(We can classify them into $2\times{\rm min}(n,m)$ classes for $n\not =m$, and $2n-1$ classes for $n=m$ with $n,m \geq 1$.)
We use the expression $N_{k,l}$ as the number of configurations including 
$k$ attractive and $l$ repulsive interactions satisfying $k+l=n+m-1$, 
and show examples of classification for $(n, m)=(3,2)$, $(n,m)=(3,3)$, $(n,m)=(4,4)$ and $(n,m)=(5,5)$ 
as following,
\begin{eqnarray}
{}_{(3+2)} C_{3}\,&=&\, N_{4,0}\,+ \,N_{3,1}\,+\, N_{2,2}\,+\, N_{1,3}
\nonumber\\
\,&=&\, 1\,+\, 4\,+\,3\,+\,2\,,
\\
{}_{(3+3)} C_{3}\,&=&\, N_{5,0}\,+ \,N_{4,1}\,+\, N_{3,2}\,+\, N_{2,3}\,+\,N_{1,4}
\nonumber\\
\,&=&\, 2\,+\, 4\,+\,8\,+\,4\,+\,2\,,
\\
{}_{(4+4)} C_{4}\,&=&\, N_{7,0}\,+ \,N_{6,1}\,+\, N_{5,2}\,+\, N_{4,3}\,+\,N_{3,4}\,+\,N_{2,5}\,+\,N_{1,6}
\nonumber\\
\,&=&\, 2\,+\, 6\,+\,18\,+\,18\,+\,18\,+\,6\,+\,2\,,
\\
{}_{(5+5)} C_{5}\,&=&\, N_{9,0}\,+ \,N_{8,1}\,+\, N_{7,2}\,+\, N_{6,3}\,+\,N_{5,4}\,+\,N_{4,5}
\,+\,N_{3,6}\,+\,N_{2,7}\,+\,N_{1,8}
\nonumber\\
\,&=&\, 2\,+\, 8\,+\,32\,+\,48\,+\,72\,+\,48\,+\,32\,+\,8\,+2\,.
\end{eqnarray}  
When calculating the amplitude of each in the $N_{k,l}$ configurations,
we have ${}_{(k+l)} C_{k}$ possible orderings of the multi moduli integrals
in relation to the subtraction scheme 
which is described in 
Sec.~\ref{sec:general} and \ref{sec:3instanton}.
After calculating these ${}_{(k+l)} C_{k}$ integrals,
we average the results and obtain a contribution from one of 
$N_{k,l}$ configurations.
As such, we calculate each contribution, and by summing up all 
the integral results,
we end up with the semi-classical amplitude of the configuration 
with $n$ instantons and $m$ anti-instantons.


\section{Comparison to uniform-WKB}
\label{sec:WKB}

\subsection{General formalism}

A systematic method called the uniform-WKB method has been 
applied extensively to study quantum mechanics for potentials 
with degenerate minima, such as the sine-Gordon quantum 
mechanics~\cite{Dunne:2013ada, Dunne:2014bca}. 
In this method, one combines the ansatz in Eq.(\ref{Dansatz}) 
with a global boundary condition to take account of the other 
minima than the one to do perturbative computation. 
This boundary condition enables one to go beyond 
the ordinary perturbative computation, and obtain all 
nonperturbative contributions leading to the resurgence. 
Defining the even $f_1$ and odd $f_2$ functions, and the 
Wronskian $\mathcal{W}$, 
\begin{align}
f_{1}(y) \,&=\, {1\over{\sqrt{u'(y)}}}\left[ D_{\nu}(u(y)/g)\,
+\, D_{\nu}(-u(y)/g) \right]\,,
\\
f_{2}(y) \,&=\, {1\over{\sqrt{u'(y)}}}\left[ D_{\nu}(u(y)/g)\,
-\, D_{\nu}(-u(y)/g) \right]\,,
\\
\mathcal{W}\,&=\, f_{1}f'_{2}-f'_{1}f_{2}\,
=\,-\sqrt{8\pi\over{g^{2}}}{1\over{\Gamma(-\nu)}}\,,
\end{align}
the following uniform-WKB boundary condition 
\cite{Dunne:2013ada, Dunne:2014bca} is imposed at the midpoint 
$y=\pi/2$ of the adjacent minima of the sine-Gordon potential  
\begin{equation}
\cos \theta \,=\, {1\over{\mathcal{W}}}
\left(
f'_{1}(\pi/2)f_{2}(\pi/2)\,+\, f'_{2}(\pi/2)f_{1}(\pi/2)
\right)\,,
\end{equation}
where parameter $\theta$ is the Bloch angle in 
Eq.(\ref{eq:bloch-angle}) to label the state within the energy band.
This boundary condition can be rewritten as 
\begin{equation}
{1\over{\Gamma(-\nu)}}\left({2\over{g^{2}}}\right)^{-\nu}\,\pm\,
{i\pi\over{2}}\left({2e^{\pm i\pi}\over{g^{2}}}\right)^{\nu}
{\xi^{2}H_{0}^{2}(\nu,g^{2})\over{\Gamma(1+\nu)}}
\,=\,
\xi H_{0}(\nu,g^{2})\cos \theta
\,,
\label{UWKBC}
\end{equation}
where $\xi$ is the instanton factor with the instanton action 
$S_{I}={1\over{2g^{2}}}$ in Eq.(\ref{eq:instanton_action}) 
\begin{equation}
\xi \equiv {1\over{\sqrt{\pi g^{2}}}}\exp\left[-S_{I} \right]\,,
\end{equation}
The function $H_{0}(\nu,g^{2})$ describing the perturbative 
fluctuations around the instanton is defined in terms of 
hypergeometric functions in Ref.~\cite{Dunne:2014bca}. 
We just show the necessary expression at the zeroth order 
of $g^{2}$ ($|g^{2}|\ll1$):
\begin{equation}
H_{0}(\nu, |g^2| \ll 1) = 2,\,\,\,\,\,\,\,{dH_{0}\over{d\nu}}=0,
\,\,\,\,\,\,\,\,{d^2H_{0}\over{d\nu^{2}}}=0,\cdot\cdot\cdot\,.
\end{equation}  
This boundary condition of the uniform-WKB approximation is also derived 
as the quantization condition in Ref.~\cite{ZinnJustin:2004ib}.

Since $\nu=E-1/2$ is the shift of energy eigenvalue $E$ from 
the harmonic oscillator ground state energy, 
the expansion of Eq.~(\ref{UWKBC}) up to the $\xi^4$ 
order with respect to $
\nu$ for the energy eigenvalue 
of the lowest band 
is 
\begin{align}
&\left[ -
\nu Q_{0} +
\nu
^{2}Q_{1} + 
\nu
^{3} Q_{2} +
\nu
^{4}Q_{3}\right]
\nonumber\\
&\,\,\,\,\,\,\,\,\,\,\,\,\,\,\,\pm\,{i\pi\xi^2 \over{2}}
\left[  H_{0}^2 +
\nu(2H_{0}H_{0}'+H_{0}^{2}Q_{1}^{\pm}) +
\nu
^{2}(2(H_{0}')^2 + 2H_{0}H_{0}''-H_{0}^{2}Q_{2}^{\pm})
\right]
\nonumber\\
&\,\,\,\,\,\,\,\,\,\,\,\,\,\,\,=\xi\cos\theta \left[H_{0} +
\nu H_{0}' + 
\nu
^{2}H_{0}''+
\nu
^{3}H_{0}'''
\right]\,,
\label{expansion}
\end{align}
with
\begin{align}
&Q_{0}=1
\nonumber\\
&Q_{1}^{\pm}\,=\, \gamma+\sigma_{\pm}
\nonumber\\
&Q_{2}^{\pm}\,=\, -{1\over{2}}(\gamma+\sigma_{\pm})^{2}+{\pi^2\over{12}}
\nonumber\\
&Q_{3}^{\pm}\,=\, {1\over{6}}(\gamma+\sigma_{\pm})^{3}
-{\pi^{2}\over{12}}(\gamma+\sigma_{\pm})
-{1\over{6}}\psi^{(2)}(1)\,,
\end{align}
with $\sigma_{\pm}=\gamma+\log(2/g^{2})\pm\pi$.
For $Q_{1}$, $Q_{2}$ and $Q_{3}$, we just replace 
$\sigma_{\pm}$ by $\sigma=\gamma +\log(2/g^{2})$.
We iteratively solve the expansion in Eq.~(\ref{expansion}) 
in the semi-classical limit ($|g^{2}|\ll 1,\,H_{0}' = H_{0}''=0$), 
then obtain the expression of $\nu$ as
\begin{align}
\nu\,=\,&-\xi H_{0} \cos \theta
\nonumber\\
&\,+\, \xi^{2}H_{0}^{2}\left[ Q_{1} \cos^{2}\theta \pm {i\pi\over{2}}  \right]
\nonumber\\
&\,+\,\xi^{3}H_{0}^{3}\left[ - (2Q_{1}^2 +Q_{2})\cos^{3}\theta
\,+  \left( {\pi^{2}\over{2}} \mp i\pi{3\over{2}}Q_{1} \right) \cos \theta
\right]
\nonumber\\
&\,+\,
\xi^{4}H_{0}^{4}\Big[
\left(5Q_{1}^{3}+5Q_{1}Q_{2}+Q_{3}
\right)\cos^{4}\theta
\nonumber\\
&\,\,\,\,\,\,\,\,+\,
\left(
-2\pi^{2}Q_{1} \pm i\pi({9\over{2}}Q_{1}^{2}+Q_{2}-{\pi^{2}\over{4}})
\right)
\cos^{2}\theta 
\,-\,
\left({\pi^{2}\over{2}}Q_{1}\pm i\pi{\pi^{2}\over{4}} \right)
\Big]\,.
\label{delnu}
\end{align}
For $|g^{2}|\ll 1 $, the lowest energy level $E(\nu, g^{2})$ 
is expressed as 
\begin{equation}
E(\nu, |g^{2}|\ll1)\,=\, {1\over{2}}\,+\,\nu\,.
\end{equation}

\subsection{Uniform-WKB results at each $\xi$ order}

From Eq.~(\ref{delnu}), we can calculate the contributions from each order of $\xi$ and $e^{i\theta}$.
The order of $\xi$ is interpreted as the number of instanton 
and anti-instanton constituents in the corresponding 
configurations while the order of $e^{i\theta}$ is interpreted 
as the instanton number.
In the followings, we show that the coefficients at each order 
of $\xi$ and $e^{i\theta}$ are completely
consistent with the contributions derived from the instanton calculations in the previous section.

We first start with the $\xi^{2}$ order, where we have two 
instanton constituents.
In the $\xi^{2}$ order, the terms proportional to $e^{2i\theta}$ are
\begin{equation}
{e^{2i\theta}\xi^{2}H_{0}^{2}\over{2^{2}}}\left[ \gamma\,+\,\log{2\over{g^{2}}}
 \right]\,.
\end{equation}
It agrees with the contribution from the two-instanton amplitude 
 in Eq.~(\ref{II}).
On the other hand, the terms proportional to $e^{0\times i\theta}=1$ are given by
\begin{equation}
{\xi^{2}H_{0}^{2}\over{2^{2}}}\times \left[ 2\left(\gamma\,+\,\log{2\over{g^{2}}}\right)\pm 2i\pi
 \right]\,.
\end{equation}
These terms composed of the real and imaginary 
parts precisely 
agree with 
the result of the one-instanton and one--anti-instanton 
amplitude in Eq.~(\ref{IIbar}).

We next look into the $\xi^{3}$ order.
In the $\xi^{3}$ order, the terms with $e^{3i\theta}$ are
\begin{equation}
-{e^{3i\theta}\xi^{3}H_{0}^{3}\over{2^{3}}}\left[ 
{3\over{2}}\left(\gamma+\log{2\over{g^{2}}}\right)^2 \,+ \,{\pi^2\over{12}} 
 \right]\,,
\end{equation}
which agree with the contributions from the three-instanton 
amplitude in Eq.~(\ref{III}).
The terms proportional to $e^{i\theta}$ are given by
\begin{equation}
-{e^{i\theta}\xi^{3}H_{0}^{3}\over{2^{3}}}\times 
{9\over{2}}\left[\left(\gamma+\log{2\over{g^{2}}}\right)^2 \,- \,{7\pi^2\over{18}} 
\pm {4\over{3}}i\pi\left(\gamma+\log{2\over{g^{2}}}\right)\right]\,.
\end{equation}
The real and imaginary parts in this result are identical to
the results from the two-instanton and one--anti-instanton 
amplitudes in Eq.~(\ref{IIIbar}).

We finally investigate the $\xi^{4}$ order.
In this order, the terms proportional to $e^{4i\theta}$ are given by
\begin{equation}
{e^{4i\theta}\xi^{4}H_{0}^{4}\over{2^{4}}} \times {8\over{3}}
\left[
\left(\gamma+\log{2\over{g^{2}}}\right)^3 \,+ \,{\pi^2\over{8}}\left( \gamma+\log{2\over{g^{2}}}\right) 
-{1\over{16}}\psi^{(2)}(1)
\right]\,,
\end{equation}
which agrees with the four-instanton contribution in Eq.~(\ref{IIII}).
Next, the terms with $e^{2i\theta}$ are 
\begin{equation}
{e^{2i\theta}\xi^{4}H_{0}^{4}\over{2^{4}}} \left[
 {32\over{3}}\left(\gamma+\log{2\over{g^{2}}}\right)^3 \,- \,{20\over{3}}\pi^{2}\left( \gamma+\log{2\over{g^{2}}}\right) 
-{2\over{3}}\psi^{(2)}(1)\,\pm\,i\pi\Big[16\left(\gamma+\log{2\over{g^{2}}}\right)^2 
-{2\pi^2\over{3}}\Big] \right]\,.
\end{equation}
These terms are the same ones as the contribution obtained 
from the three-instanton and one--anti-instanton amplitudes 
in Eq.~(\ref{IIIIbar}).
Finally, the terms proportional to $e^{0\times i\theta}=1$ 
are given by
\begin{equation}
{\xi^{4}H_{0}^{4}\over{2^{4}}}  \left[
16\left(\gamma+\log{2\over{g^{2}}}\right)^3 \,- \,22\pi^{2}\left( \gamma+\log{2\over{g^{2}}}\right) 
-\psi^{(2)}(1)\,\pm\,i\pi\Big[32\left(\gamma+\log{2\over{g^{2}}}\right)^2 
-{16\pi^2\over{3}}\Big] \right]\,,
\end{equation}
which 
precisely 
agree with the contribution from 
two-instanton and two--anti-instanton amplitudes 
shown in Eq.~(\ref{IIIbarIbar}).


\section{Neutral bions in the ${\mathbb C}P^{N-1}$ model on 
${\mathbb R}^1\times S^{1}$}
\label{sec:CPN}

In this section, we discuss 
fractional instantons and neutral bions 
in the ${\mathbb C}P^{N-1}$ model on 
${\mathbb R}^1\times S^{1}$ with the ${\mathbb Z}_{N}$ twisted 
boundary conditions in comparison to the sine-Gordom quantum 
mechanics. 
The ${\mathbb C}P^{N-1}$ model is described in terms of the 
$N$-component complex field $H$ satisfying the constraint 
$HH^\dagger=v^2$ with a constant $v^2$, whose euclidean 
Lagrangian is given by 
\begin{equation}
\mathcal{L}_{{\mathbb C}P^{N-1}} 
={\rm Tr}\left[{\cal D}_\mu H{\cal D}_\mu H^\dagger\right], 
\quad {\cal D}_\mu=\partial_\mu+i A_\mu, 
\quad 
A_\mu
= \frac{i}{v^2} 
{\partial}_\mu H H^{\dagger}. 
\label{eq:cpn_lag}
\end{equation}
The parameter $1/v$ serves as a coupling constant
\footnote{
The size of the ${\mathbb C}P^{N-1}$ manifold is denoted as 
$v$, and is related to the notation of Dunne and Unsal notation 
\cite{Dunne:2012ae} as $v^2=2/g_{\rm DU}^2$, $H^\dagger/v=n_{\rm DU}$. }
of the ${\mathbb C}P^{N-1}$ model, which is asymptotically free.  
The BPS solutions are obtained if a complex $N$-component vector 
$H_0$ called the moduli matrix is holomorphic (independent of 
$\bar z$) \cite{Eto:2006pg} 
\begin{equation}
H=S^{-1}H_0(z), \quad 
SS^\dagger =H_0H_0^\dagger/v^2, 
\label{eq:moduli_matrix}
\end{equation}
whereas the moduli matrix $H_0$ for anti-BPS 
solutions should be anti-holomorphic ($H_0(\bar z)$ independent 
of $z$). 
Here we have defined the complex coordinate 
$z \equiv x_1+ix_2$ with 
$-\infty<x_1<\infty, 0\le x_2 < L$.
The ${\mathbb Z}_N$ symmetric twisted boundary condition can be written as 
\begin{equation}
 H_0 (z+iL) = H_0(z) \,{\rm diag.} \left(1,\exp i{2\pi\over N}, 
\exp i{4\pi \over N},
\cdots \exp i{2\pi (N-1) \over N}\right).
\end{equation}
The moduli matrix for a fractional instanton as a BPS solution is given by 
\begin{equation}
H_0^{\cal I}(z)
 = \left(0, \cdots, 0,
\lambda e^{i\phi}e^{-2\pi z/(NL)} , 1, 0,
\cdot\cdot\cdot,0 \right),
\label{eq:1instanton}
\end{equation}
for which we find the topological charge to be fractional $1/N$ \cite{Eto:2004rz,Eto:2006mz,Eto:2006pg} 
(see also subsequent works \cite{Bruckmann:2007zh}). 
This solution can be regarded as a kink connecting two neighboring vacua \cite{Abraham:1992vb,Gauntlett:2000ib}, 
where real parameters $\lambda, \phi$ are their modulus, 
representing the relative phase $e^{i\phi}$ of the neighboring 
vacua and the position of the kink at 
$x_1=\frac{NL}{2\pi}\log\lambda$. 
The action and topological charge densities of the single 
fractional instanton solution has no dependence on the 
coordinate $x_2$ of the compactified dimension, thus at the first 
glance the situation seems quite similar to that of the 
(euclidean) 
quantum mechanics with the periodic potential, namely the 
sine-Gordon quantum mechanics.

\begin{figure}[htbp]
\begin{center}
 \includegraphics[width=0.2\textwidth]{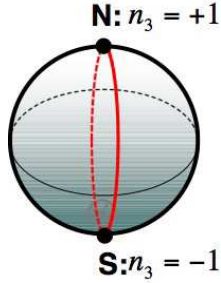}
\end{center}
\caption{The embedding of the sine-Gordon model into 
the ${\mathbb C}P^1$ model. 
The great circle $S^1$ denotes the target space of the 
sine-Gordon model.
N and S are two vacua $\Theta=0$ and $\Theta=\pi$ 
modulo $2\pi$ of the sine-Gordon model. 
}
\label{fig:SG-CP1}
\end{figure}

To clarify the similarities and differences 
with the sine-Gordon 
quantum mechanics, we consider a small compactification length 
limit ($L\to 0$) of the two-dimensional ${\mathbb C}P^{N-1}$ model. 
Let us first take the ${\mathbb C}P^{1}$ model for simplicity. 
By a stereographic projection, we can parametrize 
${\mathbb C}P^{1}=S^2$ target space in terms of two fields 
$\Theta(x_1,x_2)$ and $\Phi(x_1,x_2)$ corresponding to the 
zenith and azimuth angles of $S^2$ 
\begin{equation}
H(x_1,x_2)
 = v\left(
\cos\frac{\Theta(x_1,x_2)}{2} e^{i\frac{\Phi(x_1,x_2)}{2}}\,
, 
\sin\frac{\Theta(x_1,x_2)}{2} e^{-i\frac{\Phi(x_1,x_2)}{2}}
\right).
\label{eq:frac_inst}
\end{equation}
One should note that only the ratio of the first and second 
components $H_1/H_2$ is needed to parametrize $S^2$ as the 
inhomogeneous coordinate of the ${\mathbb C}P^{1}$ field space 
\begin{equation}
\frac{H_1(x_1,x_2)}{H_2(x_1,x_2)}
 =  \cot\left(\frac{\Theta(x_1,x_2)}{2}\right) e^{i\Phi(x_1,x_2)}.
\label{eq:inhom_coord}
\end{equation}
Choosing $N=2$ in Eq.(\ref{eq:cpn_lag}), 
the Lagrangian of the two-dimensional ${\mathbb C}P^{1}$ model 
can be rewritten in terms of $\Theta(x_1,x_2)$ and 
$\Phi(x_1,x_2)$ 
\begin{equation}
\mathcal{L}_{{\mathbb C}P^1} 
 = \frac{v^2}{4}\left[(\partial_\mu\Theta)^2 + 
(\sin\Theta)^2 \left(\partial_\mu\Phi\right)^2
\right].
\label{eq:cp1_lag}
\end{equation}
The Scherk-Schwarz dimensional reduction assumes the 
following ansatz of a particular $x_2$ dependence for the 
fields 
\begin{equation}
\Theta(x_1, x_2) = \Theta(x_1), 
\qquad \Phi(x_1, x_2)=\phi - \kappa x_2, 
\label{eq:SS_ansatz}
\end{equation}
with constants $\phi, \kappa$. 
One should note that we have restricted $\Theta$ to $x_2$ 
independent field and ignored the fluctuation of field $\Phi$. 
By inserting the ansatz (\ref{eq:SS_ansatz}) and integrating 
over $x_2$, we obtain the euclidean action $S$ as an integral of 
the euclidean Lagrangian $L$ over the euclidean time $t$ 
as 
\begin{equation}
S = \int dx_1 \int_0^{L}dx_2 \mathcal{L}_{{\mathbb C}P^1} 
= \int dx_1 \frac{Lv^2\kappa^2}{4}\left[
\left(\frac{1}{\kappa}\frac{d\Theta}{dx_1}\right)^2 + 
(\sin\Theta)^2 \right]
=\int dt L.
\label{eq:SSred_cp1_lag}
\end{equation}
We notice that the euclidean Lagrangian $L$ is identical to that 
of the sine-Gordon quantum mechanics in Eq.(\ref{eq:y_lagrangian}) 
with the identification 
$y=\Theta$, $t=\kappa x_1$, and the sine-Gordon coupling $g$ as 
\begin{equation}
g= \frac{1}{v \sqrt{2L\kappa}}. 
\label{eq:coupling_SSred}
\end{equation}
Since the ${\mathbb Z}_{N}$-twisted boundary condition 
($N=2$ for ${\mathbb C}P^{1}$ model) requires 
\begin{equation}
\kappa=\frac{2\pi}{LN}, 
\label{eq:SSphase}
\end{equation}
we finally find the coupling of the sine-Gordon quantum mechanics 
$g$ in terms of the parameters of the 
${\mathbb C}P^{N-1}$ model 
\begin{equation}
g=\frac{1}{2v}\sqrt{\frac{N}{\pi}}, 
\label{eq:coupling_SSred2}
\end{equation}
which is independent of the compactification scale $L$. 
In Fig.~\ref{fig:SG-CP1} we show 
how the sine-Gordon model is embedded into 
the ${\mathbb C}P^1$ model. 
The target space of the ${\mathbb C}P^1$ model 
is ${\mathbb C}P^1 \simeq S^2$ in which 
the $S^1$ as a great circle of $S^2$ 
is the target space of the sine-Gordon model.
The two fixed points of the action of the twisted boundary 
conditions,  the north and south poles denoted by N and S, 
respectively, corresponding to 
$\Theta = 0$ and $\pi$ (modulo $2\pi$) in the Lagrangian in 
Eq.~(\ref{eq:SSred_cp1_lag}),  
are two vacua in the reduced sine-Gordon model 
(see Fig.~\ref{fig:SG-CP1}): 
\begin{eqnarray}
 {\rm N}:&&  \quad \Theta = 0  \;{\rm mod} \; 2\pi\,,
 \nonumber\\
 {\rm S}:&&  \quad \Theta = \pi \; {\rm mod} \; 2\pi\,.
\end{eqnarray}

\begin{figure}[htbp]
\begin{center}
 \includegraphics[width=0.7\textwidth]{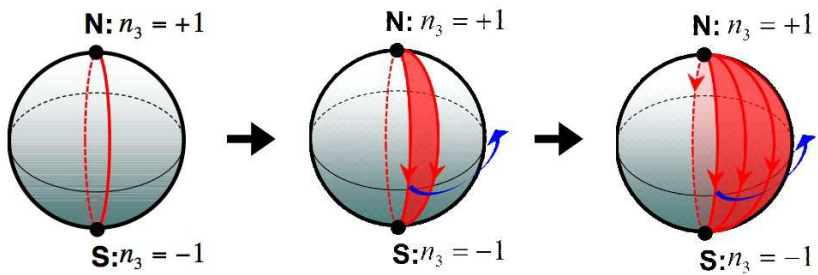}
  \end{center}
\caption{Fractional instanton configuration on $S^{2}$ in the 
reduced quantum mechanics is depicted.
It corresponds to a single line from the north to
the south pole which is rotated over the half of $S^{2}$ homogeneously.
The figure depicts the rotation of the line around the half sphere.
The red arrows denote paths depending on $x_2$ 
with a constant $x_1$, while the blue arrows denote 
the $x_1$ dependence of such paths.
}
\label{SG123}
\end{figure}

To see the correspondence between instantons 
in the ${\mathbb C}P^1$ model 
and reduced sine-Gordon model,
let us first examine the BPS one-instanton solution in 
Eq.(\ref{eq:1instanton}), whose inhomogeneous coordinate 
is given as 
\begin{equation}
\frac{H_1}{H_2}=\frac{H_{0,1}^{\mathcal I}}{H_{0,2}^{\mathcal I}}
 = \lambda e^{i\phi}e^{-\frac{\pi z}{L}}
=\lambda e^{-\frac{\pi x_1}{L}+i(\phi-\frac{\pi x_2}{L})} ,
\label{eq:ratio_1instanton}
\end{equation}
\begin{equation}
 \Phi(x_1,x_2)=\phi -\frac{\pi x_2}{L}, \quad 
\Theta(x_1, x_2)=2\arctan \left(e^{\frac{\pi}{L}
(x_1-\frac{L}{\pi}\log\lambda)}\right). 
\label{eq:inhomo_1instanton}
\end{equation}
We see that this solution is consistent with the assumption 
of the Scherk-Schwarz reduction in Eq.(\ref{eq:SS_ansatz}) 
with Eq.(\ref{eq:SSphase}). 
This solution gives the vertical path starting from N 
at $x_1=-\infty$ and reaching S at $x_1=\infty$ with 
 $\Phi=0$ when $x_2=0$. 
With varying from $x_2=0$ to $x_2=L$, this vertical path 
revolves in $\Phi$ to end up at $x_2=L$ with another vertical 
path connecting N (at $x_1=-\infty$) to S (at 
$x_1=\infty$) with $\Phi=\pi$. 
Thus the paths sweep precisely half 
of $S^2$, as illustrated in Fig.~\ref{SG123}. 
The solution $\Theta$ in Eq.~(\ref{eq:inhomo_1instanton}) is 
identical to the single instanton solution of the sine-Gordon 
quantum mechanics in Eq.(\ref{eq:instanton}). 
Therefore the fractional instanton solution of the 
${\mathbb C}P^{1}$ model is captured correctly by the 
sine-Gordon quantum mechanics as the instanton solution. 
The second homotopy group $\pi_2$ for the ${\mathbb C}P^1$ 
fractional instanton is 1/2, 
while the first homotopy group $\pi_1$ for the reduced 
sine-Gordon model is also 1/2 because it 
corresponds to a half orbit of the great circle.

\begin{figure}[htbp]
\begin{center}
\begin{tabular}{ccccc}
&
 \includegraphics[width=0.23\textwidth]{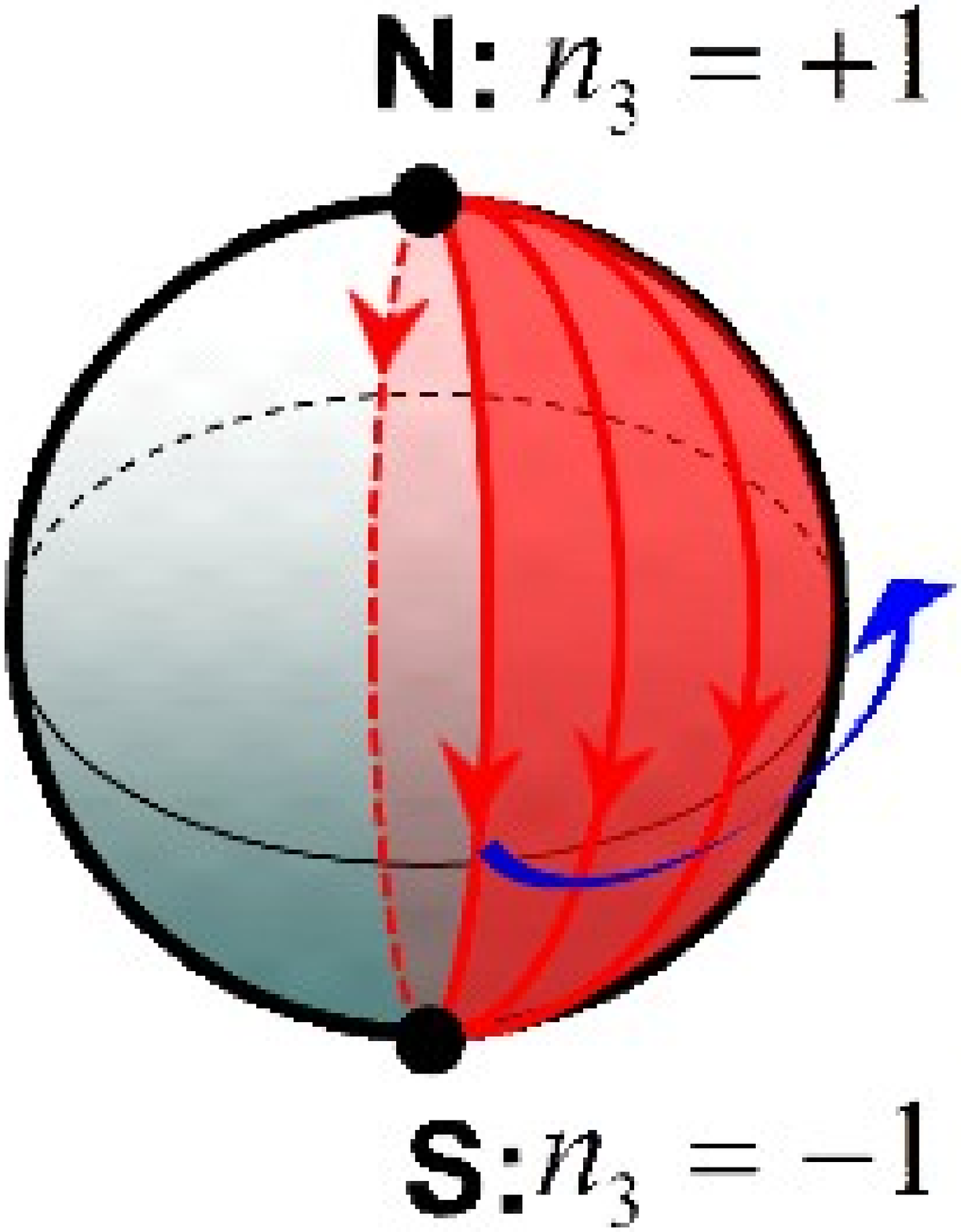}
& 
 \includegraphics[width=0.23\textwidth]{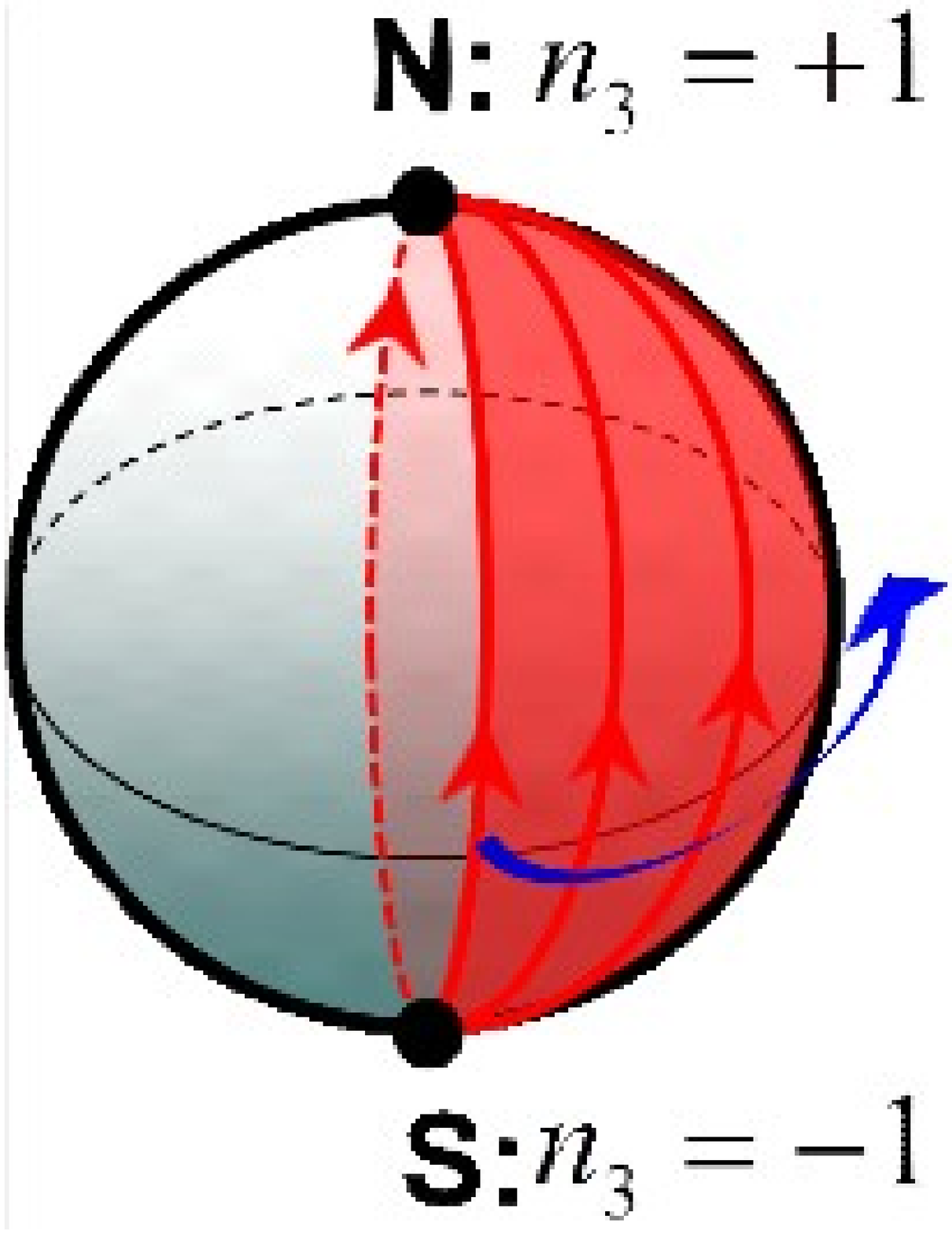}
& 
 \includegraphics[width=0.23\textwidth]{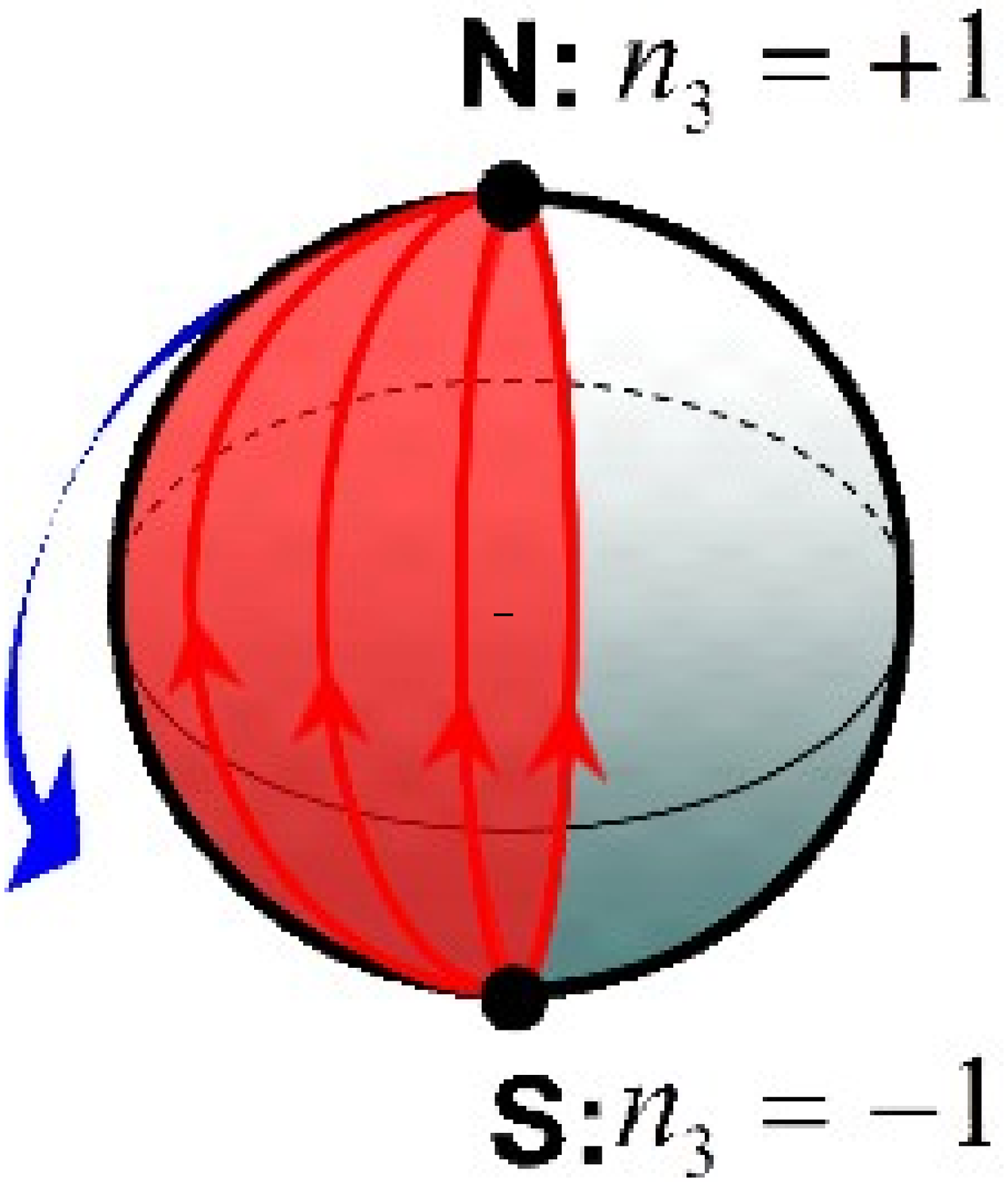}
& 
 \includegraphics[width=0.23\textwidth]{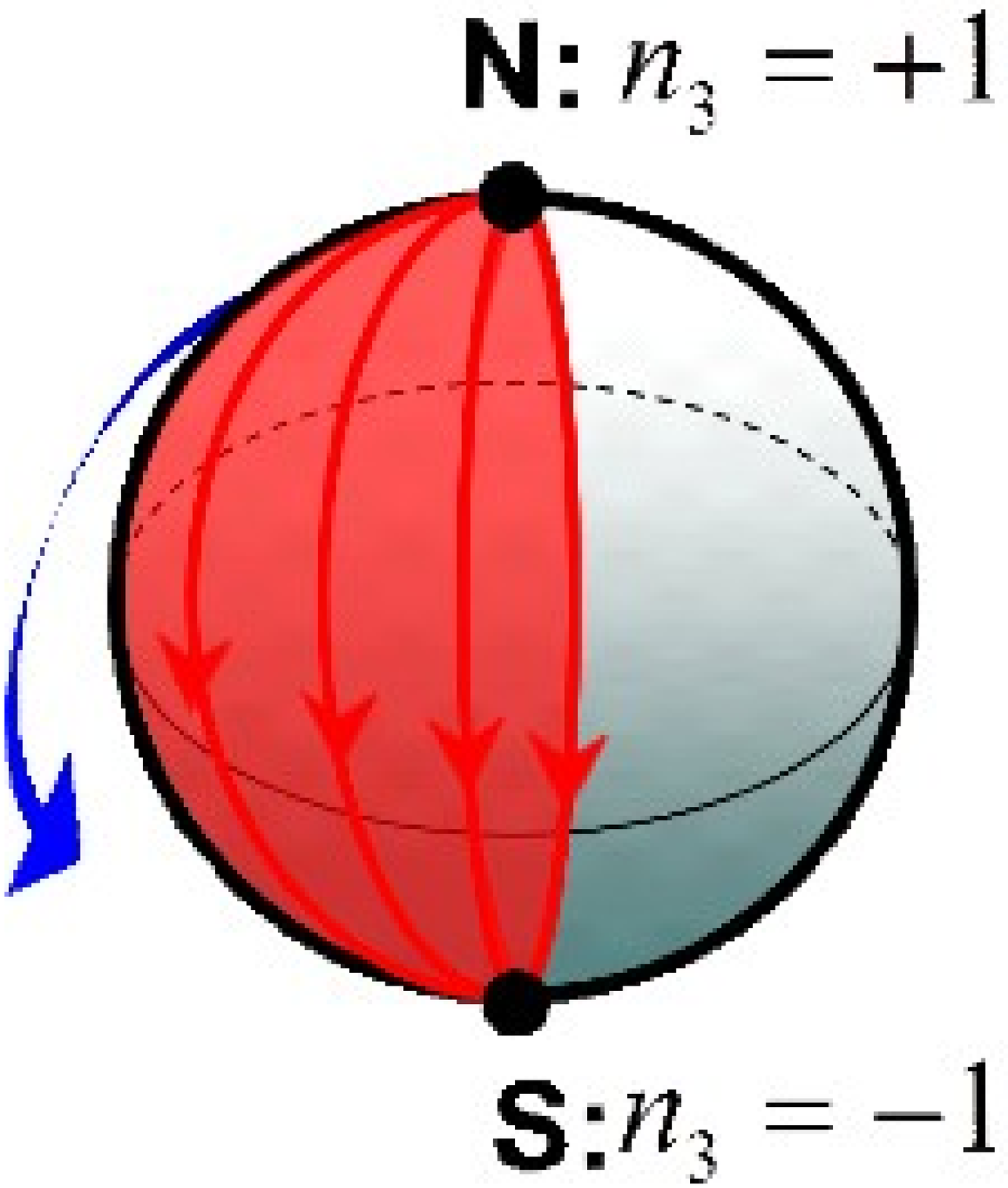}\\
$\pi_1$ & $+1/2$ & $-1/2$ & $+1/2$ & $-1/2$ \\
$\pi_2$ & $+1/2$ & $-1/2$ & $-1/2$ & $+1/2$ \\
& (a) & (b) & (c) & (d) 
\end{tabular}
  \end{center}
\caption{Fractional (anti-)instanton configurations  in the 
reduced quantum mechanics is depicted 
on the $S^{2}$ target space of the ${\mathbb C}P^1$ model.  
The first and second homotopy groups 
for instantons in the sine-Gordon model and 
${\mathbb C}P^1$ model are shown.
Configurations with positive values
of the second homotopy class $\pi_{2}$
are BPS while those with negative values are anti-BPS
both in the ${\mathbb C}P^1$ model and reduced sine-Gordon model.
Thus, (a) and (d) are BPS while
(b) and (c) are anti-BPS.
}
\label{fig:fractional}
\end{figure}

In Fig.~\ref{fig:fractional}, 
we classify how all possible (anti-)instanton configurations 
in the reduced sine-Gordon model correspond to 
the fractional instantons 
in the original ${\mathbb C}P^1$ model.
As shown in Eqs.~(\ref{eq:bps-bound}) and (\ref{eq:anti-bps-bound}),
a sine-Gordon soliton connecting from a vacuum labeled by even $n$
to a vacuum labeled by odd $n$
is BPS while that from odd $n$ to even $n$ is anti-BPS.
When the sine-Gordon quantum mechanics is embedded into
the ${\mathbb C}P^1$ model,
a soliton from N to S on $S^2$ is BPS while that from S to N is anti-BPS, consistent with BPS and anti-BPS kinks
in the ${\mathbb C}P^1$ model.
In other words, configurations with positive values
of the second homotopy class $\pi_{2}$
are BPS while those with negative values are anti-BPS
both in the ${\mathbb C}P^1$ model and reduced sine-Gordon model.
Thus, (a) and (d) in Fig.~\ref{fig:fractional} are BPS while
(b) and (c) are anti-BPS. 


Here we point out that 
the BPS solution of two fractional-instantons
in the ${\mathbb C}P^{1}$ model 
cannot be described in the reduced sine-Gordon model.
The BPS solution of two fractional-instantons, 
which contains 
the ordinary one-instanton BPS solution ($Q=1$) in the limit 
of small separation of two fractional instantons is given by 
\begin{equation}
H_0^{\cal II}(z)
 = \left(\lambda_1 e^{i\phi_1}e^{-\frac{\pi z}{L}} 
+\lambda_2 e^{i\phi_2}e^{\frac{\pi z}{L}} , 1 \right).
\end{equation}
This is a composite of (a) and (d) in Fig.\ref{fig:fractional}.
The inhomogeneous coordinate of ${\mathbb C}P^{1}$ now reads 
\begin{equation}
\frac{H_1^{\cal II}}{H_2^{\cal II}}=\frac{H_{0,1}^{\cal II}}{H_{0,2}^{\cal II}}
 = \lambda_1 e^{-\frac{\pi x_1}{L}} e^{i(\phi_1-\frac{\pi x_2}{L})}
+\lambda_2 e^{\frac{\pi x_1}{L}}e^{i(\phi_2+\frac{\pi x_2}{L})} ,
\label{eq:ratio_2instanton}
\end{equation}
which cannot satisfy the assumption (\ref{eq:SS_ansatz}) 
of the Scherk-Schwarz reduction. 
This is because 
in the reduced sine-Gordon model
a configuration starting from N and ending at S 
[(a) in Fig.~\ref{fig:fractional}] 
cannot be connected to another configuration starting from N and ending at S [(d) in Fig.~\ref{fig:fractional}].
The former can be connected only to a configuration 
starting from S and ending at N.
Therefore the BPS two fractional instanton solution cannot be 
described by the sine-Gordon quantum mechanics even in the limit 
of small $L$. 
More generally, all the BPS multi-fractional-instanton solutions 
are inconsistent with the Scherk-Schwarz reduction and hence the 
sine-Gordon quantum mechanics fails to capture them. 
This is consistent with the fact that configurations containing 
$n$-instantons ($n \ge 2$) are always non-BPS in the 
sine-Gordon quantum mechanics.

\begin{figure}[htbp]
\begin{center}
 \includegraphics[width=0.55\textwidth]{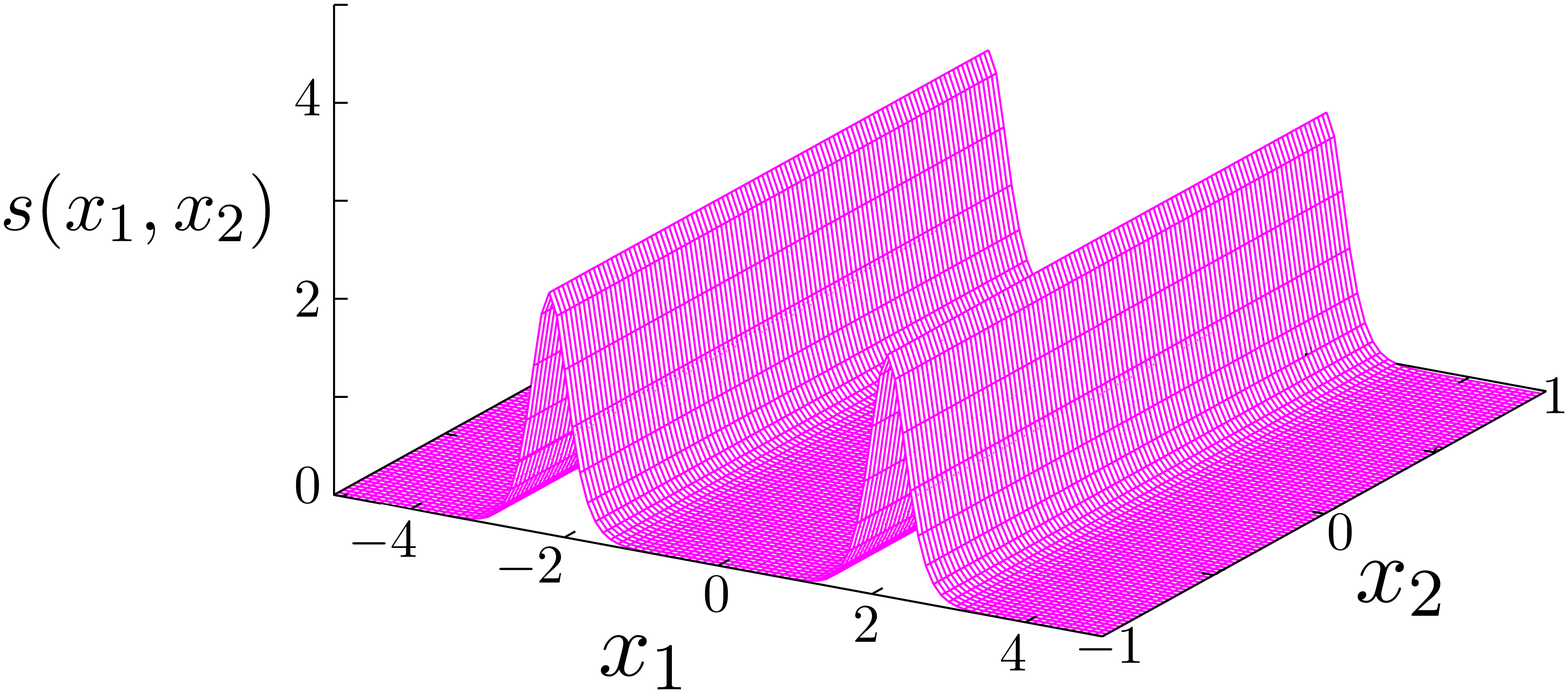}
  \includegraphics[width=0.4\textwidth]{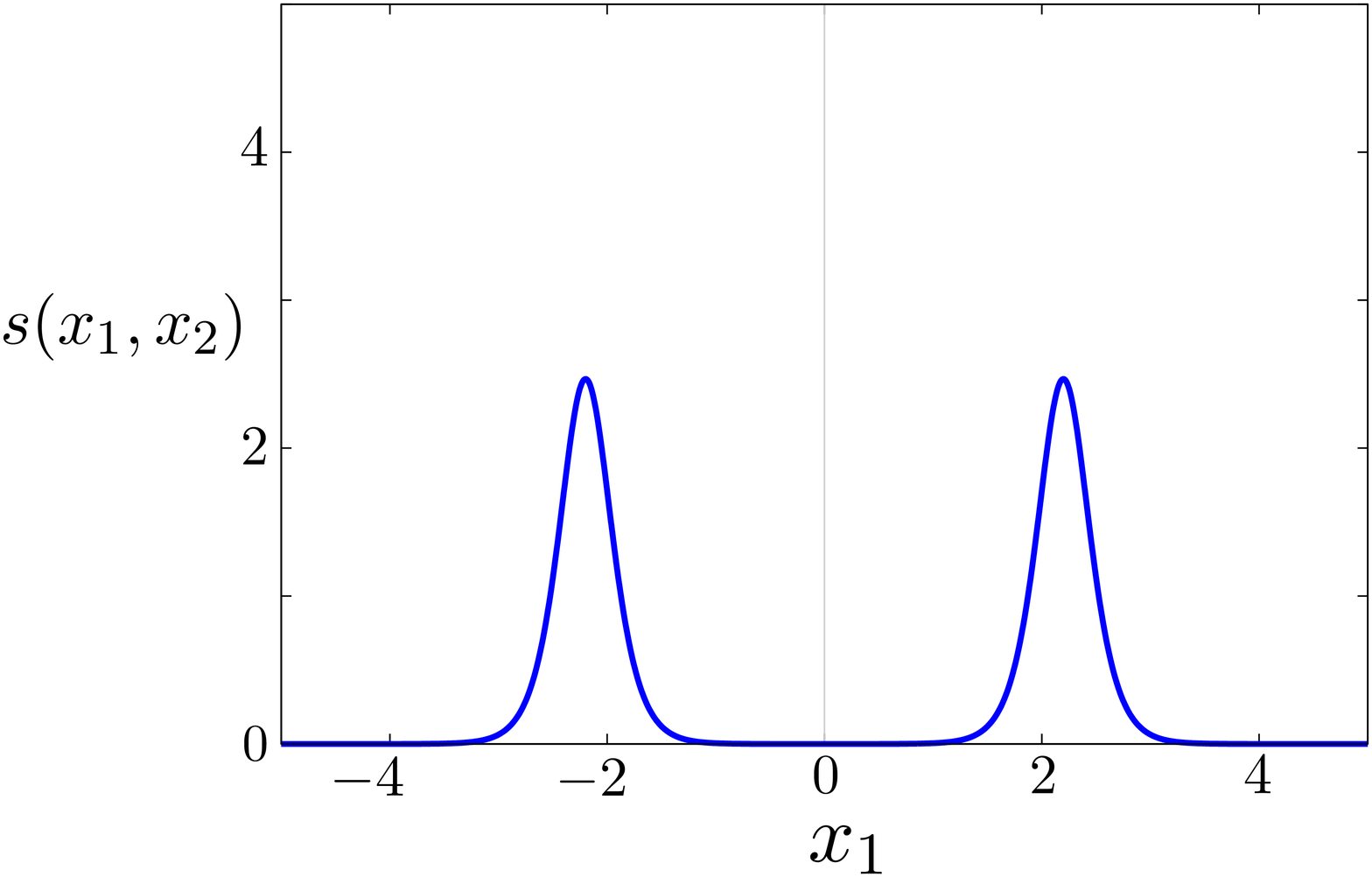}
\end{center}
\caption{
The euclidean action density $s(x_1,x_2)$ of 
neutral bion configurations for 
$\lambda_{1}=1/1000, \lambda_{2}=1/1000$ and $\phi = \pi/4$ 
in the ${\mathbb C}P^{1}$ model on ${\mathbb R}^1\times S^{1}$.
The same action density is depicted in two ways, 
as a function of $x_1,x_2$ (left) and $x_1$ (right).
There is no $x_2$ dependence in the action density, with $x_2$ 
being a coordinate of the compactied dimension.}
\label{025pi}
\end{figure}

\begin{figure}[htbp]
\begin{center}
  \includegraphics[width=0.8\textwidth]{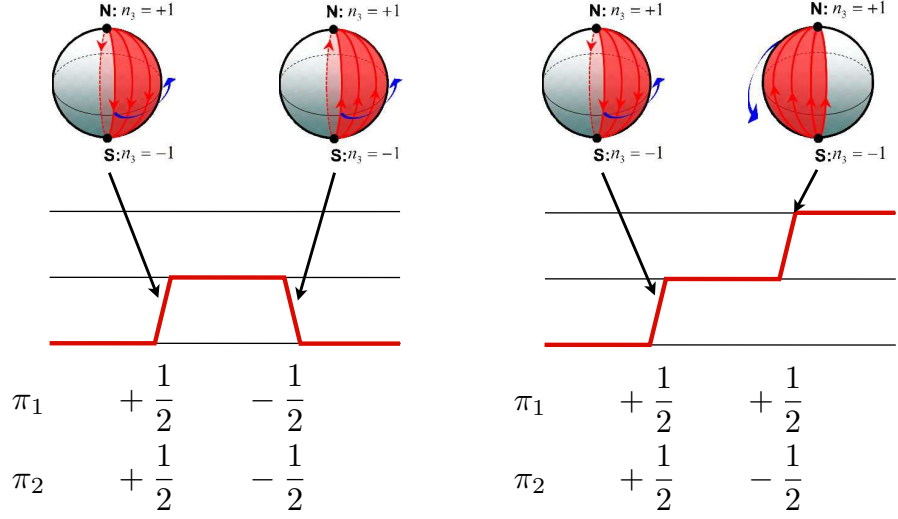} 
\end{center}
\caption{One fractional instanton and anti-instanton(left) and two fractional instantons (right)
in the ${\mathbb C}P^{1}$ model are depicted.}
\label{SGII}
\end{figure}

Next let us consider the non-BPS configuration 
of neutral bion, which is a composite of a fractional instanton 
and a fractional anti-instanton as depicted in Fig.~\ref{025pi}. 
We can write down an Ansatz for the moduli matrix $H_0$ 
\begin{equation}
H_0^{\cal I\bar I}(z, \bar z)
 = \left(\lambda_1 e^{i\phi_1}e^{-\frac{\pi z}{L}} 
+\lambda_2 e^{i\phi_2}e^{\frac{\pi \bar z}{L}} , 1 \right),
\label{eq:bion_cp1}
\end{equation}
which is guaranteed to become an exact solution of the field 
equation in the limit of large separation 
($-\frac{L}{\pi}\log(\lambda_1\lambda_2)\to \infty$) between 
constituents. 
The inhomogeneous coordinate of ${\mathbb C}P^{1}$ now reads 
\begin{equation}
\frac{H_1^{\cal I\bar I}}{H_2^{\cal I\bar I}}
=\frac{H_{0,1}^{\cal I\bar I}}{H_{0,2}^{\cal I\bar I}}
 = \left(\lambda_1 e^{-\frac{\pi x_1}{L}} 
+\lambda_2 e^{\frac{\pi x_1}{L}}e^{i(\phi_2-\phi_1)}\right)
e^{i(\phi_1-\frac{\pi x_2}{L})} ,
\label{eq:ratio_2instanton}
\end{equation}
which satisfy the assumption (\ref{eq:SS_ansatz}) 
of the Scherk-Schwarz reduction if and only if 
$e^{i(\phi_2-\phi_1)}=\mp 1$. 
In that case, we obtain the angular coordinate fields of $S^2$ as 
\begin{equation}
 \Phi(x_1,x_2)=\phi_1 -\frac{\pi x_2}{L}, \quad 
\cot\frac{\Theta(x_1, x_2)}{2}=
\lambda_1 e^{-\frac{\pi x_1}{L}} 
\mp \lambda_2 e^{\frac{\pi x_1}{L}}. 
\label{eq:inhomo_IbarI}
\end{equation}
This configuration starts from N at $x_1=-\infty$. 
For the upper sign, it goes through S at 
$x_1=-\frac{L}{\pi}\log(\lambda_1\lambda_2)$ and 
reaches to N with $\Theta =2\pi$ at $x_1=\infty$, namely 
it winds once around the great circle. 
The configuration represents the double instanton 
configuration of the sine-Gordon quantum mechanics 
as shown in Fig.\ref{fig:II}. 
For the lower sign, the configuration returns back to 
N with $\Theta=0$ at $x_1=\infty$ approaching but never reaching 
S at any point in $-\infty < x_1 < \infty$. 
This clearly represents the instanton and anti-instanton 
configuration $[\mathcal{I}\bar{\mathcal{I}}]$ of the sine-Gordon 
quantum mechanics, as shown in the left panel of Fig.~\ref{fig:IIbar}. 
The sine-Gordon quantum mechanics captures only field 
configurations that can cover the (part of) $S^2$ in the following 
specific fashion : When $x_1$ is varied with fixed $x_2$, $\Theta$ 
goes along the great circle (namely fixed $\Phi$), whereas $x_2$ 
variation with fixed $x_1$ makes a rotation of $\Phi$ with the 
constant velocity by an amount $\pi$ at fixed $\Theta$.  
The first homotopy group $\pi_1$ for the sine-Gordon model 
is one for the upper sign and zero for the lower sign, 
but the second homotopy group $\pi_2$ for 
the ${\mathbb C}P^1$ model is zero for the both cases.
In Fig.~\ref{SGII}, we show the instanton--anti-instanton 
and instanton-instanton configurations
in the sine-Gordon quantum mechanics 
corresponding to $e^{i(\phi_2-\phi_1)}=\mp 1$ in Eq.~(\ref{eq:ratio_2instanton}), 
and how the corresponding configuration of the 
${\mathbb C}P^{1}$ model in Eq.(\ref{eq:inhomo_IbarI}) cover 
the sphere $S^2$. 
Here, each of fractional instanton again corresponds to the 
line between the north and south poles sweeping around the 
half sphere.

On the other hand, generic configurations of the neutral bion 
of the ${\mathbb C}P^{1}$ model in Eq.~(\ref{eq:ratio_2instanton}) 
exhibit complicated ways of covering (part of) $S^2$ in terms of 
$\Theta, \Phi$. 
In terms of the $H(x_1, x_2)$ field, however, they are quite 
similar to those special configurations with $e^{i(\phi_2-\phi_1)}=\mp 1$ 
describable by the sine-Gordon quantum mechanics except for an 
important new feature : they have the relative phase 
 $e^{i(\phi_2-\phi_1)}$ between fractional instanton and 
anti-instanton as an additional moduli of the neutral bion 
configuration. 
This relative phase moduli introduces a striking physical 
effect into interactions between fractional instanton 
and anti-instanton constituents even in the limit of $L\to 0$, 
as we see immediately.

\begin{figure}[htbp]
\begin{center}
 \includegraphics[width=0.32\textwidth]{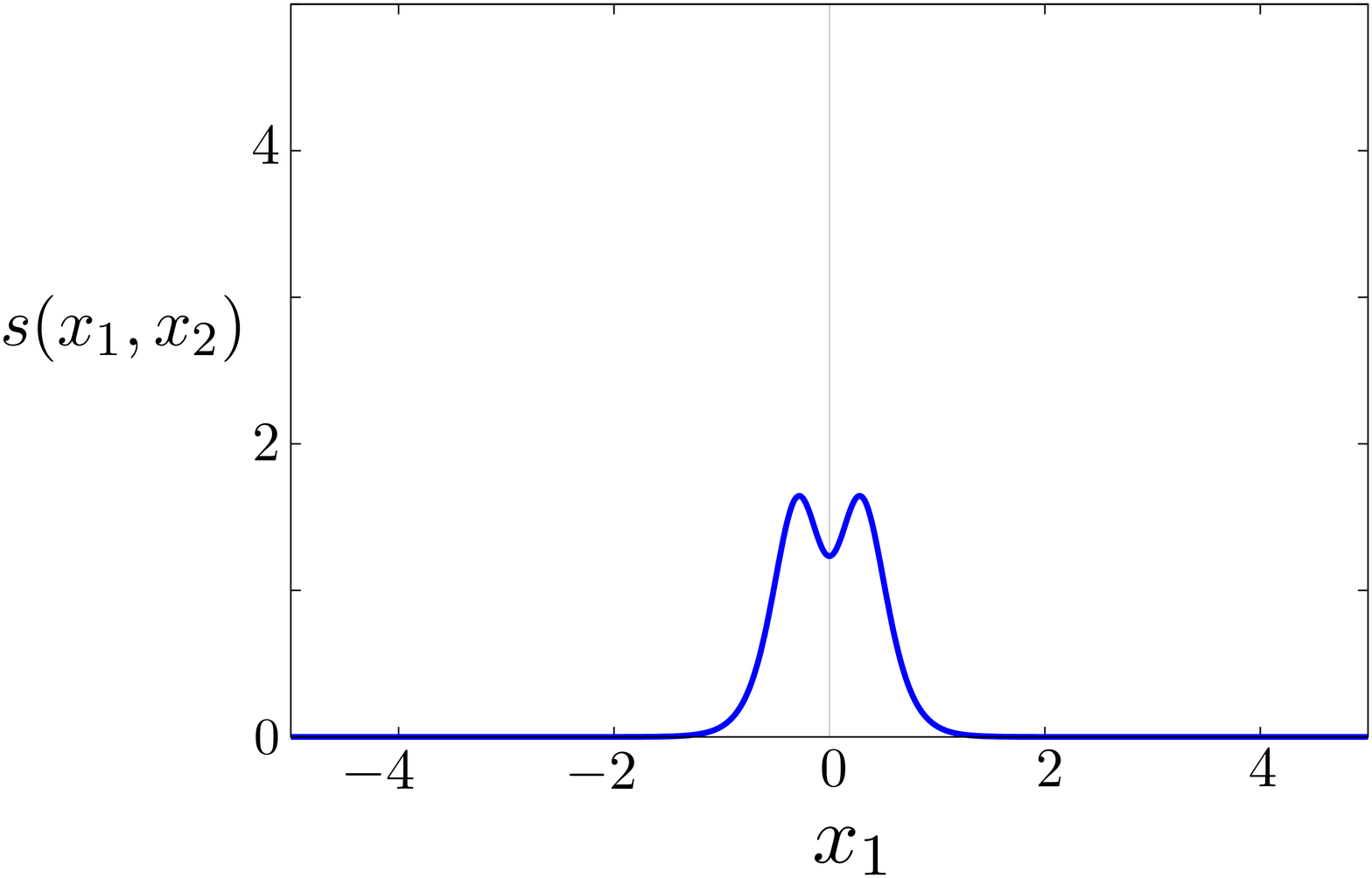}
  \includegraphics[width=0.32\textwidth]{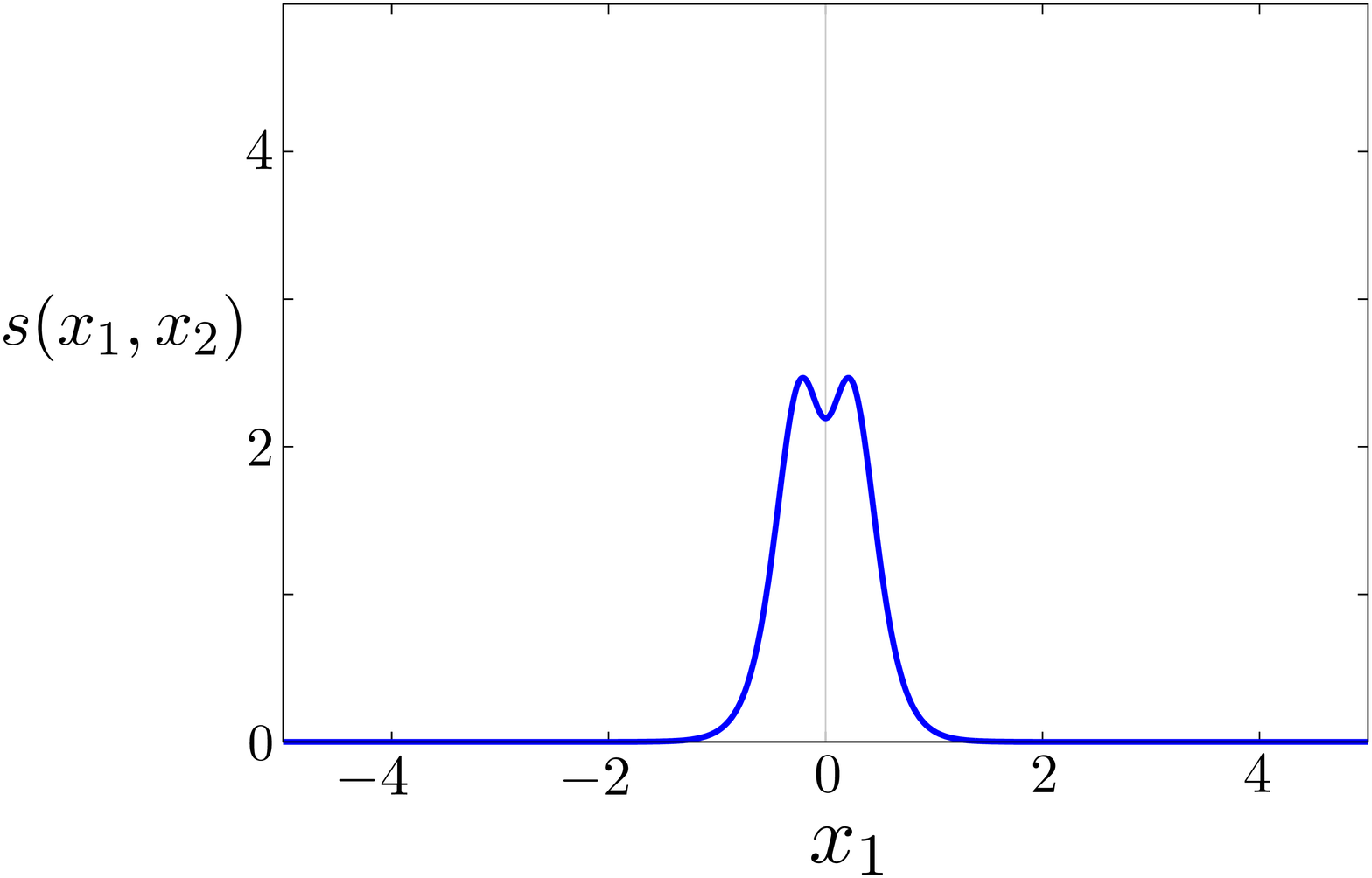}
   \includegraphics[width=0.32\textwidth]{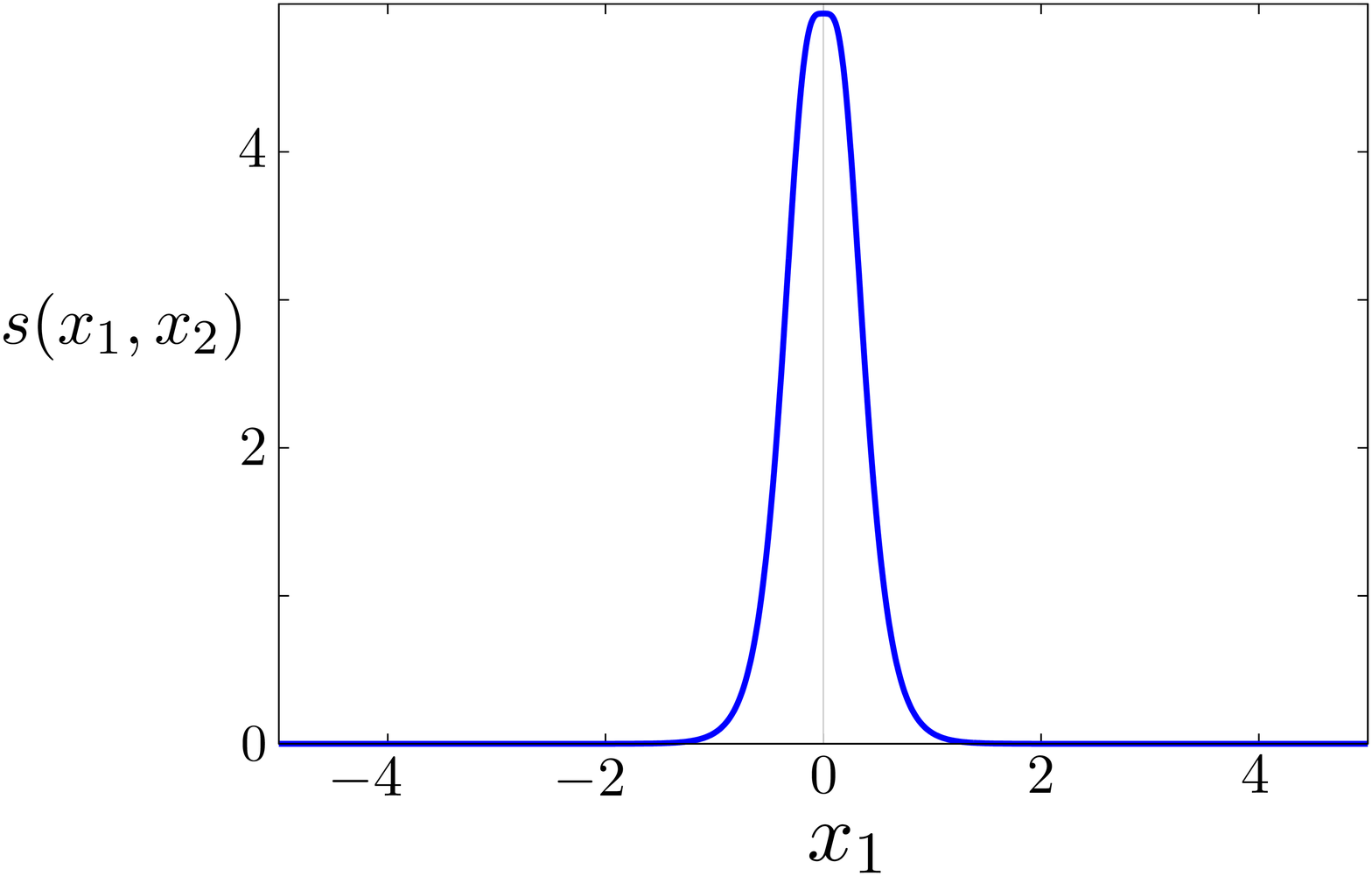}
  \end{center}
\caption{
The euclidean action density $s(x_1,x_2)$ of 
neutral bion configurations for $\lambda_{1}=1/2, 
\lambda_{2}=1/2$ are shown for several values of relative phase 
moduli $\phi$ : $\phi = 0$ (left),
$\pi/2$ (center) and $\pi$ (right). 
The total action increases 
as the relative phase $\phi$ is varied from $0$ to $\pi$, 
while configurations in a far-separated situation are almost 
the same for the three cases.  
The attractive force decreases in the region from $\phi=0$ 
to $\phi=\pi/2$, then becomes repulsive for $\pi/2<\phi<\pi$.}
\label{thetadep}
\end{figure}

The absence of the relative phase moduli between the instanton 
constituents in the bion configuration 
is a crucial drawback of the sine-Gordon quantam mechanical description. 
To clarify how crucial it is, we discuss the neutral bion 
configuration in the ${\mathbb C}P^{N-1}$ model 
on ${\mathbb R}^1\times S^{1}$ with the ${\mathbb Z}_{N}$ 
twisted boundary conditions. 
Generalizing Eq.(\ref{eq:bion_cp1}) in the ${\mathbb C}P^{1}$ 
model, the neutral bion configuration with the relative phase 
moduli between the two constituents is given by 
\cite{Misumi:2014jua} 
\begin{equation}
H_0^{\mathcal{I}\bar{\mathcal{I}}} \,=\, 
\left( \lambda_{1}\,e^{-{2\pi\over{NL}}z} 
+e^{i\phi}\lambda_{2} e^{{2\pi\over{NL}}\bar{z}}, 1, 0...,0   \right)\,,
\label{bion}
\end{equation}
where $\lambda_{1},\lambda_{2}$ are real parameters 
corresponding to positions of constituent fractional instanton 
and fractional anti-instanton, and $\phi$ is the relative phase.
Here, the separation between the fractional instanton and 
anti-instanton is given by $R=-(NL/2\pi)\log\lambda_{1}\lambda_{2}$. 
We study the interaction potential as a function of separation 
$R$ of constituents for the fixed relative phase $\phi$. 
As we vary this relative phase $\phi$ from $0$ to $2\pi$, 
we find that the interaction between the instanton and 
anti-instanton constituents is attractive for $0\le \phi < \pi/2$ 
and $3\pi/2 < \phi \le 2\pi$, 
whereas it is repulsive for $\pi/2 < \phi < 3\pi/2$. 
We compare the energy density for various relative phases 
in Fig.~\ref{thetadep}.
The interaction potential between fractional instanton and 
fractional anti-instanton is derived by both numerically 
\cite{Misumi:2014jua} and analytically as 
\begin{equation}
V[R] \,=\, -{4\kappa L v^2\cos \phi 
}\,
\exp(-\kappa R)\,,
\label{potNB}
\end{equation}
where we note $\kappa\equiv 2\pi/(NL)$ and the coupling 
of ${\mathbb C}P^{N-1}$ model is denoted as $1/v$.
We here follow the definition in our previous work \cite{Misumi:2014jua}
: Our coupling $1/v$ is related to the coupling $g_{\rm DU}$ 
adopted in \cite{Dunne:2012ae} as $v^{2}=2/g_{\rm DU}^{2}$. 
We note that the corresponding configuration in the sine-Gordon quantum 
mechanics has no relative phase moduli and the interaction 
potential between the instanton and anti-instanton constituents 
is given by Eq.~(\ref{pot2}) with the minus sign (attraction). 
Therefore the neutral bion contribution in the 
${\mathbb C}P^{N-1}$ model cannot be calculated 
in the reduced quantum mechanics correctly.

The neutral bion amplitude has been computed previously 
by ignoring the relative phase moduli ($\phi=0$). 
The result can be given in terms of the dimensionless 
separation variable $\tau=\kappa R$ 
\cite{Dunne:2012ae,Misumi:2014bsa} 
\begin{eqnarray}
[{\mathcal B}_{ii}]^{\phi=0} &=& C e^{-2S_{I}/N}I(v^2), 
\nonumber \\ 
I(v^2)&=&\int_0^\infty d\tau \exp\left(4\kappa L v^2e^{-\tau}
-\epsilon \tau\right)
=\int_0^\infty d\tau \exp\left(\frac{8\pi v^2}{N}e^{-\tau}
-\epsilon \tau\right)
\nonumber \\ 
&=&
-\left(\gamma +\log \left(\frac{8\pi v^2}{N}\right)\right) \mp i\pi, 
\label{eq:bion_no_phase}
\end{eqnarray}
where $S_{I}=2\pi v^2$ is the instanton action and $C$ denotes 
the numerical coefficient when the relative phase moduli is 
ignored ($\phi=0$). 
Based on the interaction potential we obtained in 
Eq.~(\ref{potNB}), we find that the corrected contributions 
of the neutral bion is given by taking account of the relative 
phase moduli as 
\begin{equation}
[{\mathcal B}_{ii}] = C e^{-2S_{I}/N}\int_0^{2\pi} d\phi \,I(v^2 \cos\phi). 
\end{equation}
Since the moduli integrals for $[0,\pi]$ and $[\pi,2\pi]$ are 
identical, we will double the result for $[0,\pi]$. 
We apply the BZJ prescription to the integral in the case of 
the attractive interaction for $[0,\pi/2]$, 
while we simply perform the integral for the repulsive case 
$[\pi/2, \pi]$.  
Using Eqs.(\ref{IIa}) and (\ref{eq:I-barI-1}), we obtain 
\begin{align}
\int_0^{2\pi} d&\phi\,I(v^2\cos\phi)
\nonumber\\
&=2\left[\int_0^{\pi/2} d\phi 
\left\{-\left(\gamma 
+\log \left(\frac{8\pi v^2}{N}\cos\phi\right)\right) \mp i\pi\right\}
+\int_{\pi/2}^{\pi} d\phi \left\{-\left(\gamma 
+\log \left(-\frac{8\pi v^2}{N}\cos\phi\right)\right) \right\}\right] 
\nonumber \\
&=
-2\pi\left(\gamma +\log \left(\frac{4\pi v^2}{N}\right) \right)
\,\mp\, i\pi^{2}. 
\end{align}
Thus we obtain the neutral bion contribution as 
\begin{equation}
\left[{\mathcal B}_{ii}\right]
\,=\, 
Ce^{-\frac{2S_{I}}{N}}
\left[-2\pi\left(\gamma +\log \left(\frac{4\pi v^2}{N}\right) \right)
\,\mp\, i\pi^{2} \right]
\,,
\label{E0NB}
\end{equation}
in contrast to the result in Eq.(\ref{eq:bion_no_phase}) 
where the relative phase is ignored. 
We have obtained quantitative corrections to that in 
Ref.~\cite{Dunne:2012ae}, 
by taking account of the effects of the integral of 
the relative phase moduli.
To establish the absolute magnitude of the instanton 
contribution definitely, it is desirable to examine the one-loop 
determinant around the fractional instanton and anti-instanton 
background for the two-dimensional ${\mathbb C}P^{N-1}$ model 
which has an explicit (weak) dependence on $x_2$. 
We consider these as future works.

Before closing this section, we 
note an alternative 
possibility to relate the sine-Gordon quantum mechanics 
and the ${\mathbb C}P^{N-1}$ model without compactification.
Even if the compactification length is nonzero, the ${\mathbb C}P^1$ 
model can be deformed so as to produce the sine-Gordon instanton 
solutions. 
To clarify this point, we move to 
the $O(3)$ non-linear sigma model 
equivalent to the ${\mathbb C}P^{1}$ model. 
In this model, we have three real scalar fields 
$n_{1}, n_{2}, n_{3}$ subjected by the constraint 
$(n_1)^2 + (n_2)^2 + (n_3)^2 = 1$. 
We introduce the two potential terms $V_{1}=m^{2}(1-n_{3}^{2})$ 
and $V_{2}=-\triangle m^{2} n_{1}$,
with a mass hierarchy $\triangle m^{2} \ll m^{2}$.
For  the parameter region $\triangle m^2 \sim 0$, 
the potential $V_1$ admits two discrete vacua 
$n_3 = \pm 1$ and 
a ${\mathbb C}P^1$ domain wall solution interpolating 
these two vacua \cite{Abraham:1992vb} 
with the width $m^{-1}$. 
Let us place it  perpendicularly to the $x_{2}$ direction.
With a small  $\triangle m^2 (\neq 0) \ll m^{2}$, the above 
vacua are shifted and the domain wall is deformed 
accordingly. 
In this case, the sine-Gordon model is induced 
on the ${\mathbb C}P^1$ domain wall 
as the effective theory \cite{Nitta:2012xq}. 
Then, an $O(3)$ ($\mathbb{C}P^{1}$) instanton is 
 restricted to the ${\mathbb C}P^1$ domain wall 
and becomes  a sine-Gordon instanton 
with the width $\triangle m^{-1}$ (in the $x_{1}$ direction), 
in the domain wall effective theory \cite{Nitta:2012xq}. 
This setting gives a precise relation between 
the ${\mathbb C}P^1$ model and the sine-Gordon quantum mechanics.  
By sending $m, \triangle m$ to zero, we may be able to 
investigate how the contribution of the relative complex phase 
disappears in the reduction process from the ${\mathbb C}P^1$ 
model to the sine-Gordon quantum mechanics. 
In the case of the ${\mathbb C}P^{N-1}$ model,  
multiple $N-1$ parallel domain walls in it \cite{Gauntlett:2000ib} 
play the role to connect the ${\mathbb C}P^{N-1}$ 
model instantons 
 and instantons in a sine-Gordon-like model.


\section{Summary and Discussion}
\label{sec:SD}

In this paper we have calculated multi-instanton contributions 
in the quantum mechanics with 
the sine-Gordon potential by integrating out separation 
moduli parameters between instantons and anti-instantons 
in the semi-classical limit ($|g^{2}|\ll 1$).
We have adopted an extended Bogomolnyi--Zinn-Justin prescription 
for multi-instanton configurations and the step-by-step subtraction 
scheme for divergent parts.
We show that the imaginary parts of the multi-instanton amplitudes 
cancel those arising from the Borel resummation of the 
large-order perturbation series. 
We verify that our results completely agree with those based 
on the uniform-WKB calculations \cite{Dunne:2013ada, Dunne:2014bca}
up to a four-instanton order.
We have also shown that the neutral bion amplitude in 
${\mathbb C}P^{N-1}$ model based on 
the potential including the relative phase moduli parameter 
gives corrections to the results obtained in the sine-Gordon 
quantum mechanics \cite{Dunne:2012ae}.

Our main results, that the multi-instanton amplitudes in the 
sine-Gordon quantum mechanics are consistent with the 
large-order behavior of the perturbative calculations, and 
are completely reproduced by the uniform-WKB boundary conditions, 
strongly indicate the following facts: 
the uniform-WKB boundary condition provides the correct link 
between the perturbation series around the perturbative vacuum 
and non-perturbative instanton effects, and furthermore the 
perturbative calculation {\it knows} non-perturbative aspects 
of the quantum system in the first place. 
We can take the result as an evidence at least in the quantum 
mechanics, in favor of the resurgence conjecture, which gives 
an unambiguous and self-consistent definition of the quantum 
theory \cite{ZinnJustin:1982td, ZinnJustin:1983nr, ZinnJustin:2004ib, 
 ZinnJustin:2004cg, Jentschura:2010zza, Dunne:2013ada, Dunne:2014bca}.

As for the neutral bion in the ${\mathbb C}P^{N-1}$ model 
on ${\mathbb R}^1\times S^{1}$, we have calculated, for the 
first time, the neutral bion amplitude beyond the sine-Gordon 
quantum mechanics, with the moduli integral of the relative 
phase parameter, which does not exist in the sine-Gordon quantum 
mechanics \cite{Dunne:2012ae}. 
As a future study, we will perform perturbative calculations 
and multi-instanton calculations in the ${\mathbb C}P^{N-1}$ 
model with keeping the compactified-direction dependence, 
then we may be able to show that its large-order behavior of 
the perturbations originates in the imaginary part of the 
neutral-bion and bion-molecule amplitude {\it at the field 
theory level}.

We here make a comment on the relation of the configuration 
Eq.~(\ref{bion}) to the Lefschetz thimble \cite{LT}.
In the recent attempt to perform the path integral in the 
extended theory \cite{LT}, one complexifies the coupling 
constant (${\rm arg}[g^{2}]\not=0$) and the field variables,
then the standard path integral can be replaced by the integral 
along the steepest descent curves in the complex configuration 
space, or Lefschetz thimbles with the imaginary part of the 
action being constant ${\rm Im}[S] = {\rm const.}$.  
In terms of this method, the configuration Eq.~(\ref{bion}) 
corresponds to the special thimble for ${\rm arg}[g^{2}] = 0$, 
called the Stokes line, where the two critical points, in 
which fractional instantons are infinitely-separated or 
completely compressed, are directly connected by the configuration.
Although, for a general case ${\rm arg}[g^{2}]\not=0$, the two 
critical points should belong to two different Lefschetz thimbles, 
our study on the Stokes line ${\rm arg}[g^{2}] = 0$ could be 
a good starting point for investigating the general Lefschetz 
thimbles. Part of future work will be devoted to this study. 
We also note that the relation of Lefschetz thimbles and the topologically 
neutral configurations has been investigated 
in terms of the hidden topological angles in Ref.~\cite{Behtash:2015kna}.

We also comment that the cancellation of imaginary ambiguities in the sine-Gordon
quantum mechanics, or the resurgent structure in the system, can be understood as
a special case of the generic resummation structure, called ``median resummation" \cite{Aniceto:2013fka}.
It is expected that this resummation is applicable to 
any quantum theories including quantum mechanics and field theory.
Thus, we consider that it is intriguing to study the application of the median resummation to
the sine-Gordon quantum mechanics and verify its structure as a future work.



\begin{acknowledgments}
We are grateful to Mithat \"{U}nsal and Gerald Dunne for their 
interest and valuable comments and correspondences on their 
related work during the entire course of our study. 
T.\ M.\ and N.\ S.\ thank CERN theory 
institute 2014, ``Resurgence and Transseries in quantum, 
gauge and string theories" for the fruitful discussion 
and useful correspondence. 
This work  is  supported in part 
by  the Japan Society for the 
Promotion of Science (JSPS) 
Grant-in-Aid for Scientific Research
(KAKENHI) Grant Numbers 
(26800147 (T.\ M.),  25400268 (M.\ N.) and 
25400241 (N.\ S.)).
\end{acknowledgments}


\appendix
\section{Calculation of $G_{i}$ ($i=1,...,6$)}
\label{app1}

In this appendix we show the details of calculation of the 
functions $G_{i}(g^{2})$ ($i=1,2,3,4,5,6$). 
The functions $G_{1},G_{2},G_{3}$ are based on the three subtraction 
patterns of the integral in Eq.~(\ref{G123}).
They are given by
\begin{align}
G_{1}(g^2)
&\,\overset{|g^{2}|,|\tilde{g}^{2}| \ll 1}{\longrightarrow}\,
\left({g^{2}\over{2}}\right)^{\epsilon}\Gamma(\epsilon)
\left[\left({g^{2}\over{2}}\right)^{\epsilon}\Gamma(\epsilon)
\left[\left({-\tilde{g}^{2}\over{2}}\right)^{\epsilon}\Gamma(\epsilon)-{1\over{\epsilon}}\right]  
+ {\gamma+\log(2/(-\tilde{g}^2))\over{\epsilon}}\right]
\nonumber\\[5pt] 
&\,\overset{-\tilde{g}^{2}= \tilde{g}^{2} e^{\mp i\pi}}{\longrightarrow}\,
-{8\over{3}}
\left(\gamma+\log{2\over{g^{2}}}\right)^3 \,+ \,{7\over{6}}\pi^{2}\left( \gamma+\log{2\over{g^{2}}}\right) 
+{1\over{6}}\psi^{(2)}(1)
\nonumber\\
&\qquad\qquad\qquad\,\mp\,4i\pi\left(\gamma+\log{2\over{g^{2}}}\right)^2
\,+\, O\left(\frac{1}{\epsilon}\right)\,+\,O(\epsilon)\,,
\end{align}

\begin{align}
G_{2}(g^2)
&\,\overset{|g^{2}|,|\tilde{g}^{2}| \ll 1}{\longrightarrow}\,
\left({g^{2}\over{2}}\right)^{\epsilon}\Gamma(\epsilon)
\left[\left({-\tilde{g}^{2}\over{2}}\right)^{\epsilon}\Gamma(\epsilon)
\left[\left({g^{2}\over{2}}\right)^{\epsilon}\Gamma(\epsilon)-{1\over{\epsilon}}\right]  
+ {\gamma+\log(2/g^2)\over{\epsilon}}\right]
\nonumber\\[5pt] 
&\,\overset{-\tilde{g}^{2}= \tilde{g}^{2} e^{\mp i\pi}}{\longrightarrow}\,
-{8\over{3}}
\left(\gamma+\log{2\over{g^{2}}}\right)^3 \,+ \,{1\over{6}}\pi^{2}\left( \gamma+\log{2\over{g^{2}}}\right) 
+{1\over{6}}\psi^{(2)}(1)
\nonumber\\
&\qquad\qquad\qquad\,\mp\,i\pi\Big[{5\over{2}}\left(\gamma+\log{2\over{g^{2}}}\right)^2 +{\pi^2\over{12}}\Big]
\,+\, O\left(\frac{1}{\epsilon}\right)\,+\,O(\epsilon)\,,
\end{align}

\begin{align}
G_{3}(g^2)
&\,\overset{|g^{2}|,|\tilde{g}^{2}| \ll 1}{\longrightarrow}\,
\left({-\tilde{g}^{2}\over{2}}\right)^{\epsilon}\Gamma(\epsilon)
\left[\left({g^{2}\over{2}}\right)^{\epsilon}\Gamma(\epsilon)
\left[\left({g^{2}\over{2}}\right)^{\epsilon}\Gamma(\epsilon)-{1\over{\epsilon}}\right]  
+ {\gamma+\log(2/g^2)\over{\epsilon}}\right]
\nonumber\\[5pt] 
&\,\overset{-\tilde{g}^{2}= \tilde{g}^{2} e^{\mp i\pi}}{\longrightarrow}\,
-{8\over{3}}
\left(\gamma+\log{2\over{g^{2}}}\right)^3 \,- \,{1\over{3}}\pi^{2}\left( \gamma+\log{2\over{g^{2}}}\right) 
+{1\over{6}}\psi^{(2)}(1)
\nonumber\\
&\qquad\qquad\qquad\mp\,i\pi\Big[{3\over{2}}\left(\gamma+\log{2\over{g^{2}}}\right)^2 +{\pi^2\over{12}}\Big]
\,+\, O\left(\frac{1}{\epsilon}\right)\,+\,O(\epsilon)\,.
\end{align}

The functions $G_{4},G_{5},G_{6}$ are based on the three patterns 
of subtraction in the integral Eq.~(\ref{G456}).
They are given by

\begin{align}
G_{4}(g^2)
&\,\overset{|g^{2}|,|\tilde{g}^{2}| \ll 1}{\longrightarrow}\,
\left({-\tilde{g}^{2}\over{2}}\right)^{\epsilon}\Gamma(\epsilon)
\left[\left({-\tilde{g}^{2}\over{2}}\right)^{\epsilon}\Gamma(\epsilon)
\left[\left({g^{2}\over{2}}\right)^{\epsilon}\Gamma(\epsilon)-{1\over{\epsilon}}\right]  
+ {\gamma+\log(2/g^2)\over{\epsilon}}\right]
\nonumber\\[5pt] 
&\,\overset{-\tilde{g}^{2}= \tilde{g}^{2} e^{\mp i\pi}}{\longrightarrow}\,
-{8\over{3}}
\left(\gamma+\log{2\over{g^{2}}}\right)^3 \,+ \,{7\over{6}}\pi^{2}\left( \gamma+\log{2\over{g^{2}}}\right) 
+{1\over{6}}\psi^{(2)}(1)
\nonumber\\
&\qquad\qquad\qquad\,\mp\,i\pi\Big[4\left(\gamma+\log{2\over{g^{2}}}\right)^2 +{\pi^2\over{6}}\Big]
\,+\, O\left(\frac{1}{\epsilon}\right)\,+\,O(\epsilon)\,,
\end{align}

\begin{align}
G_{5}(g^2)
&\,\overset{|g^{2}|,|\tilde{g}^{2}| \ll 1}{\longrightarrow}\,
\left({-\tilde{g}^{2}\over{2}}\right)^{\epsilon}\Gamma(\epsilon)
\left[\left({g^{2}\over{2}}\right)^{\epsilon}\Gamma(\epsilon)
\left[\left({-\tilde{g}^{2}\over{2}}\right)^{\epsilon}\Gamma(\epsilon)-{1\over{\epsilon}}\right]  
+ {\gamma+\log(2/(-\tilde{g}^2))\over{\epsilon}}\right]
\nonumber\\[5pt] 
&\,\overset{-\tilde{g}^{2}= \tilde{g}^{2} e^{\mp i\pi}}{\longrightarrow}\,
-{8\over{3}}
\left(\gamma+\log{2\over{g^{2}}}\right)^3 \,+ \,{19\over{6}}\pi^{2}\left( \gamma+\log{2\over{g^{2}}}\right) 
+{1\over{6}}\psi^{(2)}(1)
\nonumber\\
&\qquad\qquad\qquad\,\mp\,i\pi\Big[{11\over{2}}
\left(\gamma+\log{2\over{g^{2}}}\right)^2 
-{5\pi^2\over{12}}\Big]\,+\, O\left(\frac{1}{\epsilon}\right)\,
+\,O(\epsilon)\,,
\end{align}

\begin{align}
G_{6}(g^2)
&\,\overset{|g^{2}|,|\tilde{g}^{2}| \ll 1}{\longrightarrow}\,
\left({g^{2}\over{2}}\right)^{\epsilon}\Gamma(\epsilon)
\left[\left({-\tilde{g}^{2}\over{2}}\right)^{\epsilon}\Gamma(\epsilon)
\left[\left({-\tilde{g}^{2}\over{2}}\right)^{\epsilon}\Gamma(\epsilon)-{1\over{\epsilon}}\right]  
+ {\gamma+\log(2/(-\tilde{g}^2))\over{\epsilon}}\right]
\nonumber\\[5pt] 
&\,\overset{-\tilde{g}^{2}= \tilde{g}^{2} e^{\mp i\pi}}{\longrightarrow}\,
-{8\over{3}}
\left(\gamma+\log{2\over{g^{2}}}\right)^3 \,+ \,{14\over{3}}\pi^{2}\left( \gamma+\log{2\over{g^{2}}}\right) 
+{1\over{6}}\psi^{(2)}(1)
\nonumber\\
&\qquad\qquad\qquad\,\mp\,i\pi\Big[{13\over{2}}\left(\gamma+\log{2\over{g^{2}}}\right)^2 -{11\pi^2\over{12}}\Big]
\,+\, O\left(\frac{1}{\epsilon}\right)\,+\,O(\epsilon)\,.
\end{align}
In the final expressions, we implicitly return $\tilde{g}^{2}$ to $g^{2}$.


\begin{thebibliography}{99}


\bibitem{Unsal:2007vu} 
  M.~\"{U}nsal,
  ``Abelian duality, confinement, and chiral symmetry breaking in QCD(adj),''
  Phys.\ Rev.\ Lett.\  {\bf 100}, 032005 (2008)
  [arXiv:0708.1772 [hep-th]].
  
\bibitem{Unsal:2007jx} 
  M.~\"{U}nsal,
  ``Magnetic bion condensation: A New mechanism of confinement and mass gap in four dimensions,''
  Phys.\ Rev.\ D {\bf 80}, 065001 (2009)
  [arXiv:0709.3269 [hep-th]].

\bibitem{Shifman:2008ja} 
  M.~Shifman and M.~\"{U}nsal,
  ``QCD-like Theories on R(3) x S(1): A Smooth Journey from Small to Large r(S(1)) with Double-Trace Deformations,''
  Phys.\ Rev.\ D {\bf 78}, 065004 (2008)
  [arXiv:0802.1232 [hep-th]].

\bibitem{Poppitz:2009uq} 
  E.~Poppitz and M.~\"{U}nsal,
  ``Conformality or confinement: (IR)relevance of topological excitations,''
  JHEP {\bf 0909}, 050 (2009)
  [arXiv:0906.5156 [hep-th]].

\bibitem{Anber:2011de} 
  M.~M.~Anber and E.~Poppitz,
  ``Microscopic Structure of Magnetic Bions,''
  JHEP {\bf 1106}, 136 (2011)
  [arXiv:1105.0940 [hep-th]].

\bibitem{Poppitz:2012sw} 
  E.~Poppitz, T.~Schaefer and M.~\"{U}nsal,
  ``Continuity, Deconfinement, and (Super) Yang-Mills Theory,''
  JHEP {\bf 1210}, 115 (2012)
  [arXiv:1205.0290 [hep-th]].

\bibitem{Argyres:2012vv} 
  P.~Argyres and M.~\"{U}nsal,
  ``A semiclassical realization of infrared renormalons,''
  Phys.\ Rev.\ Lett.\  {\bf 109}, 121601 (2012)
  [arXiv:1204.1661 [hep-th]].
  
  \bibitem{Argyres:2012ka} 
  P.~C.~Argyres and M.~\"{U}nsal,
  ``The semi-classical expansion and resurgence in gauge theories: new perturbative, instanton, bion, and renormalon effects,''
  JHEP {\bf 1208}, 063 (2012)
  [arXiv:1206.1890 [hep-th]].

\bibitem{Dunne:2012ae} 
  G.~V.~Dunne and M.~\"{U}nsal,
  ``Resurgence and Trans-series in Quantum Field Theory: The CP(N-1) Model,''
  JHEP {\bf 1211}, 170 (2012)
  [arXiv:1210.2423 [hep-th]].

\bibitem{Dunne:2012zk} 
  G.~V.~Dunne and M.~\"{U}nsal,
  ``Continuity and Resurgence: towards a continuum definition of the CP(N-1) model,''
  Phys.\ Rev.\ D {\bf 87}, 025015 (2013)
  [arXiv:1210.3646 [hep-th]].
  
\bibitem{Dabrowski:2013kba} 
  R.~Dabrowski and G.~V.~Dunne,
  ``Fractionalized Non-Self-Dual Solutions in the CP(N-1) Model,''
  Phys.\ Rev.\ D {\bf 88}, 025020 (2013)
  [arXiv:1306.0921 [hep-th]].
  
\bibitem{Dunne:2013ada} 
  G.~V.~Dunne and M.~\"{U}nsal,
  ``Generating Non-perturbative Physics from Perturbation Theory,''
  Phys.\ Rev.\ D {\bf 89}, 041701 (2014)
  [arXiv:1306.4405 [hep-th]].
  
\bibitem{Cherman:2013yfa} 
  A.~Cherman, D.~Dorigoni, G.~V.~Dunne and M.~\"{U}nsal,
  ``Resurgence in QFT: Unitons, Fractons and Renormalons in the Principal Chiral Model,''
  Phys.\ Rev.\ Lett.\  {\bf 112}, 021601 (2014)
  [arXiv:1308.0127 [hep-th]].
  
\bibitem{Basar:2013eka} 
  G.~Basar, G.~V.~Dunne and M.~\"{U}nsal,
  ``Resurgence theory, ghost-instantons, and analytic continuation of path integrals,''
  JHEP {\bf 1310}, 041 (2013)
  [arXiv:1308.1108 [hep-th]].

\bibitem{Dunne:2014bca} 
  G.~V.~Dunne and M.~\"{U}nsal,
  ``Uniform WKB, Multi-instantons, and Resurgent Trans-Series,''
  Phys.\ Rev.\ D {\bf 89}, 105009 (2014)
  [arXiv:1401.5202 [hep-th]].

\bibitem{Cherman:2014ofa} 
  A.~Cherman, D.~Dorigoni and M.~\"{U}nsal,
  ``Decoding perturbation theory using resurgence: Stokes phenomena, new saddle points and Lefschetz thimbles,''
  arXiv:1403.1277 [hep-th].

\bibitem{Behtash:2015kna} 
  A.~Behtash, T.~Sulejmanpasic, T.~Schaefer and M.~Unsal,
  ``Hidden topological angles and Lefschetz thimbles,''
  arXiv:1502.06624 [hep-th].

\bibitem{Bolognesi:2013tya} 
  S.~Bolognesi and W.~Zakrzewski,
  ``Clustering and decomposition for non BPS solutions of the $\mathbb{CP}^{N-1}$ models,''
  Phys.\ Rev.\ D {\bf 89}, 065013 (2014)
  [arXiv:1310.8247 [hep-th]].

\bibitem{Misumi:2014jua} 
  T.~Misumi, M.~Nitta and N.~Sakai,
  ``Neutral bions in the ${\mathbb C}P^{N-1}$ model,''
  JHEP {\bf 1406}, 164 (2014)
  [arXiv:1404.7225 [hep-th]];

  T.~Misumi, M.~Nitta and N.~Sakai,
  ``Neutral bions in the $\mathbb CP^{N-1}$ model for resurgence,''
  J.\ Phys.\ Conf.\ Ser.\  {\bf 597}, no. 1, 012060 (2015)
  [arXiv:1412.0861 [hep-th]].


\bibitem{Shermer:2014wxa} 
  S.~Shermer,
  ``Twisted CP(N-1) instanton projectors and the N-level quantum density matrix,''
  arXiv:1412.3185 [hep-th].
  

\bibitem{Misumi:2014raa} 
  T.~Misumi and T.~Kanazawa,
  ``Adjoint QCD on $\mathbb{R}^3\times S^1$ with twisted fermionic boundary conditions,''
  JHEP {\bf 1406}, 181 (2014)
  [arXiv:1405.3113 [hep-ph]].

\bibitem{Misumi:2014bsa} 
  T.~Misumi, M.~Nitta and N.~Sakai,
  ``Classifying bions in Grassmann sigma models and non-Abelian gauge theories by D-branes,''
  PTEP {\bf 2015}, 033B02 (2015)
  [arXiv:1409.3444 [hep-th]].

\bibitem{Nitta:2015tua} 
  M.~Nitta,
  ``Fractional instantons and bions in the principal chiral model on ${\mathbb R}^2\times S^1$ with twisted boundary conditions,''
  arXiv:1503.06336 [hep-th].

\bibitem{Nitta:2014vpa} 
  M.~Nitta,
  ``Fractional instantons and bions in the O$(N)$ model with twisted boundary conditions,''
  JHEP {\bf 1503}, 108 (2015)
  [arXiv:1412.7681 [hep-th]].

\bibitem{Dunne:2015ywa} 
  G.~V.~Dunne and M.~Unsal,
  ``Resurgence and Dynamics of O(N) and Grassmannian Sigma Models,''
  arXiv:1505.07803 [hep-th].

\bibitem{'tHooft:1977am} 
  G.~'t Hooft,
  ``Can We Make Sense Out of Quantum Chromodynamics?,''
  Subnucl.\ Ser.\  {\bf 15}, 943 (1979).

\bibitem{Fateev:1994ai} 
  V.~A.~Fateev, V.~A.~Kazakov and P.~B.~Wiegmann,
  ``Principal chiral field at large N,''
  Nucl.\ Phys.\ B {\bf 424}, 505 (1994)
  [hep-th/9403099].
  
\bibitem{Fateev:1994dp} 
  V.~A.~Fateev, P.~B.~Wiegmann and V.~A.~Kazakov,
  ``Large N chiral field in two-dimensions,''
  Phys.\ Rev.\ Lett.\  {\bf 73}, 1750 (1994).
    
\bibitem{Ec1}
J.~Ecalle, ``Les Fonctions Resurgentes", Vol.~I - III 
(Publ. Math. Orsay, 1981).

\bibitem{Marino:2007te} 
  M.~Marino, R.~Schiappa and M.~Weiss,
  ``Nonperturbative Effects and the Large-Order Behavior of Matrix Models and Topological Strings,''
  Commun.\ Num.\ Theor.\ Phys.\  {\bf 2}, 349 (2008)
  [arXiv:0711.1954 [hep-th]].

\bibitem{Marino:2008ya} 
  M.~Marino,
  ``Nonperturbative effects and nonperturbative definitions in matrix models and topological strings,''
  JHEP {\bf 0812}, 114 (2008)
  [arXiv:0805.3033 [hep-th]].

\bibitem{Marino:2008vx} 
  M.~Marino, R.~Schiappa and M.~Weiss,
  ``Multi-Instantons and Multi-Cuts,''
  J.\ Math.\ Phys.\  {\bf 50}, 052301 (2009)
  [arXiv:0809.2619 [hep-th]].

\bibitem{Pasquetti:2009jg} 
  S.~Pasquetti and R.~Schiappa,
  ``Borel and Stokes Nonperturbative Phenomena in Topological String Theory and c=1 Matrix Models,''
  Annales Henri Poincare {\bf 11}, 351 (2010)
  [arXiv:0907.4082 [hep-th]].

\bibitem{Drukker:2010nc} 
  N.~Drukker, M.~Marino and P.~Putrov,
  ``From weak to strong coupling in ABJM theory,''
  Commun.\ Math.\ Phys.\  {\bf 306}, 511 (2011)
  [arXiv:1007.3837 [hep-th]].

\bibitem{Aniceto:2011nu} 
  I.~Aniceto, R.~Schiappa and M.~Vonk,
  ``The Resurgence of Instantons in String Theory,''
  Commun.\ Num.\ Theor.\ Phys.\  {\bf 6}, 339 (2012)
  [arXiv:1106.5922 [hep-th]].

\bibitem{Marino:2012zq} 
  M.~Marino,
  ``Lectures on non-perturbative effects in large $N$ gauge theories, matrix models and strings,''
  Fortsch.\ Phys.\  {\bf 62}, 455 (2014)
  [arXiv:1206.6272 [hep-th]].

\bibitem{Hatsuda:2013gj} 
  Y.~Hatsuda, S.~Moriyama and K.~Okuyama,
  ``Instanton Bound States in ABJM Theory,''
  JHEP {\bf 1305}, 054 (2013)
  [arXiv:1301.5184 [hep-th]].

\bibitem{Schiappa:2013opa} 
  R.~Schiappa and R.~Vaz,
  ``The Resurgence of Instantons: Multi-Cut Stokes Phases and the Painleve II Equation,''
  Commun.\ Math.\ Phys.\  {\bf 330}, 655 (2014)
  [arXiv:1302.5138 [hep-th]].
  
\bibitem{Hatsuda:2013oxa} 
  Y.~Hatsuda, M.~Marino, S.~Moriyama and K.~Okuyama,
  ``Non-perturbative effects and the refined topological string,''
  JHEP {\bf 1409}, 168 (2014)
  [arXiv:1306.1734 [hep-th]].  

\bibitem{Aniceto:2013fka} 
  I.~Aniceto and R.~Schiappa,
  ``Nonperturbative Ambiguities and the Reality of Resurgent Transseries,''
  Commun.\ Math.\ Phys.\  {\bf 335}, no. 1, 183 (2015)
  [arXiv:1308.1115 [hep-th]].

\bibitem{Santamaria:2013rua} 
  R.~C.~Santamaria, J.~D.~Edelstein, R.~Schiappa and M.~Vonk,
  ``Resurgent Transseries and the Holomorphic Anomaly,''
  arXiv:1308.1695 [hep-th].

\bibitem{Kallen:2013qla} 
  J.~Kallen and M.~Marino,
  ``Instanton effects and quantum spectral curves,''
  arXiv:1308.6485 [hep-th].

\bibitem{Honda:2014ica} 
  M.~Honda and S.~Moriyama,
  ``Instanton Effects in Orbifold ABJM Theory,''
  JHEP {\bf 1408}, 091 (2014)
  [arXiv:1404.0676 [hep-th]].


\bibitem{Grassi:2014cla} 
  A.~Grassi, M.~Marino and S.~Zakany,
  ``Resumming the string perturbation series,''
  JHEP {\bf 1505}, 038 (2015)
  [arXiv:1405.4214 [hep-th]].

\bibitem{Sauzin}
 D. Sauzin, 
  ``Introduction to 1-summability and resurgence, "
  arXiv:1405.0356 [math.DS].

\bibitem{Kallen:2014lsa} 
  J.~Kallen,
  ``The spectral problem of the ABJ Fermi gas,''
  arXiv:1407.0625 [hep-th].

\bibitem{Couso-Santamaria:2014iia} 
  R.~Couso-Santamaria, J.~D.~Edelstein, R.~Schiappa and M.~Vonk,
  ``Resurgent Transseries and the Holomorphic Anomaly: Nonperturbative Closed Strings 
  in Local ${\mathbb{C}\mathbb{P}^2}$,''
  Commun.\ Math.\ Phys.\  {\bf 338}, no. 1, 285 (2015)
  [arXiv:1407.4821 [hep-th]].

\bibitem{Honda:2014bza} 
  M.~Honda,
  ``On Perturbation theory improved by Strong coupling expansion,''
  JHEP {\bf 1412}, 019 (2014)
  [arXiv:1408.2960 [hep-th]].

\bibitem{Aniceto:2014hoa} 
  I.~Aniceto, J.~G.~Russo and R.~Schiappa,
  ``Resurgent Analysis of Localizable Observables in Supersymmetric Gauge Theories,''
  JHEP {\bf 1503}, 172 (2015)
  [arXiv:1410.5834 [hep-th]].

\bibitem{Couso-Santamaria:2015wga} 
  R.~Couso-Santamaria, R.~Schiappa and R.~Vaz,
  ``Finite N from Resurgent Large N,''
  Annals Phys.\  {\bf 356}, 1 (2015)
  [arXiv:1501.01007 [hep-th]].

\bibitem{Honda:2015ewa} 
  M.~Honda and D.~P.~Jatkar,
  ``Interpolating function and Stokes Phenomena,''
  arXiv:1504.02276 [hep-th].

\bibitem{Hatsuda:2015owa} 
  Y.~Hatsuda and K.~Okuyama,
  ``Resummations and Non-Perturbative Corrections,''
  arXiv:1505.07460 [hep-th].
  
\bibitem{Aniceto:2015rua} 
  I.~Aniceto,
  ``The Resurgence of the Cusp Anomalous Dimension,''
  arXiv:1506.03388 [hep-th].

\bibitem{Dorigoni:2015dha} 
  D.~Dorigoni and Y.~Hatsuda,
  ``Resurgence of the Cusp Anomalous Dimension,''
  arXiv:1506.03763 [hep-th].

\bibitem{Bogomolny:1980ur} 
  E.~B.~Bogomolny,
  ``Calculation Of Instanton - Anti-instanton Contributions In Quantum Mechanics,''
  Phys.\ Lett.\ B {\bf 91}, 431 (1980).

\bibitem{ZinnJustin:1981dx} 
  J.~Zinn-Justin,
  ``Multi - Instanton Contributions in Quantum Mechanics,''
  Nucl.\ Phys.\ B {\bf 192}, 125 (1981).

\bibitem{ZinnJustin:1982td}
  J.~Zinn-Justin,
  ``Multi - Instanton Contributions in Quantum Mechanics. 2.,''
  Nucl.\ Phys.\ B {\bf 218} (1983) 333.

\bibitem{ZinnJustin:1983nr}
  J.~Zinn-Justin,
  ``Instantons in Quantum Mechanics: Numerical Evidence for a Conjecture,''
  J.\ Math.\ Phys.\  {\bf 25} (1984) 549.

\bibitem{ZinnJustin:2004ib} 
  J.~Zinn-Justin and U.~D.~Jentschura,
  ``Multi-instantons and exact results I: Conjectures, WKB expansions, and instanton interactions,''
  Annals Phys.\  {\bf 313}, 197 (2004)
  [quant-ph/0501136].

\bibitem{ZinnJustin:2004cg} 
  J.~Zinn-Justin and U.~D.~Jentschura,
  ``Multi-instantons and exact results II: Specific cases, higher-order effects, and numerical calculations,''
  Annals Phys.\  {\bf 313}, 269 (2004)
  [quant-ph/0501137].
  
\bibitem{Jentschura:2010zza} 
  U.~D.~Jentschura, A.~Surzhykov and J.~Zinn-Justin,
  ``Multi-instantons and exact results. III: Unification of even and odd anharmonic oscillators,''
  Annals Phys.\  {\bf 325}, 1135 (2010).

\bibitem{Jentschura:2011zza} 
  U.~D.~Jentschura and J.~Zinn-Justin,
  ``Multi-instantons and exact results. IV: Path integral formalism,''
  Annals Phys.\  {\bf 326}, 2186 (2011).

\bibitem{Escobar-Ruiz:2015nsa} 
  M.~A.~Escobar-Ruiz, E.~Shuryak and A.~V.~Turbiner,
  ``Three-loop Correction to the Instanton Density. I. The Quartic Double Well Potential,''
  arXiv:1501.03993 [hep-th];
  ``Three-loop Correction to the Instanton Density. II. The Sine-Gordon potential,''
  arXiv:1505.05115 [hep-th].

 \bibitem{Manton:2004tk} 
  N.~S.~Manton and P.~Sutcliffe,
  ``Topological solitons,''
  Cambridge, UK: Univ. Pr. (2004) 493 p
 
  \bibitem{alvarez}
G. \'Alvarez, 
``Langer-Cherry derivation of the multi-instanton expansion for the symmetric double well'',
J. Math. Phys. {\bf 45}, 3095 (2004).

  
  \bibitem{langer}
  R. E. Langer,  ``The Asymptotic Solutions of Certain Linear Ordinary Differential Equations of the Second Order'', Trans.  Am. Math. Soc. {\bf 36}, 90 (1934).

\bibitem{cherry}
T. M. Cherry, ``Expansions in terms of Parabolic Cylinder Functions'',
Proc.  Edinburgh Math. Soc. {\bf   8}, 50 (1948). 


\bibitem{millergood}
S. C. Miller and R. H. Good,
``A WKB-Type Approximation to the Schr\"odinger Equation'',
Phys. Rev. {\bf 91}, 174 (1953).

\bibitem{galindo}
A. Galindo and P. Pascual,
{\it  Quantum Mechanics}, Vol.  II (Springer, 1991).

 
\bibitem{ZinnJustin:1989mi} 
  J.~Zinn-Justin,
  ``Quantum field theory and critical phenomena,''
  Int.\ Ser.\ Monogr.\ Phys.\  {\bf 77}, 1 (1989). 


\bibitem{Eto:2004rz} 
  M.~Eto, Y.~Isozumi, M.~Nitta, K.~Ohashi and N.~Sakai,
  ``Instantons in the Higgs phase,''
  Phys.\ Rev.\ D {\bf 72}, 025011 (2005)
  [hep-th/0412048].


\bibitem{Eto:2006mz} 
  M.~Eto, T.~Fujimori, Y.~Isozumi, M.~Nitta, K.~Ohashi, K.~Ohta and N.~Sakai,
  ``Non-Abelian vortices on cylinder: Duality between vortices and walls,''
  Phys.\ Rev.\ D {\bf 73}, 085008 (2006)
  [hep-th/0601181]; 
  M.~Eto, T.~Fujimori, M.~Nitta, K.~Ohashi, K.~Ohta and N.~Sakai,
  ``Statistical mechanics of vortices from D-branes and T-duality,''
  Nucl.\ Phys.\ B {\bf 788}, 120 (2008)
  [hep-th/0703197].

\bibitem{Eto:2006pg} 
  M.~Eto, Y.~Isozumi, M.~Nitta, K.~Ohashi and N.~Sakai,
  ``Solitons in the Higgs phase: The Moduli matrix approach,''
  J.\ Phys.\ A {\bf 39}, R315 (2006)
  [hep-th/0602170].


\bibitem{Bruckmann:2007zh} 
  F.~Bruckmann,
  ``Instanton constituents in the O(3) model at finite temperature,''
  Phys.\ Rev.\ Lett.\  {\bf 100}, 051602 (2008)
  [arXiv:0707.0775 [hep-th]]; 
  W.~Brendel, F.~Bruckmann, L.~Janssen, A.~Wipf and C.~Wozar,
  ``Instanton constituents and fermionic zero modes in twisted CP**n models,''
  Phys.\ Lett.\ B {\bf 676}, 116 (2009)
  [arXiv:0902.2328 [hep-th]];
  D.~Harland,
  ``Kinks, chains, and loop groups in the CP**n sigma models,''
  J.\ Math.\ Phys.\  {\bf 50}, 122902 (2009)
  [arXiv:0902.2303 [hep-th]].
  F.~Bruckmann and T.~Sulejmanpasic,
  ``Nonlinear sigma models at nonzero chemical potential: breaking up instantons and the phase diagram,''
  arXiv:1408.2229 [hep-th].


\bibitem{Abraham:1992vb} 
  E.~R.~C.~Abraham and P.~K.~Townsend,
  ``Q kinks,''
  Phys.\ Lett.\ B {\bf 291}, 85 (1992);
  E.~R.~C.~Abraham and P.~K.~Townsend,
  ``More on Q kinks: A (1+1)-dimensional analog of dyons,''
  Phys.\ Lett.\ B {\bf 295}, 225 (1992);
  M.~Arai, M.~Naganuma, M.~Nitta and N.~Sakai,
  ``Manifest supersymmetry for BPS walls in N=2 nonlinear sigma models,''
  Nucl.\ Phys.\ B {\bf 652}, 35 (2003)
  [hep-th/0211103]; 
  M.~Arai, M.~Naganuma, M.~Nitta and N.~Sakai,
  ``BPS wall in N=2 SUSY nonlinear sigma model with Eguchi-Hanson manifold,''
  In *Arai, A. (ed.) et al.: A garden of quanta* 299-325
  [hep-th/0302028].
 
\bibitem{Gauntlett:2000ib} 
  J.~P.~Gauntlett, D.~Tong and P.~K.~Townsend,
  ``Multidomain walls in massive supersymmetric sigma models,''
  Phys.\ Rev.\ D {\bf 64}, 025010 (2001)
  [hep-th/0012178];
  Y.~Isozumi, M.~Nitta, K.~Ohashi and N.~Sakai,
  ``Construction of non-Abelian walls and their complete moduli space,''
  Phys.\ Rev.\ Lett.\  {\bf 93}, 161601 (2004)
  [hep-th/0404198];
  Y.~Isozumi, M.~Nitta, K.~Ohashi and N.~Sakai,
  ``Non-Abelian walls in supersymmetric gauge theories,''
  Phys.\ Rev.\ D {\bf 70}, 125014 (2004)
  [hep-th/0405194]; 
  M.~Eto, Y.~Isozumi, M.~Nitta, K.~Ohashi, K.~Ohta and N.~Sakai,
  ``D-brane construction for non-Abelian walls,''
  Phys.\ Rev.\ D {\bf 71}, 125006 (2005)
  [hep-th/0412024];
  M.~Eto, Y.~Isozumi, M.~Nitta, K.~Ohashi, K.~Ohta, N.~Sakai and Y.~Tachikawa,
  ``Global structure of moduli space for BPS walls,''
  Phys.\ Rev.\ D {\bf 71}, 105009 (2005)
  [hep-th/0503033].


\bibitem{Nitta:2012xq} 
  M.~Nitta,
  ``Josephson vortices and the Atiyah-Manton construction,''
  Phys.\ Rev.\ D {\bf 86}, 125004 (2012)
  [arXiv:1207.6958 [hep-th]];
  M.~Kobayashi and M.~Nitta,
  ``Sine-Gordon kinks on a domain wall ring,''
  Phys.\ Rev.\ D {\bf 87}, no. 8, 085003 (2013)
  [arXiv:1302.0989 [hep-th]].


\bibitem{LT}
Y.~Tanizaki,
  ``Lefschetz-thimble techniques for path integral of zero-dimensional $O(n)$ sigma models,''
  Phys.\ Rev.\ D {\bf 91}, no. 3, 036002 (2015)
  [arXiv:1412.1891 [hep-th]];
T.~Kanazawa and Y.~Tanizaki,
  ``Structure of Lefschetz thimbles in simple fermionic systems,''
  JHEP {\bf 1503}, 044 (2015)
  [arXiv:1412.2802 [hep-th]];
Y.~Tanizaki, H.~Nishimura and K.~Kashiwa,
  ``Evading the sign problem in the mean-field approximation through Lefschetz-thimble path integral,''
  Phys.\ Rev.\ D {\bf 91}, no. 10, 101701 (2015)
  [arXiv:1504.02979 [hep-th]].


\end{thebibliography}
\end{document}